\documentclass[aps, prd, twocolumn, amsmath, floats,floatfix, superscriptaddress, nofootinbib,
showpacs]{revtex4}

\usepackage{amssymb}
\usepackage{amsmath}
\usepackage{verbatim}
\usepackage{mathrsfs}
\usepackage{amsfonts}
\usepackage{latexsym}
\usepackage{epsfig}
\usepackage{color}
\usepackage{graphicx,subfigure}
\usepackage{units}
\usepackage{slashbox}
\usepackage{envmath}
\usepackage{natbib}
\begin{document}

%%%%%%%%%%%%%%%%%%
%%%   MACROS   %%%
%%%%%%%%%%%%%%%%%%

\definecolor{orange}{rgb}{0.9,0.45,0}

\newcommand{\re}{\mbox{Re}}
\newcommand{\im}{\mbox{Im}}
\newcommand{\tf}[1]{\textcolor{red}{TF: #1}}
\newcommand{\nsg}[1]{\textcolor{cyan}{#1}}
\newcommand{\ch}[1]{\textcolor{blue}{#1}}

\def\CovDev{D}
\def\Res{{\mathcal R}}
\def\Gammaflat{\hat \Gamma}
\def\metricflat{\hat \gamma}
\def\Dflat{\hat {\mathcal D}}
\def\part_n{\partial_\perp}

%=== Definition for abbreviations ===
\def\Lie{\mathcal{L}}
\def\A{\mathcal{X}}
\def\Aphi{\A_{\phi}}
\def\hAphi{\hat{\A}_{\phi}}
\def\E{\mathcal{E}}
\def\Ham{\mathcal{H}}
\def\M{\mathcal{M}}
\def\R{\mathcal{R}}
\def\p{\partial}

\def\hg{\hat{\gamma}}
\def\hA{\hat{A}}
\def\hD{\hat{D}}
\def\hE{\hat{E}}
\def\hR{\hat{R}}
\def\hcA{\hat{\mathcal{A}}}
\def\hDelt{\hat{\triangle}}

\def\na{\nabla}
\def\dif{{\rm{d}}}
\def\non{\nonumber}
\newcommand{\erf}{\textrm{erf}}
%====================================

\renewcommand{\t}{\times}

\long\def\symbolfootnote[#1]#2{\begingroup%
\def\thefootnote{\fnsymbol{footnote}}\footnote[#1]{#2}\endgroup}

%%%%%%%%%%%%%%%%%
%%%   TITLE   %%%
%%%%%%%%%%%%%%%%%

\title{Head-on collisions and orbital mergers of Proca stars}
 
\author{Nicolas Sanchis-Gual}
\affiliation{Departamento de
  Astronom\'{\i}a y Astrof\'{\i}sica, Universitat de Val\`encia,
  Dr. Moliner 50, 46100, Burjassot (Val\`encia), Spain}

    \author{Carlos Herdeiro}
\affiliation{Departamento de F\'{\i}sica da Universidade de Aveiro and 
Centre for Research and Development in Mathematics and Applications (CIDMA), 
Campus de Santiago, 
3810-183 Aveiro, Portugal}
\affiliation{Centro de Astrof\'\i sica e Gravita\c c\~ao - CENTRA, Departamento de F\'\i sica,
Instituto Superior T\'ecnico - IST, Universidade de Lisboa - UL, Avenida
Rovisco Pais 1, 1049-001, Portugal}

\author{Jos\'e A. Font}
\affiliation{Departamento de
  Astronom\'{\i}a y Astrof\'{\i}sica, Universitat de Val\`encia,
  Dr. Moliner 50, 46100, Burjassot (Val\`encia), Spain}
\affiliation{Observatori Astron\`omic, Universitat de Val\`encia, C/ Catedr\'atico 
  Jos\'e Beltr\'an 2, 46980, Paterna (Val\`encia), Spain}

 \author{Eugen Radu}
\affiliation{Departamento de F\'{\i}sica da Universidade de Aveiro and 
Centre for Research and Development in Mathematics and Applications (CIDMA), 
Campus de Santiago, 
3810-183 Aveiro, Portugal}

\author{Fabrizio Di Giovanni}
\affiliation{Departamento de
  Astronom\'{\i}a y Astrof\'{\i}sica, Universitat de Val\`encia,
  Dr. Moliner 50, 46100, Burjassot (Val\`encia), Spain}

%%%%%%%%%%%%%%%%
%%%   DATE   %%%
%%%%%%%%%%%%%%%%

\date{November 2018}

%%%%%%%%%%%%%%%%%%%%
%%%   ABSTRACT   %%%
%%%%%%%%%%%%%%%%%%%%

\begin{abstract} 
Proca stars, \textit{aka} vector boson stars, are self-gravitating Bose-Einstein condensates obtained as numerical stationary solutions of the Einstein-(complex)-Proca system. These solitonic objects can achieve a compactness comparable to that of black holes, thus yielding an example of a black hole mimicker, which, moreover, can be both stable and form dynamically from generic initial data by the mechanism of gravitational cooling. In this paper we further explore the dynamical properties of these solitonic objects by performing both head-on collisions and orbital mergers of equal mass Proca stars, using fully non-linear numerical evolutions. For the head-on collisions, we show that the end point and the gravitational waveform from these collisions depends on the compactness of the Proca star. Proca stars with sufficiently small compactness collide emitting gravitational radiation and leaving a stable Proca star remnant. But more compact Proca stars collide to form a transient {\it hypermassive} Proca star, which ends up decaying into a black hole, albeit temporarily surrounded by Proca quasi-bound states. The unstable intermediate stage can leave an imprint in the waveform, making it distinct from that of a head-on collision of black holes. The final quasi-normal ringing matches that of Schwarzschild black hole, even though small deviations may occur, as a signature of sufficiently non-linear and long-lived Proca quasi-bound states. For the orbital mergers, the outcome also depends on the compactness of the stars. For the binaries with the most compact stars, the binary merger forms a Kerr black hole which retains part of the initial orbital angular momentum, being surrounded by a transient Proca field remnant; in cases with lower compactness, the binary merger forms a massive Proca star with angular momentum, but out of equilibrium. As in previous studies of (scalar) boson stars, the angular momentum of such objects appears to converge to zero as a final equilibrium state is approached.
\end{abstract}

%%%%%%%%%%%%%%%%
%%%   PACS   %%%
%%%%%%%%%%%%%%%%


%%%%%%%%%%%%%%%%%%%%%%
%%%   MAKE TITLE   %%%
%%%%%%%%%%%%%%%%%%%%%%

\maketitle

%{\bf \large Charged spherical boson stars in a cavity}
\vspace{0.8cm}

%%%%%%%%%%%%%%%%%%%%%%
\section{Introduction}
%%%%%%%%%%%%%%%%%%%%%%
%%%%%%%%%%%%%%%%%%%%%%
The recent spectacular detections of gravitational waves \cite{Abbott2016, Abbott:2016nmj, Abbott:2017vtc, Abbott:2017oio, Abbott:2017gyy,TheLIGOScientific:2017qsa} opened up a new window into the strong-field regime of gravity~\cite{Barack:2018yly}. Even though the data so far is well fitted by the expected physics -- $i.e.$ collision of Kerr black holes (BHs) or neutron stars -- it is important to understand if alternative, non-conventional models of compact objects can also fit the data ($i.e.$ the level of degeneracy) or how much these can be ruled out by current/future data - see~\cite{Johnson-McDaniel:2018uvs} for such a discussion. In the best-case scenario, obtaining gravitational waveforms of such non-conventional objects could lead to their future discovery, and the exciting prospect of unveiling new surprising physics via the gravitational-wave window.

Amongst such exotic models, self-gravitating solitons composed of complex (boson stars~\cite{Schunck:2003kk}) or real (oscillatons~\cite{Seidel:1991zh}) scalar fields are some of the most dynamically studied cases - see $e.g.$~\cite{Liebling:2012fv}. Dynamical studies include the generation of waveforms for head-on collisions of boson stars~\cite{palenzuela2007head,cardoso2016gravitational} and oscillatons~\cite{brito2015accretion,brito2016interaction,helfer2018gravitational} as well as orbital mergers of boson stars~\cite{Bezares:2017mzk,Palenzuela:2017kcg,dietrich2018full,clough2018axion,Bezares:2018qwa}. Even if the orbital mergers describe the astrophysically more likely scenario~\cite{palenzuela2008orbital,bezares2017final}, and constitute a possible mechanism to form spinning boson stars or Kerr BHs with scalar hair~\cite{Herdeiro:2014goa,Herdeiro:2015gia}, the head-on collisions represent a first step to compute the gravitational waveforms produced by these objects, allowing a simpler comparison to those produced in head-on collisions of BHs or neutron stars.

In has been recently found that a complex Proca field can form \textit{Proca stars} (PSs),  vector analogues of the scalar boson stars. PSs were constructed as static or stationary solutions of the Einstein-(complex)Proca system~\cite{Brito:2015pxa} - see also~\cite{Herdeiro:2016tmi,Garcia:2016ldc,Duarte:2016lig,Minamitsuji:2018kof} for generalizations. Similarly to their scalar cousins, they can be regarded as single frequency, macroscopic, self-gravitating (vector) Bose-Einstein  condensates. This frequency appears as a harmonic time dependence for the Proca potential and the  domain of existence of PSs is very similar to that of scalar boson stars -- see $e.g.$ Fig. 1 in~\cite{Cunha:2017wao} - except that the latter have a smaller maximal mass and a wider frequency range. 

Several dynamical aspects of the gravitational interaction of Proca fields have been addressed in recent years, using numerical studies. Long-lived, quasi-bound states of Proca fields around Schwarzschild BHs have been considered in~\cite{witek2013superradiant,Zilhao:2015tya} (see also~\cite{Rosa:2011my}). The superradiant instability of Kerr BHs was triggered by Proca fields in~\cite{east2017superradiant1,east2017superradiant2},  leading, in particular, to the formation of Kerr BH with Proca hair~\cite{Herdeiro:2016tmi,Herdeiro:2017phl}. In~\cite{sanchis2017numerical} fully non-linear evolutions of PSs were performed, to assess their stability. The simulations showed the existence of a stable branch (connecting the vacuum with 
the solution with maximal ADM mass) and an unstable branch. Solutions belonging to the latter may have different fates, depending on the initial perturbation and the sign of their binding energy. As their scalar cousins, with and without a self-interacting term~\cite{Seidel:1990jh,Balakrishna:1997ej,Guzman:2004jw,escorihuela2017quasistationary}, unstable solutions with positive binding energy migrate 
to the stable branch, whereas unstable solutions with a negative binding energy (excess energy) undergo fission; $i.e.$ they disperse entirely. Both cases can also collapse to form a Schwarzschild BH. More recently, additional numerical simulations~\cite{di2018dynamical} have shown that PS can also form dynamically from generic initial data describing a non-compact ``cloud" of the Proca field, through the so called gravitational cooling mechanism, first described in the 1990s in the context of scalar boson stars~\cite{seidel1994formation}.

In this work we shall continue the exploration of the dynamics of PSs by performing numerical evolutions describing both head-on collisions and orbital mergers of PS binaries and computing the gravitational radiation emitted. In the case of head-on collisions, the results we obtain parallel qualitatively those for head-on collisions of scalar boson (or even oscillaton) stars. For sufficiently small compactness, or equivalently, low mass in units of the Proca field mass -- see Fig.~\ref{fig1} below -- the collisions form a more massive, but still stable, PS, not a BH. The final star is, however, perturbed and in the timescale of our simulations it only partially relaxes to equilibrium. Sufficiently compact PSs, on the other hand, form a horizon when colliding, but only after an intermediate phase that could be described as a \textit{hypermassive} PS. This intermediate stage leaves an imprint in the waveform, making it distinct from that of a head-on collision of Schwarzschild BHs. After horizon formation the BH ringdown can be seen, which matches well that of a Schwarzschild BH. But in the cases where a larger Proca remnant remains outside the horizon, in the form of Proca quasi-bound states, we observe a difference with the BH ringdown. A similar observation was reported recently in the study of head-on collisions of oscillaton stars~\cite{helfer2018gravitational}. 

For the orbital mergers, similar results to the boson case are also found. The merger can lead to the formation of a Kerr BH wherein part of the initial orbital angular momentum of the configuration is deposited. In such cases, a Proca field  remnant can be seen outside the horizon. This remnant is a quasi-bound state that decays exponentially. Thus, with the initial parameters chosen we do not see the formation of a Kerr BH with Proca hair. On the other hand, we obtain the formation of a massive PS with angular momentum for small compactness. The solitonic remnant is, however, out of equilibrium, and the angular momentum decreases significantly during and after the merger, approaching  zero. In this process, the system emits gravitational waves continuously.

This paper is organized as follows. In Section~\ref{sec2} we present the equations of the Einstein-(complex)Proca model that will be used for 
the numerical evolutions. In Section~\ref{sec3} we present the initial data that will be used in 
our numerical evolutions. A brief description of the numerical techniques is 
given in Section~\ref{sec:numerics} and our results are presented in 
Section~\ref{results}. Final remarks are presented in Section~\ref{secconclusions}. 
A brief assessment of the numerical code is given in Appendix~\ref{appendix}.

%%%%%%%%%%%%%%%%%%%%%%
\section{Basic equations}
\label{sec2}
%%%%%%%%%%%%%%%%%%%%%%
%%%%%%%%%%%%%%%%%%%%%%
We shall investigate the dynamics of a complex Proca field by solving numerically the Proca equations coupled to the Einstein equations. The system is described by the action $\mathcal{S}=\int d^4x \sqrt{-g}\mathcal{L}$, where the Lagrangian density depends on the Proca potential $\mathcal{A}$, and field strength $\mathcal{F}=d\mathcal{A}$; it is given by:
\begin{equation}
\mathcal{L}=\frac{R}{16\pi 
G}-\frac{1}{4}\mathcal{F}_{\alpha\beta}\bar{\mathcal 
{F}}^{\alpha\beta}-\frac{1 } {2}\mu^2\mathcal{A}_\alpha\bar{\mathcal{A}}^\alpha 
\ ,
\label{model}
\end{equation}
where the bar denotes complex conjugation, $R$ is the Ricci scalar, $G$ is Newton's constant and $\mu$ is the Proca field mass. The stress-energy tensor of the Proca field reads
\begin{eqnarray} 
T_{ab}&=& -\mathcal{F}_{c(a}  \bar {\mathcal{F}}_{b)}^{\,\,c}-\frac{1}{4}g_{ab}\mathcal{F}_{cd}\bar{\mathcal{F}}^{cd} 
\nonumber \\
&+& \mu^2 \left[
\mathcal{A}_{(a}\bar{\mathcal{A}}_{b)}-\frac{1}{2}g_{ab}\mathcal{A}_c\bar{\mathcal{A}}^{c}
\right]\, .
\end{eqnarray}
Using the standard 3+1 split (see $e.g.$~\cite{sanchis2017numerical}  for more details) the Proca field is split into 3+1 quantities:
\begin{eqnarray}
\mathcal{A}_{\mu}&=&\mathcal{X}_{\mu}+n_{\mu}\mathcal{X}_{\phi},\\
\mathcal{X}_{i}&=&\gamma^{\mu}_{\,i}\mathcal{A}_{\mu},\\
\mathcal{X}_{\phi}&=&-n^{\mu}\mathcal{A}_{\mu},
\end{eqnarray}
where $\mathcal{X}_{i}$ is the vector potential and $\mathcal{X}_{\phi}$ is the scalar potential.  
%In the following we briefly describe the main equations used for the numerical implementation, in order to perform time evolutions. 
The fully non-linear Einstein-Proca system reads:
%
%
%\subsection{The Einstein-Proca system} 
%We solve the fully non-linear Einstein-Proca system using the time evolution numerical code from~\cite{Zilhao:2015tya}. The system reads:
%
\begin{eqnarray}
\label{eq:dtgamma}
\p_{t} \gamma_{ij} & = & - 2 \alpha K_{ij} + \Lie_{\beta} \gamma_{ij}
,\\
\label{eq:dtAi}
\p_{t} \A_{i}      & = & - \alpha \left( E_{i} + D_{i} \Aphi \right) - \Aphi D_{i}\alpha + \Lie_{\beta} \A_{i}
,\\
\label{eq:dtE}
\p_{t} E^{i}       & = &
        \alpha \left( K E^{i} + D^{i} Z + \mu^2_{\rm V} \A^{i}
                + \epsilon^{ijk} D_{j} B_{k} \right)\nonumber\\ &&
        - \epsilon^{ijk} B_{j} D_{k}\alpha
        + \Lie_{\beta} E^{i},\\
\label{eq:dtKij}
\p_{t} K_{ij}      & = & - D_{i} D_{j} \alpha
        + \alpha \left( R_{ij} - 2 K_{ik} K^{k}{}_{j} + K K_{ij} \right)
\nonumber \\ & &
        + 2 \alpha \biggl( E_{i} E_{j} - \frac{1}{2} \gamma_{ij} E^{k} E_{k} 
        + B_{i} B_{j} \nonumber\\ &&
          - \frac{1}{2} \gamma_{ij} B^{k} B_{k} - \mu^{2}_{\rm V} \A_{i} \A_{j}
          \biggl) + \Lie_{\beta} K_{ij} ,\\
\label{eq:dtAphi}
\p_{t} \Aphi  & = & - \A^{i} D_{i} \alpha
        + \alpha \left( K \Aphi - D_{i} \A^{i} - Z \right)\nonumber\\&&
        + \Lie_{\beta} \Aphi ,\\
\label{eq:dtZ}
\p_{t} Z          & = & \alpha \left( D_{i} E^{i} + \mu^{2}_{\rm V} \Aphi - \kappa Z \right)
        + \Lie_{\beta} Z\,,
\end{eqnarray}
where $\alpha$ is the lapse function, $\beta$ is the shift vector, $\gamma_{ij}$ is the spatial metric, $K_{ij}$ is the extrinsic curvature (with $K=K^{i}_{\,\,i}$), $D_i$ is the covariant 3-derivative,  and $\Lie_{\beta}$ is the Lie derivative. Moreover, the three-dimensional ``electric" $E^{i}$ and ``magnetic" $B^{i}$
fields are also introduced in the previous equations. The system is solved using the time-evolution numerical code from~\cite{Zilhao:2015tya} (see Section~\ref{sec:numerics}).

%%%%%%%%%%%%%%%%%%%%%%
\section{Initial Data}
\label{sec3}
%%%%%%%%%%%%%%%%%%%%%%
%%%%%%%%%%%%%%%%%%%%%%
%%%%%%%%%%%%%%%%%%%%%%%%%%%%%%%%%%%%%%%%%%%%%%%%%%%%%%%%%%%%%%
 
\begin{figure}[h!]
\centering
\includegraphics[height=2.45in]{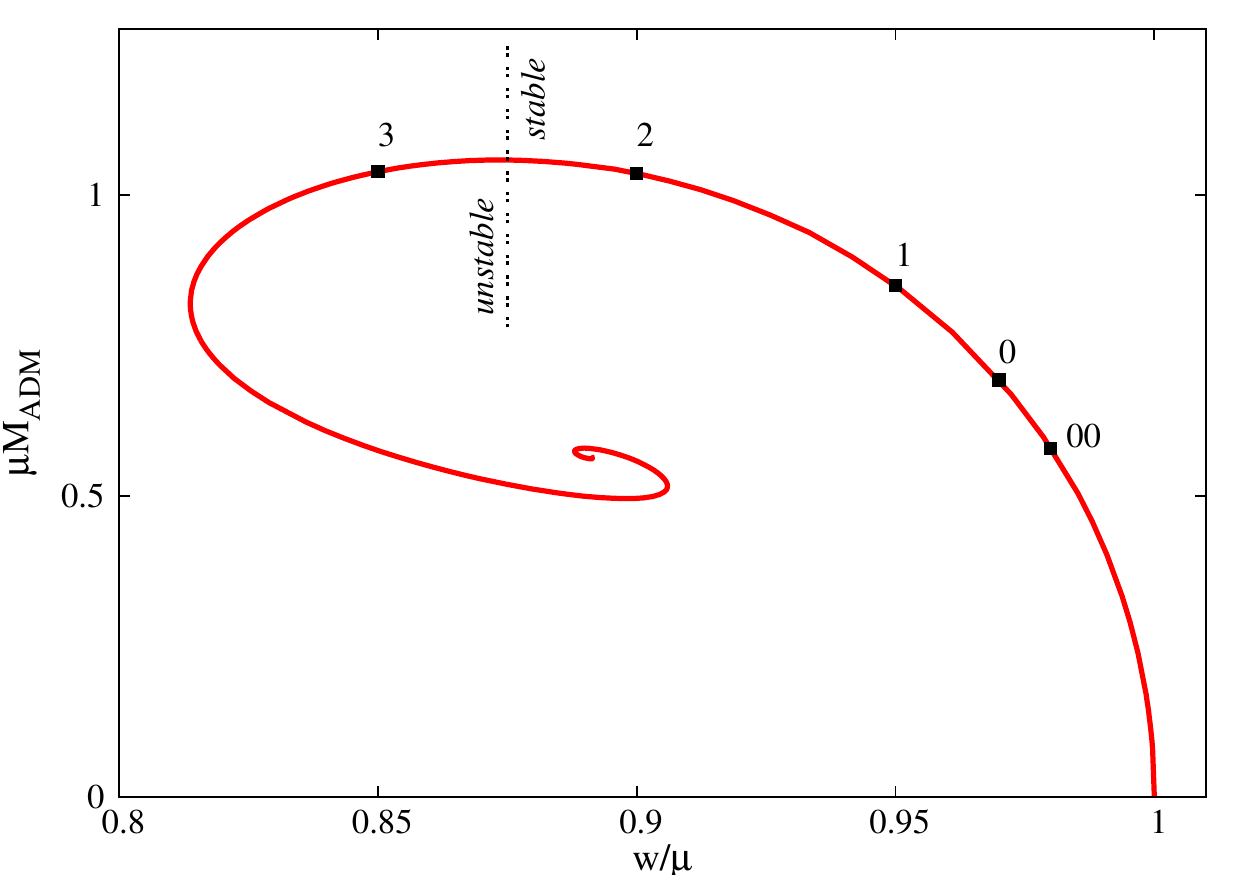}  
\caption{Domain of existence of the spherical (fundamental) PS solutions (solid line) in an ADM mass $vs.$ vector field frequency
diagram.  We highlight the five solutions used in the head-on collisions and the orbital mergers in this work.
}
\label{fig1}
\end{figure}

PSs were obtained in~\cite{Brito:2015pxa} as stationary solutions to 
the model described by the action~\eqref{model}. Five illustrative examples of spherically symmetric PSs  will be taken as the initial data for our time evolutions. Their basic physical properties, frequency, $w$, ADM mass, $M_{\rm ADM}$, Noether charge  $Q$ and the Proca ``electric" potential at the origin, $\Phi_c$,  all in units of the vector field mass,  can be found in Table~\ref{tab:mod1} and their distribution in an ADM mass $vs.$ Proca field frequency diagram is shown in Fig.~\ref{fig1}. 

\begin{table}[h!]
\caption{Spherically symmetric Proca star models.}
\label{tab:mod1}
\begin{ruledtabular}
\begin{tabular}{ccccc}
Model&$w/\mu$&$\mu M_{\rm ADM}$&$\mu^2Q$&$\Phi_c(r=0)$\\
\hline
PS00&0.98&0.580&0.584&0.0046 \\
PS0&0.97&0.693&0.702&0.0087\\
PS1&0.95&0.849&0.864 &0.0214\\
PS2&0.90&1.036&1.063 &0.0779\\
PS3&0.85&1.039&1.065   &0.2121\\

\end{tabular}
\end{ruledtabular}
\end{table}

When computed as static solutions~\cite{Brito:2015pxa}, spherically symmetric PSs are given by the line element 
\begin{eqnarray}
\label{ansatz1}
ds^2=-e^{2F_0} dt^2+e^{2F_1}\left[dr^2+r^2 (d\theta^2+  \sin^2\theta  d\varphi^2) \right],\,\,
\end{eqnarray} 
where $F_0,F_1$ are radial functions and $r,\theta,\varphi$ correspond  to 
isotropic coordinates.
The Proca field ansatz is given in terms of another two 
real functions $(V,H_1)$ which depend also on $r$
\begin{eqnarray}
\label{Paxial}
\mathcal{A}=e^{-iw t}\left( iVdt+
\frac{H_1}{r}dr   
\right) \ ,
  \ \
\end{eqnarray}
where $w>0$ is the frequency of the field. The translation between the four radial functions above, $F_0,F_1,V,H_1$, and 
the initial value for the metric and the 3+1 Proca field variables described is given as follows:
\begin{eqnarray}
\alpha &=& e^{F_{0}},\\
\gamma_{rr}&=&e^{2F_{1}},\gamma_{\theta\theta}=e^{2F_{1}}\,r^{2},\gamma_{\phi\phi}=e^{2F_{1}}\,r^{2}\sin^{2}\theta,\\
\mathcal{X}_{\phi}&=&-n^{\mu}\mathcal{A}_{\mu}\  , \label{propot}\\
\mathcal{X}_{i}&=&\gamma^{\mu}_{i}\mathcal{A}_{\mu}\ ,\\
E^{i}&=&-i\,\frac{\gamma^{ij}}{\alpha}\,\biggl(D_{j} (\alpha\mathcal{X}_{\phi})+\partial_{t}\mathcal{X}_{j}\biggl) \ .
\end{eqnarray}

\begin{figure}%[h!]
\centering
\includegraphics[height=1.75in]{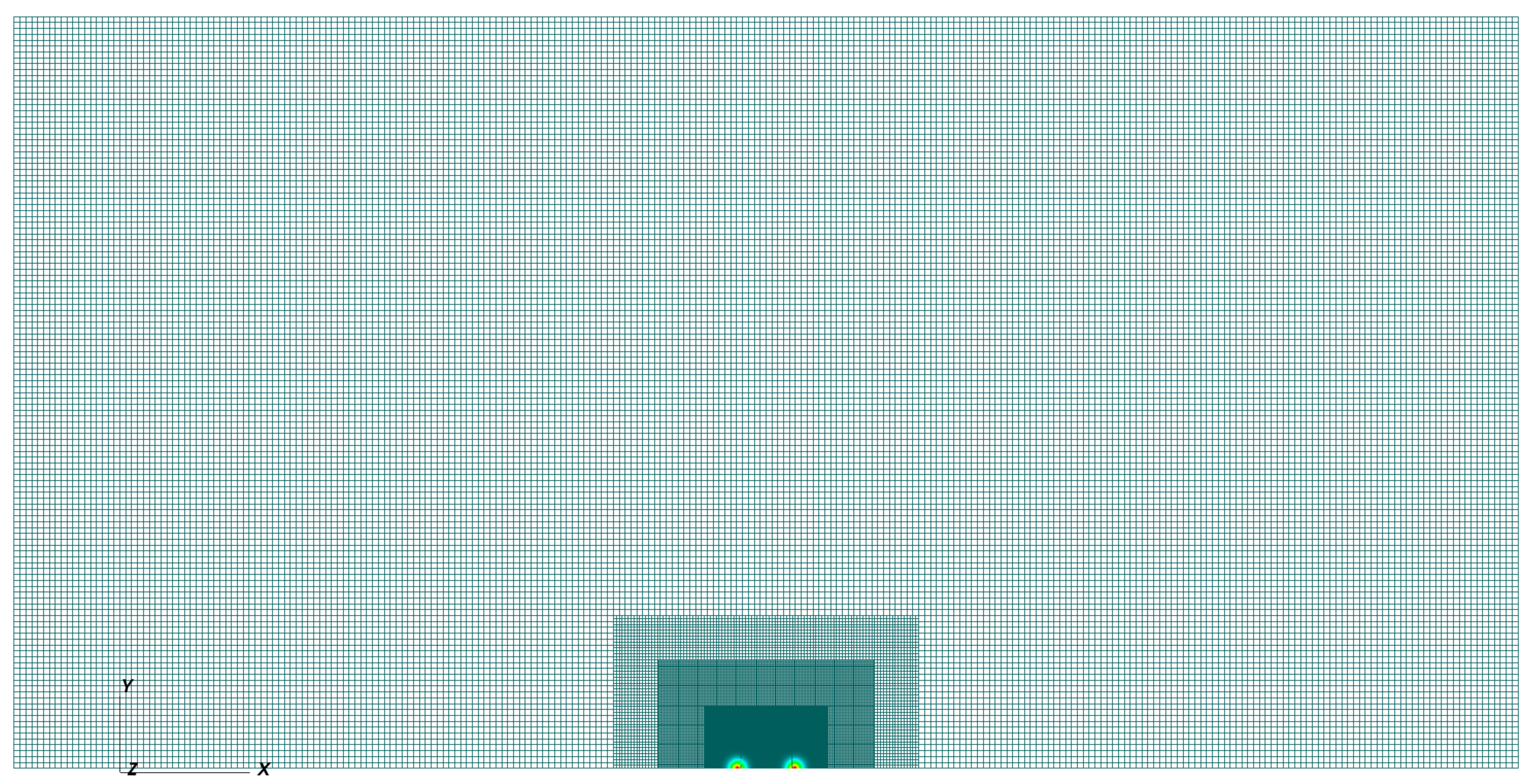}  
\includegraphics[height=1.79in]{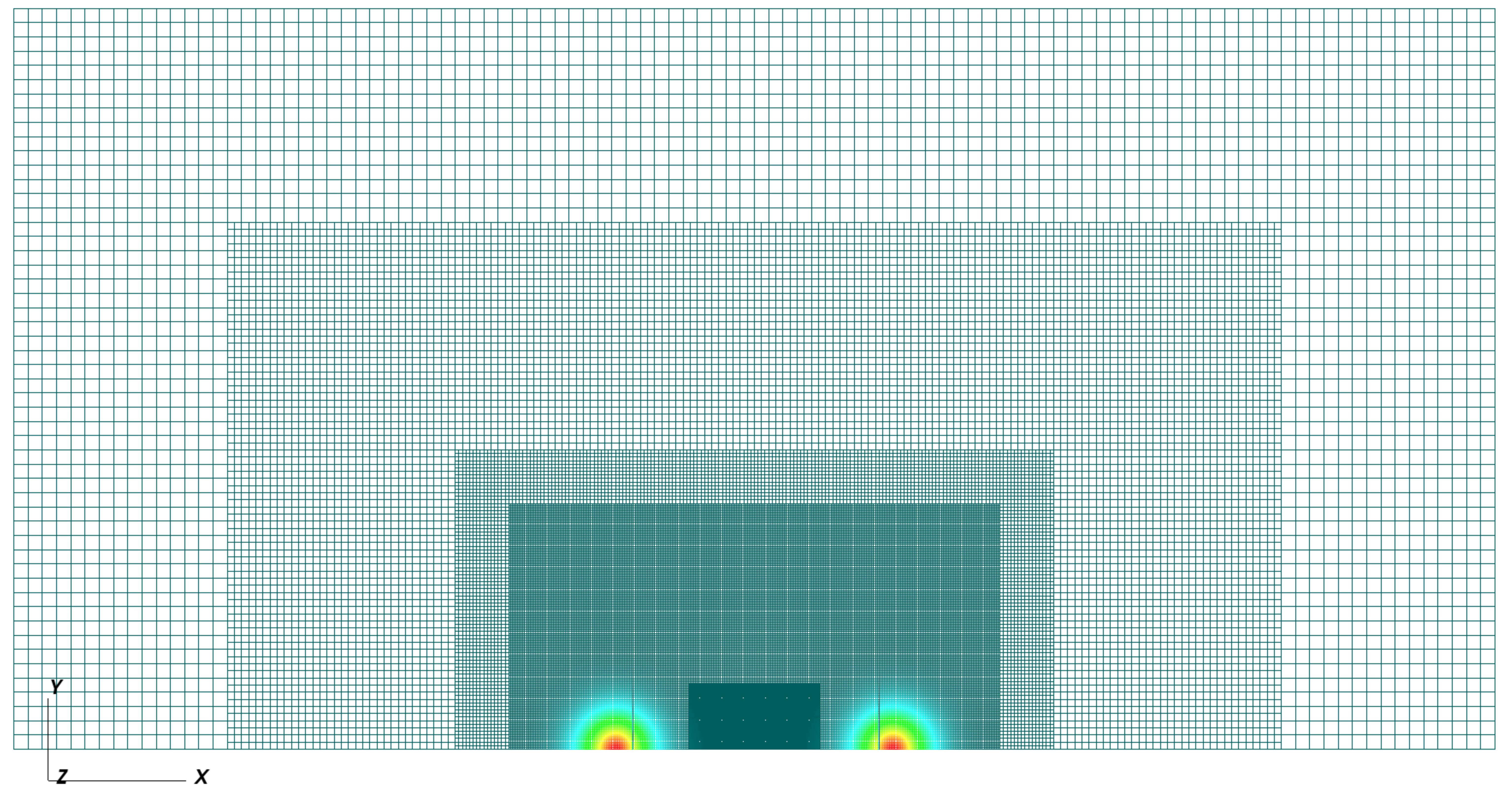}  
\caption{Computational mesh-refined grid used for the evolutions of the head-on collisions. The top panel shows the entire domain while the bottom panel is a magnification of the
inner region.}
\label{fig1b}
\end{figure} 

\begin{figure}%[h!]
\centering
\includegraphics[height=2.3in]{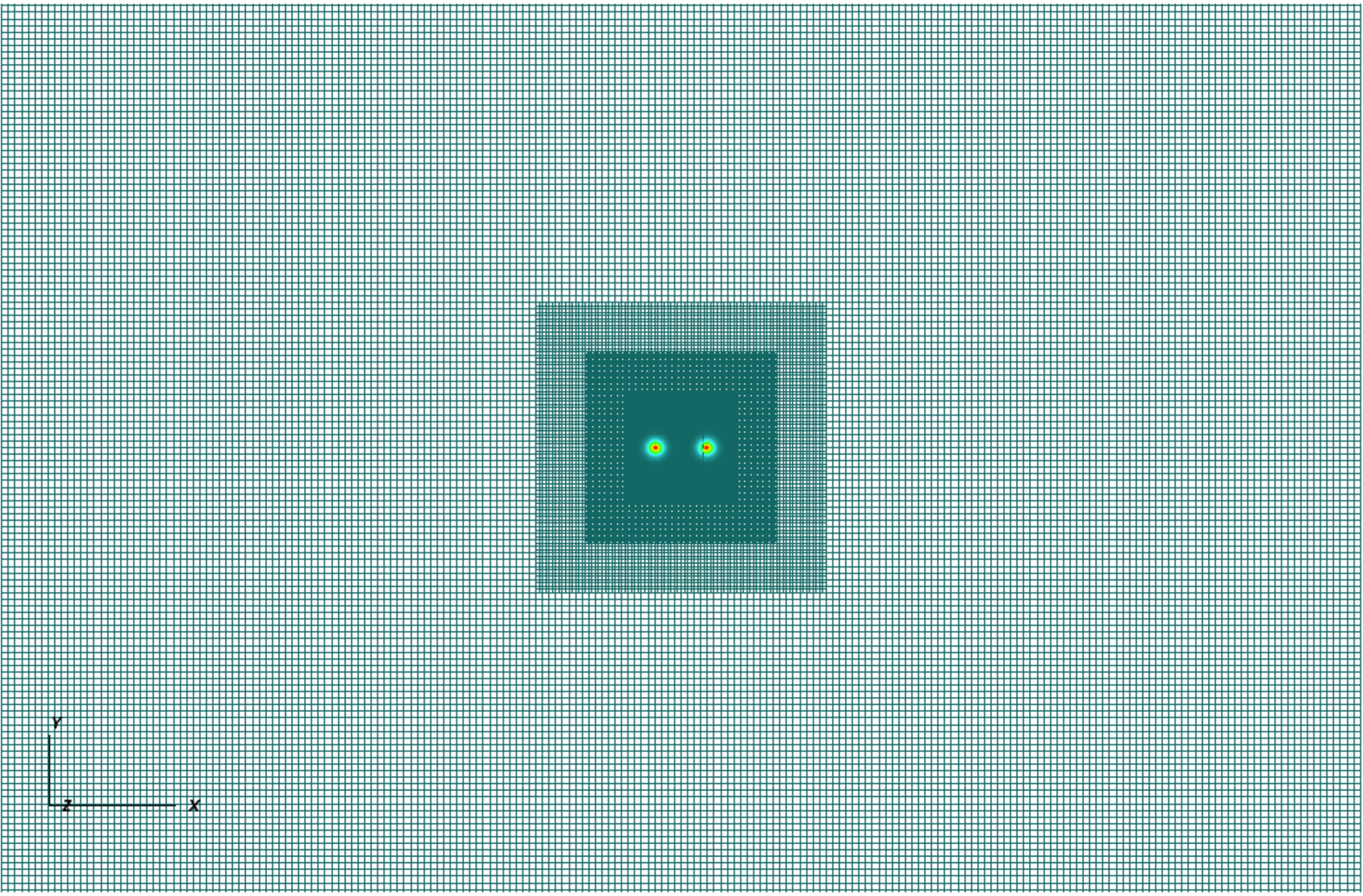}  
\includegraphics[height=2.3in]{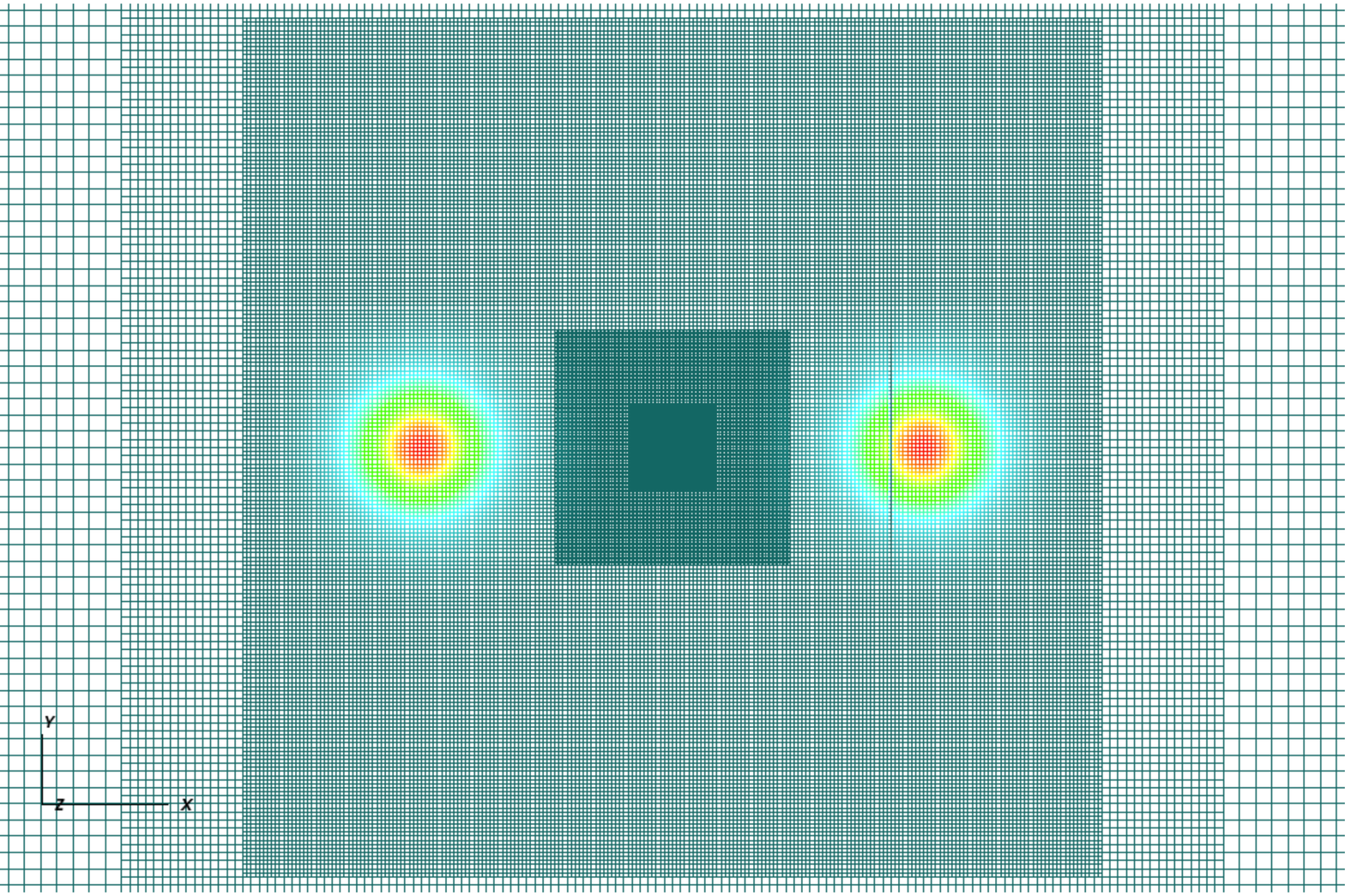}  
\caption{Computational mesh-refined grid used for the evolutions of the orbital merger case. The top panel shows the entire domain while the bottom panel is a magnification of the
inner region.}
\label{fig1c}
\end{figure} 

We follow~\cite{palenzuela2007head} to construct appropriate initial data to 
study the head-on collision of these compact objects. We take a superposition of two PS solutions:
\begin{itemize}
\item $\mathcal{A}(x_{i}) = \mathcal{A}^{(1)}(x_{i}-x_{0}) + \mathcal{A}^{(2)}(x_{i}+x_{0})$,
\item $\gamma_{ij}(x_{i}) = \gamma_{ij}^{(1)}(x_{i}-x_{0}) + \gamma_{ij}^{(2)}(x_{i}+x_{0})-\gamma_{ij}^{\rm flat}(x_{i}),$
\item $\alpha(x_{i}) = \alpha^{(1)}(x_{i}-x_{0}) + \alpha^{(2)}(x_{i}+x_{0}) - 1$,
\end{itemize}
where superindex $(i)$ labels the stars and $\pm x_{0}$ indicates their initial positions. The stars are initially separated by $\Delta x =  39$, which corresponds to $x_{0}=\pm19.5$ (in $G=c=1$ units). The solutions are not boosted. These initial data introduce constraint violations. However, they are small and do not grow during the evolution, as we discuss in~Appendix \ref{appendix}.

For the construction of initial data describing orbiting PS binary we extended the same method of the superposition of two isolated PSs solution, with the only difference that in this case the two stars are boosted along the $y$-axis which is perpendicular to the line segment linking them, with velocity $\pm v_y$, following \cite{shibata2008high,bezares2017final}. In this case, 5he stars are initially separated by $\Delta x =  30$, which corresponds to $x_{0}=\pm15.0$. We call $\mathcal{B}_b^{(i)}=v_y^{(i)}/c$ and $\Gamma_b^{(i)}= \bigl(1-\mathcal{B}_b^{(i)2}\bigl)^{-1/2}$ the Lorentz factor, and the matrix associated with the transformation has the following form
\begin{equation}\label{matrix}
\Lambda^{(i)} =
\begingroup
\setlength\arraycolsep{7pt}
\begin{pmatrix}
     \Gamma_b^{(i)} & -\Gamma_b^{(i)}\mathcal{B}_b^{(i)} &  0 & 0 \\
     -\Gamma_b^{(i)}\mathcal{B}_b^{(i)} & \Gamma_b^{(i)} &  0 &  0 \\
    0 & 0 &  1 & 0 \\
    0 & 0 &  0 &  1 
\end{pmatrix},
\endgroup
\end{equation}
 with the features that $\Lambda^{T}=\Lambda$ and $\Lambda^{-1}$ can be obtained using opposite velocity $v_y$. Note that $\Gamma_b^{(1)}=\Gamma_b^{(2)}$ and $\mathcal{B}_b^{(1)}=-\mathcal{B}_b^{(2)}$.

The line element of each star is in Cartesian coordinates given by
\begin{eqnarray}
\label{ansatz2}
ds^2=-\alpha_{0}^{2} dt^2+\psi^{4}_{0}\left[dx_{0}^{2}+dy_{0}^{2}+dz_{0}^{2}) \right],\,\,
\end{eqnarray} 
with $r_{0}=\sqrt{x^{2}_{0}+y^{2}_{0}+z^{2}_{0}}$. We perform a Lorentz transformation $t=\Gamma_{b}(t_{0}+v_{y}\,y_{0})$ and $y=\Gamma_{b}(y_{0}+v_{y}\,t_{0})$ and obtain from (\ref{ansatz2})

\begin{eqnarray}
\label{ansatz3}
ds^2=&-&\Gamma_{b}^{2}(\alpha_{0}^{2}-\psi^{4}_{0}v_{y}^{2}) dt^2+ 2\Gamma_{b}^{2}v_{y}(\alpha_{0}^{2}-\psi^{4}_{0})dtdy\nonumber\\
&+&\psi^{4}_{0}\left[dx^{2}+B_{0}^{2}dy^{2}+dz^{2}) \right],\,\,
\end{eqnarray} 

then
\begin{equation}
\alpha=\frac{\alpha_{0}}{B_{0}},\quad \beta^{y}=\biggl(\frac{\alpha_{0}^2-\psi^{4}_{0}}{\psi^{4}_{0}-\alpha_{0}^{2}v_{y}^{2}}\biggl)v_{y},
\end{equation}
where $B_{0}=\Gamma_{b}\sqrt{1-\frac{v_{y}^{2}\alpha_{0}^{2}}{\psi^{4}_{0}}}$.

The extrinsic curvature is computed from:
\begin{equation}
K_{ij}=\frac{1}{2\alpha}\biggl(\mathcal{L}_{\beta}\gamma_{ij}-\partial_{t}\gamma_{ij}\biggl)
\end{equation}
where $\mathcal{L}_{\beta}$ is the Lie derivative and $\partial_{t}\gamma_{ij}=-\Gamma_{b}v_{y}\partial_{y_{0}}\gamma_{ij}$.

%The transformation for the metric 
%\begin{equation}\label{metric_boosted}
%\gamma_{\mu\nu}^{(i) (boost)}=(\Lambda^{-1})_{\mu}^{\alpha} \gamma_{\alpha\beta}^{(i)} (\Lambda^{-1})_{\nu}^{\beta}
%\end{equation}
%takes the explicit form
%\begin{eqnarray}
%\gamma_{tt}^{(i)(boost)}&=& \Gamma_b^{(i)2} \gamma_{tt}^{(i)} + \Gamma_b^{(i)2}\mathcal{B}_b^{(i)2}\gamma_{yy}^{(i)},\\
%\gamma_{yy}^{(i)(boost)}&=& \Gamma_b^{(i)2} \gamma_{yy}^{(i)} + \Gamma_b^{(i)2}\mathcal{B}_b^{(i)2}\gamma_{tt}^{(i)},\\
%\gamma_{yt}^{(i)(boost)}&=& \Gamma_b^{(i)2}\mathcal{B}_b^{(i)} (\gamma_{tt}^{(i)} + \gamma_{yy}^{(i)}),\\
%\gamma_{ty}^{(i)(boost)}&=& \Gamma_b^{(i)2}\mathcal{B}_b^{(i)} (\gamma_{tt}^{(i)} + \gamma_{yy}^{(i)}).
%\end{eqnarray}

We have to transform the Proca fields. The Lorentz transformation resembles the one for the common electromagnetic fields. We consider that in the rest frame the magnetic field is zero, due to the spherical symmetry. The boosted fields are obtained as follows:
\begin{eqnarray}
E_x^{(i)({\rm boost})}&=&\Gamma_b^{(i)} E_x^{(i)}, \\
E_y^{(i)({\rm boost})}&=&E_y^{(i)}, \\
E_z^{(i)({\rm boost})}&=&\Gamma_b^{(i)} E_z^{(i)}, \\
\mathcal{X}_{\phi}^{(i)({\rm boost})}&=&\Gamma_b^{(i)} ( \mathcal{X}_{\phi}^{(i)}+\mathcal{B}_b^{(i)} \mathcal{X}_y^{(i)} ), \\
\mathcal{X}_x^{(i)({\rm boost})}&=&\mathcal{X}_x^{(i)}, \\
\mathcal{X}_y^{(i)({\rm boost})}&=&\Gamma_b^{(i)}(\mathcal{X}_y^{(i)}+\mathcal{B}_b^{(i)}\mathcal{X}_{\phi}^{(i)}), \\
\mathcal{X}_z^{(i)({\rm boost})}&=&\mathcal{X}_z^{(i)}.
\end{eqnarray}
The initial data for the PS binary is a superposition of the two boosted solution as in the head-on case.

%%%%%%%%%%%%%%%%%%%%%%%%%%%%%%%%%%%%%%%%%%%%%%%%%%%%
\section{Numerics} 
\label{sec:numerics}
%%%%%%%%%%%%%%%%%%%%%%%%%%%%%%%%%%%%%%%%%%%%%%%%%%%%
To perform the numerical evolutions we use the freely available \texttt{Einstein Toolkit}~\cite{toolkit2012open,loffler2012f}, which uses the \texttt{Cactus} framework and mesh refinement. The method-of-lines is employed to integrate the time-dependent differential equations. In particular, we use a fourth-order Runge-Kutta scheme for this task. The left-hand-side of the Einstein equations is solved using the \texttt{MacLachlan} code~\cite{brown2009turduckening,reisswig2011gravitational}, which is based on the 3+1 Baumgarte-Shapiro-Shibata-Nakamura (BSSN) formulation.

\begin{figure*}[t!]
\centering
\includegraphics[height=1.2in]{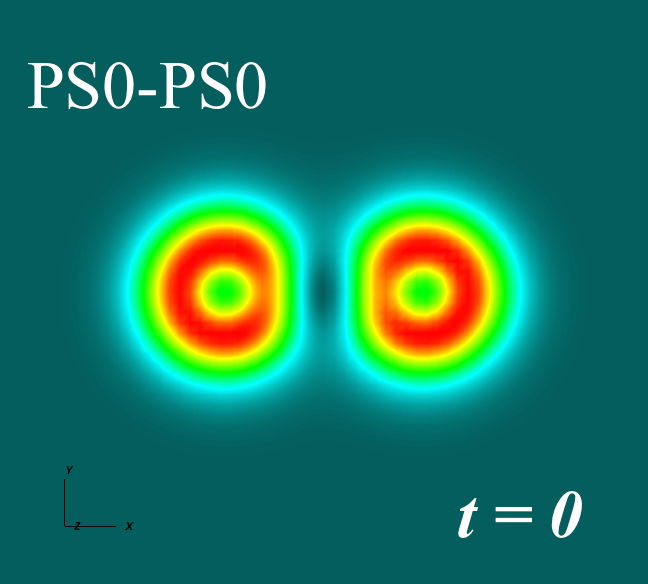} \includegraphics[height=1.2in]{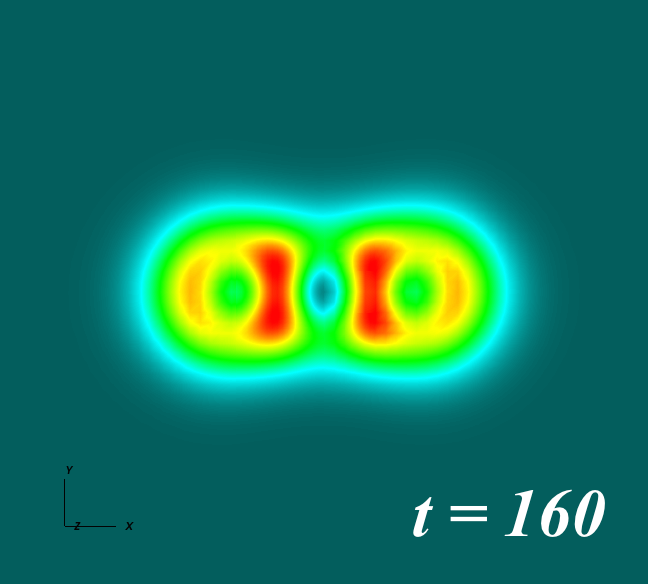} \includegraphics[height=1.2in]{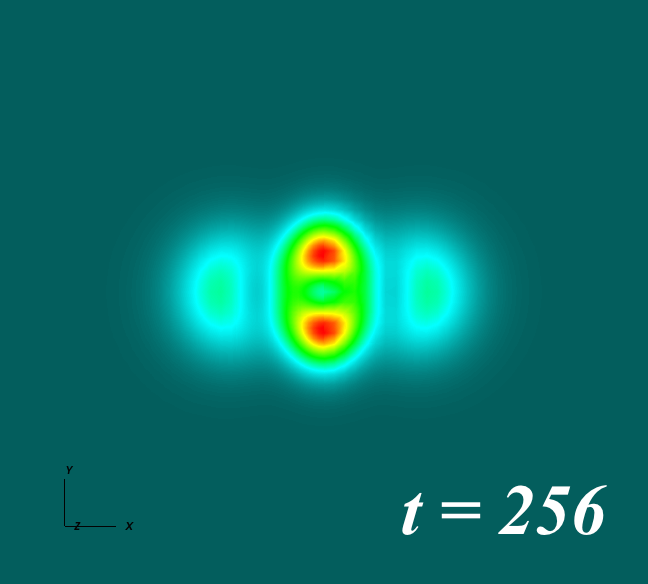} \includegraphics[height=1.2in]{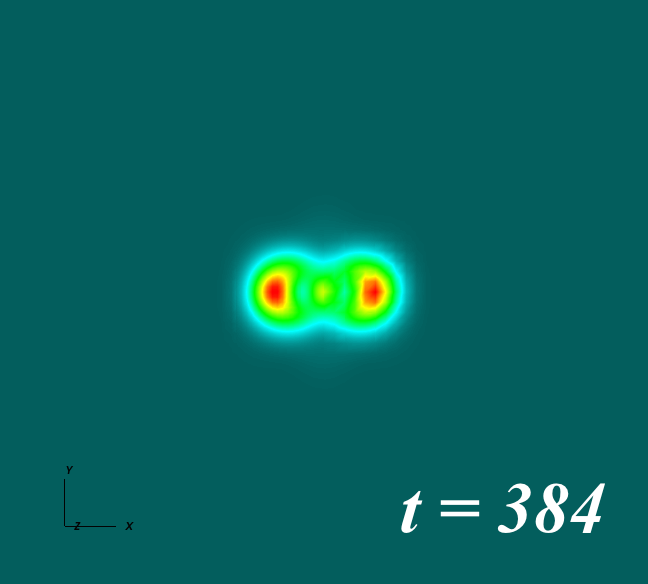} \includegraphics[height=1.2in]{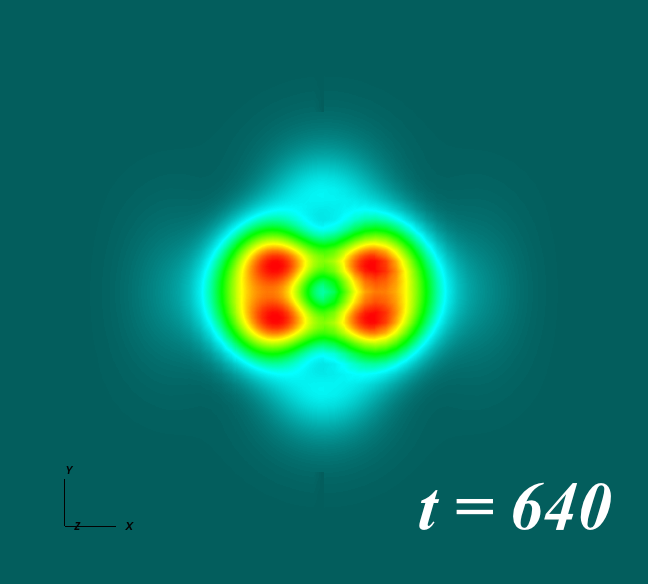}\\
\smallskip
\includegraphics[height=1.135in]{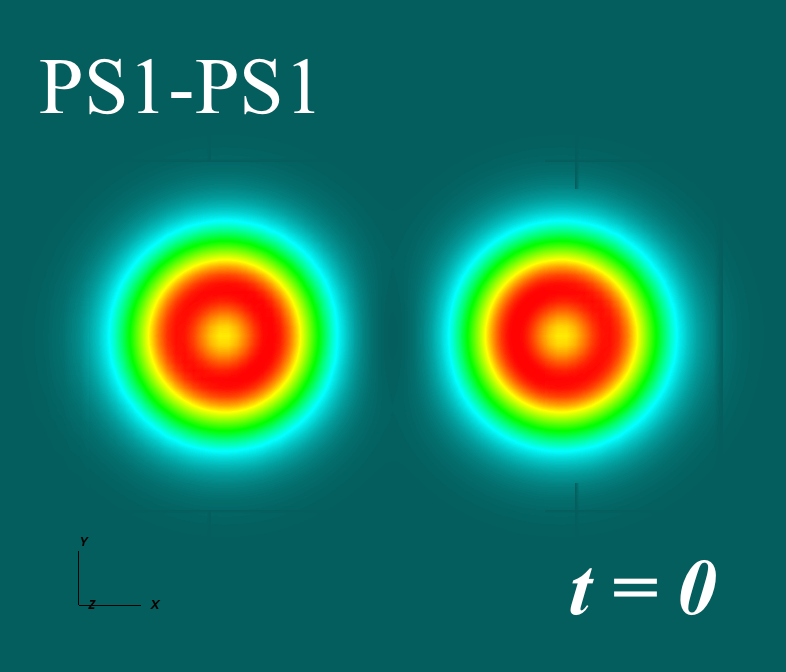} \includegraphics[height=1.142in]{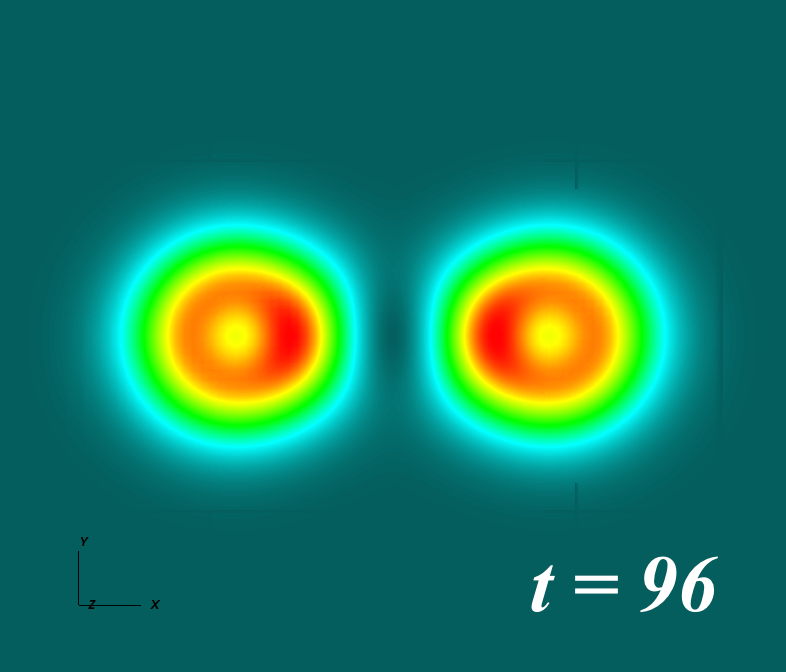}  \includegraphics[height=1.142in]{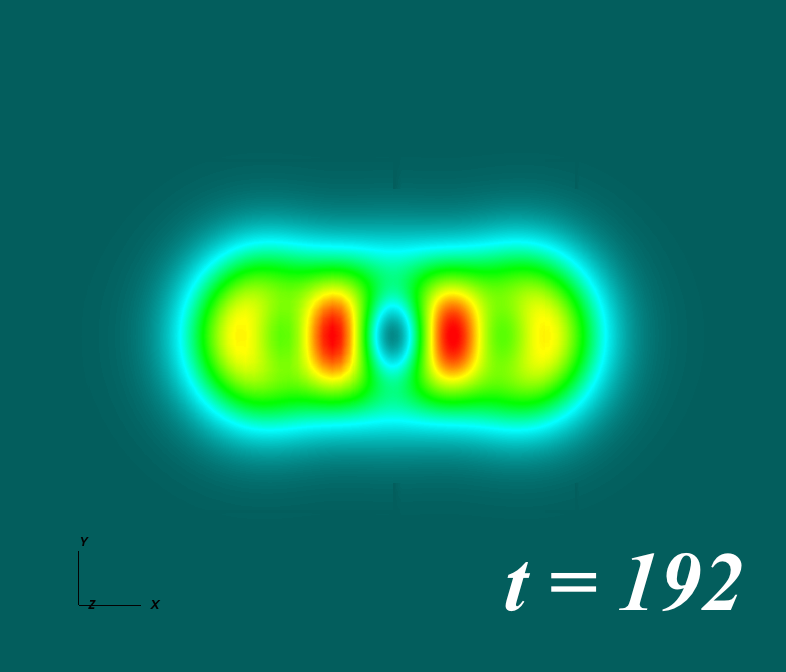}  \includegraphics[height=1.142in]{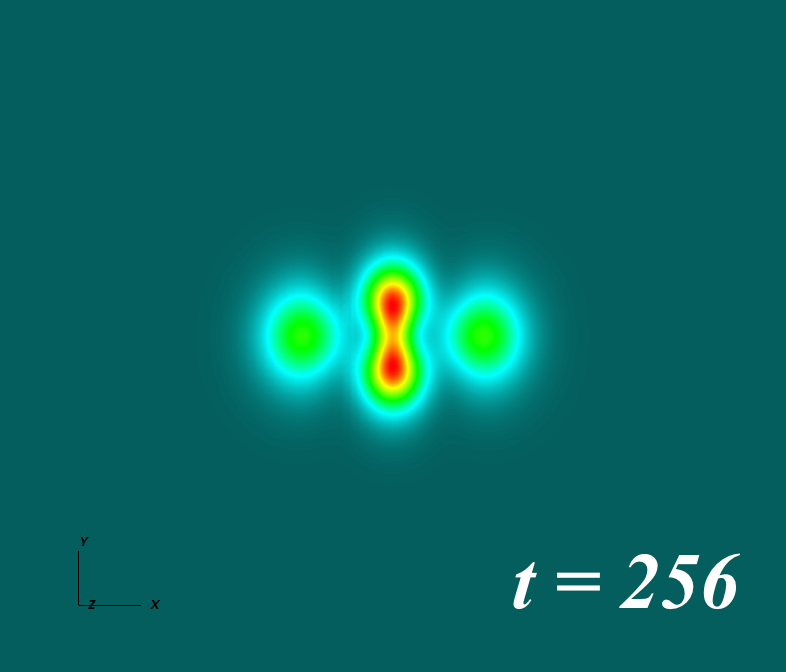} \includegraphics[height=1.135in]{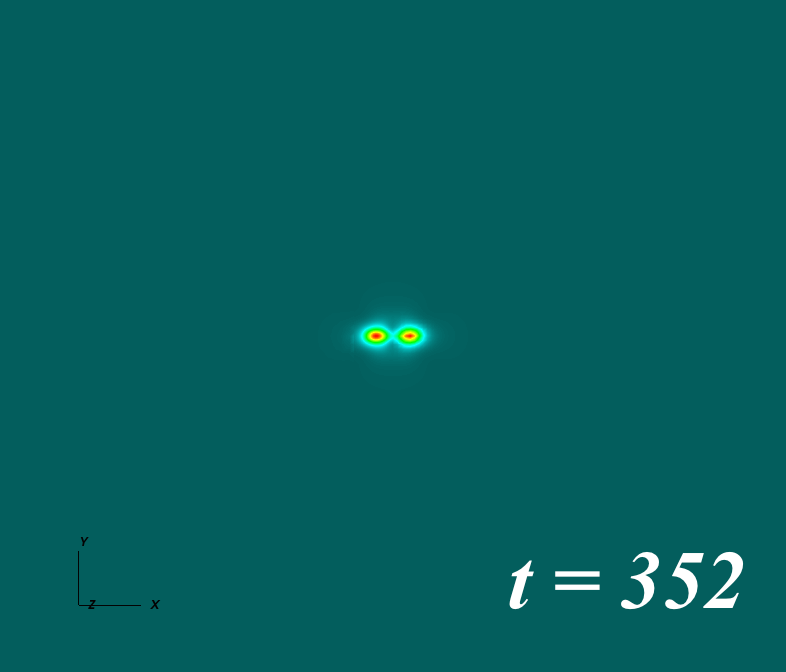} \\
\smallskip
\includegraphics[height=1.2in]{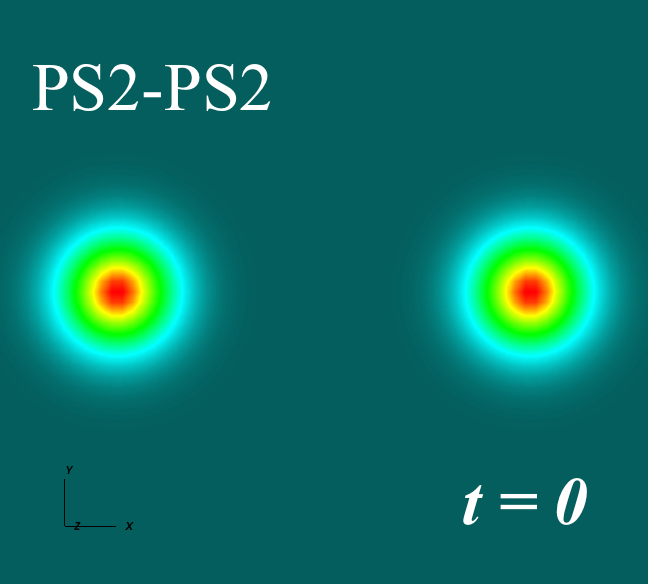} \includegraphics[height=1.2in]{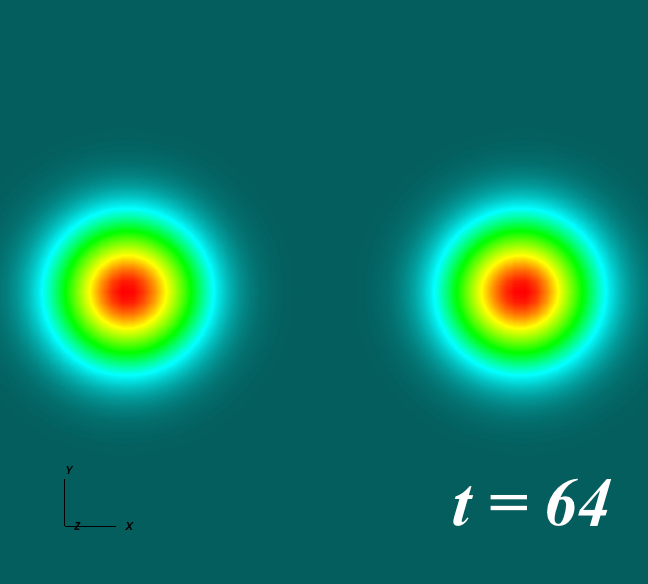}  \includegraphics[height=1.2in]{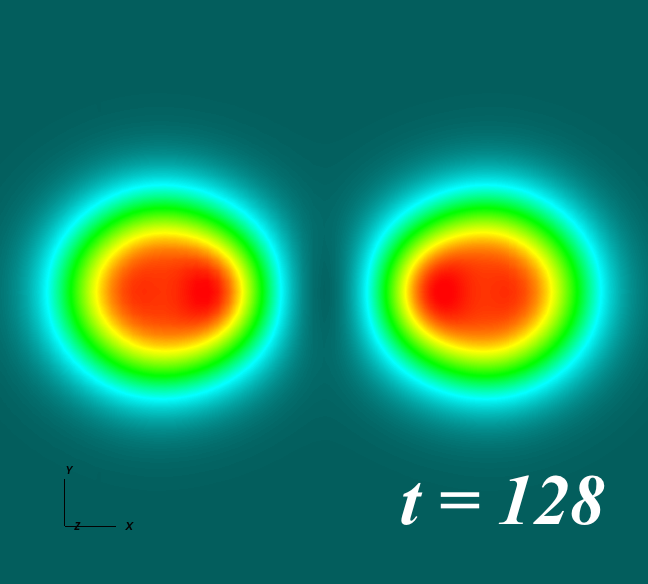} \includegraphics[height=1.2in]{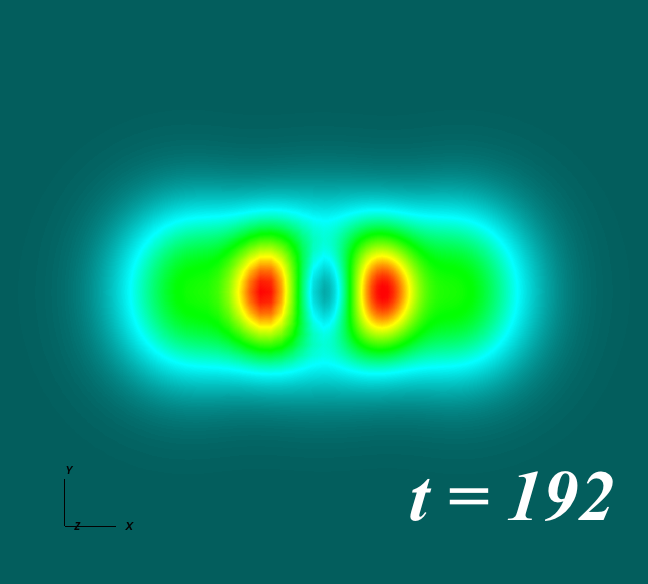} \includegraphics[height=1.2in]{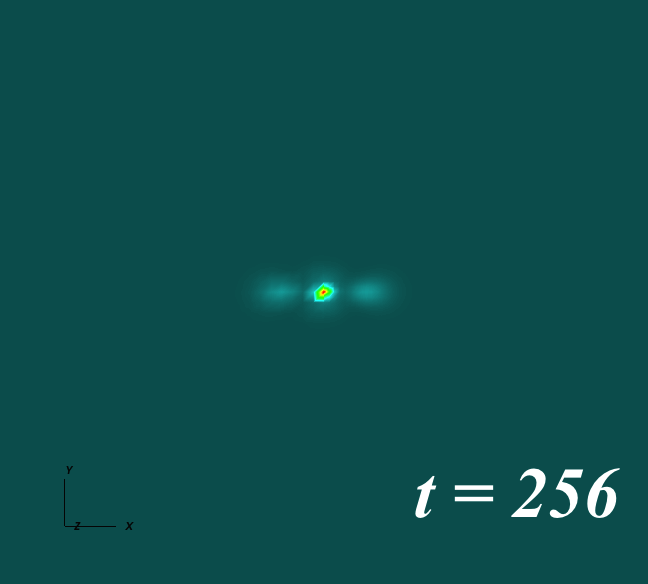} \\
\smallskip
\includegraphics[height=1.2in]{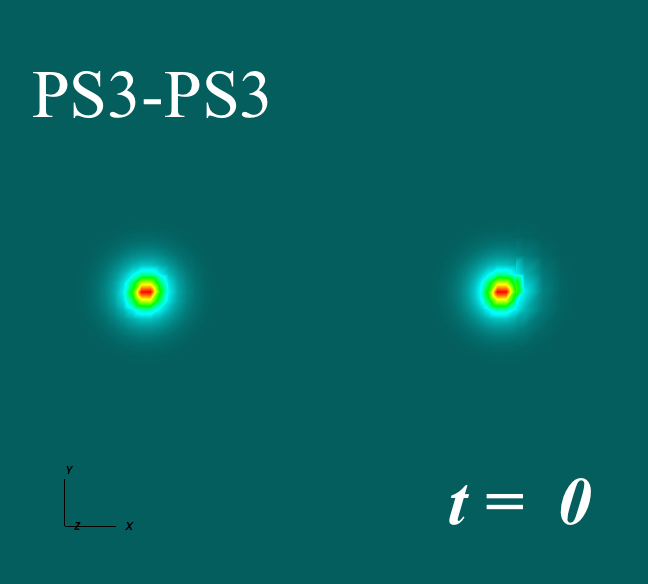}  \includegraphics[height=1.2in]{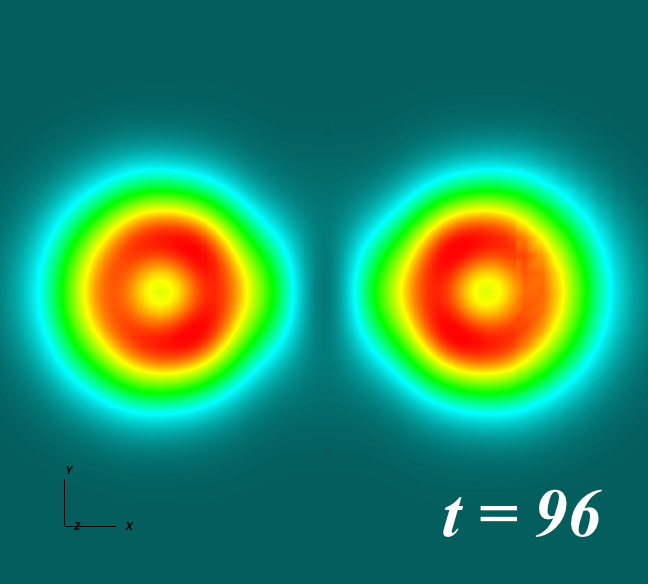}  \includegraphics[height=1.2in]{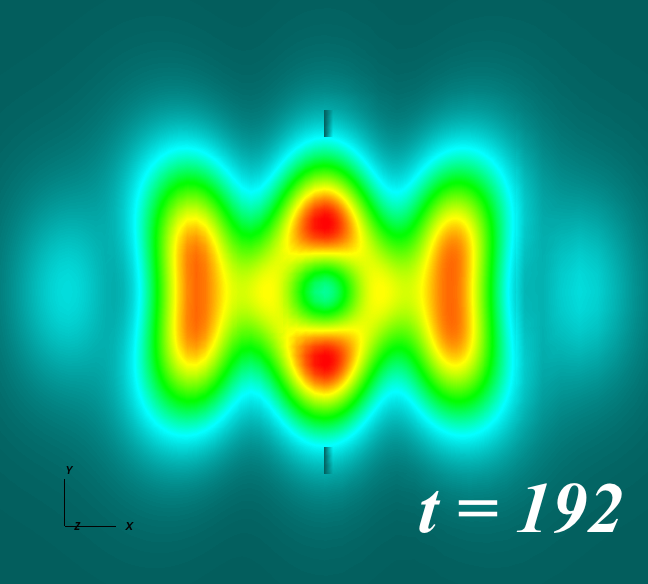} \includegraphics[height=1.2in]{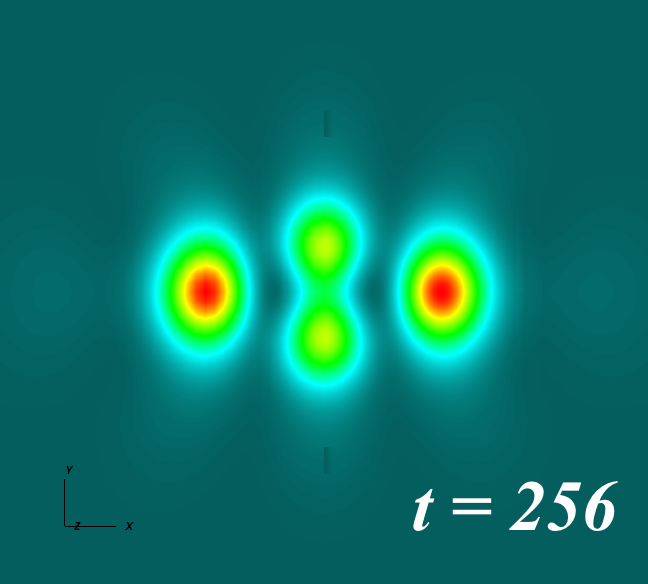} \includegraphics[height=1.2in]{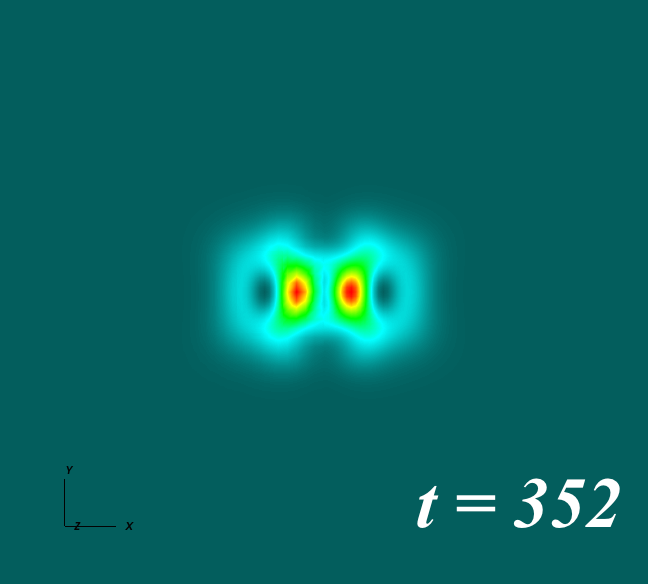} \\

\caption{Snapshots of the energy density for PS0-PS0, PS1-PS1, PS2-PS2 and PS3-PS3 in the equatorial plane. The vertical axis correspond to the $y$ direction and the horizontal to the $x$ direction.
}
\label{fig6}
\end{figure*}

The Proca evolution equations, Eqs.~(\ref{eq:dtgamma})-(\ref{eq:dtZ}), are solved using the code developed independently by M.~Zilh\~ao and H.~Witek~\cite{Zilhao:2015tya}. We have extended this code to take into account a complex field. All technical details, assessment of the code and convergence tests can be found in~\cite{Zilhao:2015tya}.

We use a numerical grid with 6 refinement levels for the head-on collisions, its structure being $\lbrace(512, 64, 64, 32, 32, 8)$, $(4, 2, 1, 0.5, 0.25, 0.125)\rbrace$, 
where the first set of numbers indicates the spatial domain of each level and the second set indicates the resolution. Fig.~\ref{fig1b} shows the grid structure at $t=0$. The outer boundary is located at $\Delta x=392$ from the gravitational-wave extraction radii. This ensures that the total time of the simulations is shorter than twice the light-crossing time, which prevents numerical reflections at the boundary from affecting the extraction. Due to the geometry of head-on collisions, we consider equatorial-plane symmetry and reflection symmetry with respect to the $x$-$z$ plane.

Correspondingly, Fig.~\ref{fig1c} shows the grid structure we employ for the orbital mergers, which contains 7 refinement levels with spatial extent and resolution given by $\lbrace(412, 64, 64, 32, 32, 8, 4)$, $(4, 2, 1, 0.5, 0.25, 0.125, 0.0625)\rbrace$. In this case there is a reflection symmetry along the equatorial-plane  but no reflection symmetry with respect to the $x$-$z$ plane.

 %%%%%%%%%%%%%%%%%%%%%%%%%%%%%%%%%%%%%%%%%%%%%%%%%%%%%%%%%%%%%
\section{Results}
\label{results}
%%%%%%%%%%%%%%%%%%%%%%%%%%%%%%%%%%%%%%%%%%%%%%%%%%%%%%%%%%%%%%

 %%%%%%%%%%%%%%%%%%%%%%%%%%%%%%%%%%%%%%%%%%%%%%%%%%%%%%%%%%%%%
\subsection{Head-on collisions}
 %%%%%%%%%%%%%%%%%%%%%%%%%%%%%%%%%%%%%%%%%%%%%%%%%%%%%%%%%%%%%

We evolve the initial data described in Section~\ref{sec3} for four equal-mass PS binaries, using the four types of PSs described in Table~\ref{tab:mod1} and highlighted in Fig.~\ref{fig1}. We label these four models as PS0-PS0, PS1-PS1, PS2-PS2, and PS3-PS3. We also evolve the latter including an initial perturbation which consists in multiplying by a number slightly larger than one the initial values of the Proca variables, as we did in \cite{sanchis2017numerical}, a model that we dub PS3b-PS3b. By including a perturbation the stars are forced to collapse before the collision, since the PS3 model is unstable against radial perturbations.

 %%%%%%%%%%%%%%%%%%%%%%%%%%%%%%%%%%%%%%%%%%%%%%%%%%%%%%%%%%%%%
\subsubsection{Visualisation of the collisions}
 %%%%%%%%%%%%%%%%%%%%%%%%%%%%%%%%%%%%%%%%%%%%%%%%%%%%%%%%%%%%%

Fig.~\ref{fig6} shows equatorial-plane snapshots of the evolutions of the PS0-PS0, PS1-PS1, PS2-PS2 and PS3-PS3 binaries, from top to bottom. Our computational grid only extends from $y=0$ to 512, therefore the data for negatives values of $y$ are mirrored by the corresponding positive $y$ values, due to axisymmetry. Animations illustrating these collisions can be found in~\cite{webpage}.

The collision of the two Proca stars happens at $t\sim160$ for models PS0-PS0 and PS3-PS3, and at $t\sim192$ for PS1-PS1 and PS2-PS2. The merged object oscillates in the $x$ and $y$ directions. For PS0-PS0, part of the field is ejected from the polar caps and approaches spherical symmetry. Model PS2-PS2 collapses quickly after the collision at $t_{\rm collapse}\sim245$, while models PS1-PS1 and PS3-PS3 oscillate during a short period of time before an apparent horizon (AH) appears at $t_{\rm collapse}\sim375$ and $t_{\rm collapse}\sim425$, respectively. Moreover, the PS1-PS1 collision forms a perturbed massive PS that oscillates and produces the first part of the gravitational-wave signal, as we show below. In the case of PS3-PS3, the stars migrate and expand before the collision. For all models, the morphology and amplitude of the gravitational waves are clearly influenced by the dynamics during the collision (cf.~Section~\ref{waveforms}). 

\begin{figure}[t!]
\centering
\includegraphics[height=2.5in]{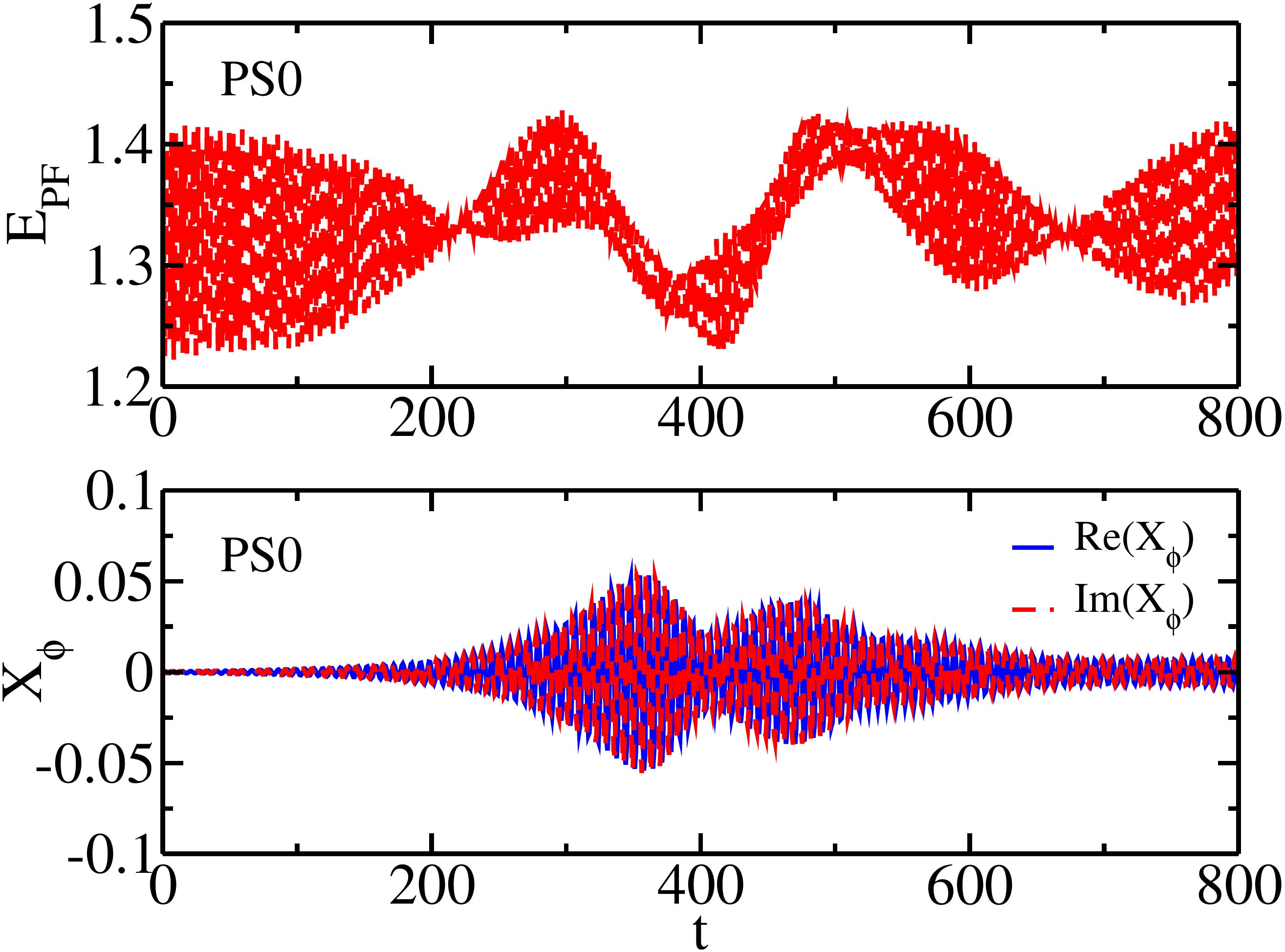} 
\caption{Model PS0-PS0. Time evolution of the Proca field energy (top panel) and of the amplitude of the real and imaginary parts of the scalar potential extracted at $r=0$ (bottom panel).
}
\label{fig2a}
\end{figure}

 %%%%%%%%%%%%%%%%%%%%%%%%%%%%%%%%%%%%%%%%%%%%%%%%%%%%%%%%%%%%%
\subsubsection{Proca energy and horizon formation}
 %%%%%%%%%%%%%%%%%%%%%%%%%%%%%%%%%%%%%%%%%%%%%%%%%%%%%%%%%%%%%

In Fig.~\ref{fig2a} we plot the time evolution of the Proca field energy
\begin{equation}
E_{\rm PF}=-\int_{\Sigma}drd\theta 
d\varphi\left(2T^t_t-T_\alpha^\alpha\right)\alpha\sqrt{\gamma} \ ,
\label{energy}
\end{equation}   
and of the amplitude of the real and imaginary parts of the time component of the Proca 4-potential (hereafter the ``scalar" potential) for model PS0-PS0. In this case we see no horizon formation. The stars have sufficiently low mass so that the final state is still a star. The amplitude of the scalar potential stabilises within the timescale of our simulation. It has been reported in~\cite{brito2015accretion,helfer2018gravitational} that some collapsed oscillatons from head-on collisions can lose part of the field and form a stable, less compact oscillaton, not collapsing to a BH even if the total mass is larger than the maximum mass of the equilibrium configuration. This seems to be precisely what is occurring for model PS0-PS0. The mass of the PS0 Proca star is $\mu M_{\rm ADM}=0.693$, so that twice that mass is larger than the maximum mass for spherical PSs, which is $\mu M_{\rm ADM}^{\rm max}=1.058$~\cite{Brito:2015pxa}. Still, a PS forms, as a result of the inelasticity of the collision.

\begin{figure}%[t!]
\centering
\includegraphics[height=2.5in]{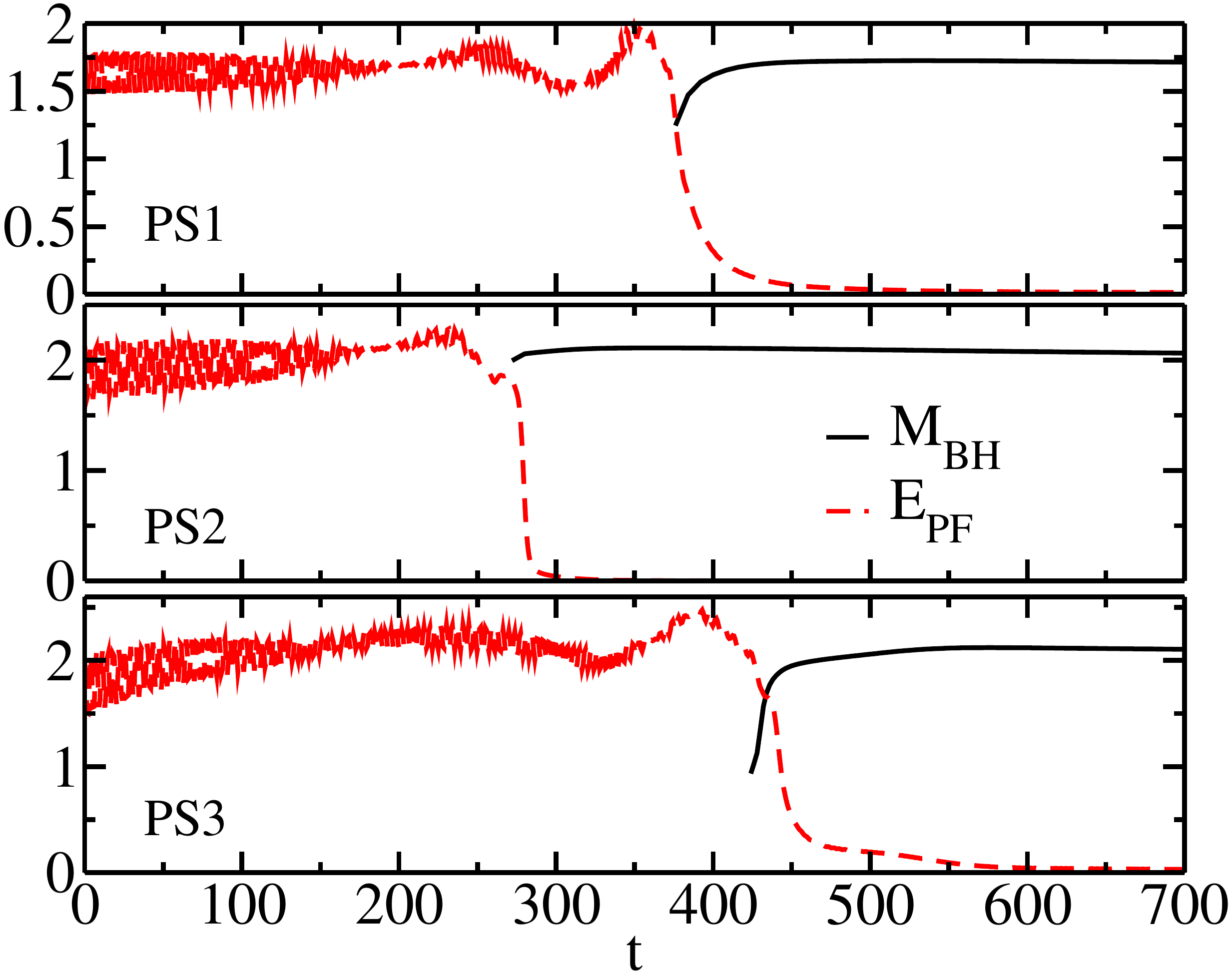} 
\caption{ Time evolution of the Proca field energy and BH mass for models PS1-PS1 (top panel), PS2-PS2 (middle panel) and PS3-PS3 (bottom panel).
}
\label{fig2b}
\end{figure}

\begin{figure}%[h!]
\centering
\includegraphics[height=2.5in]{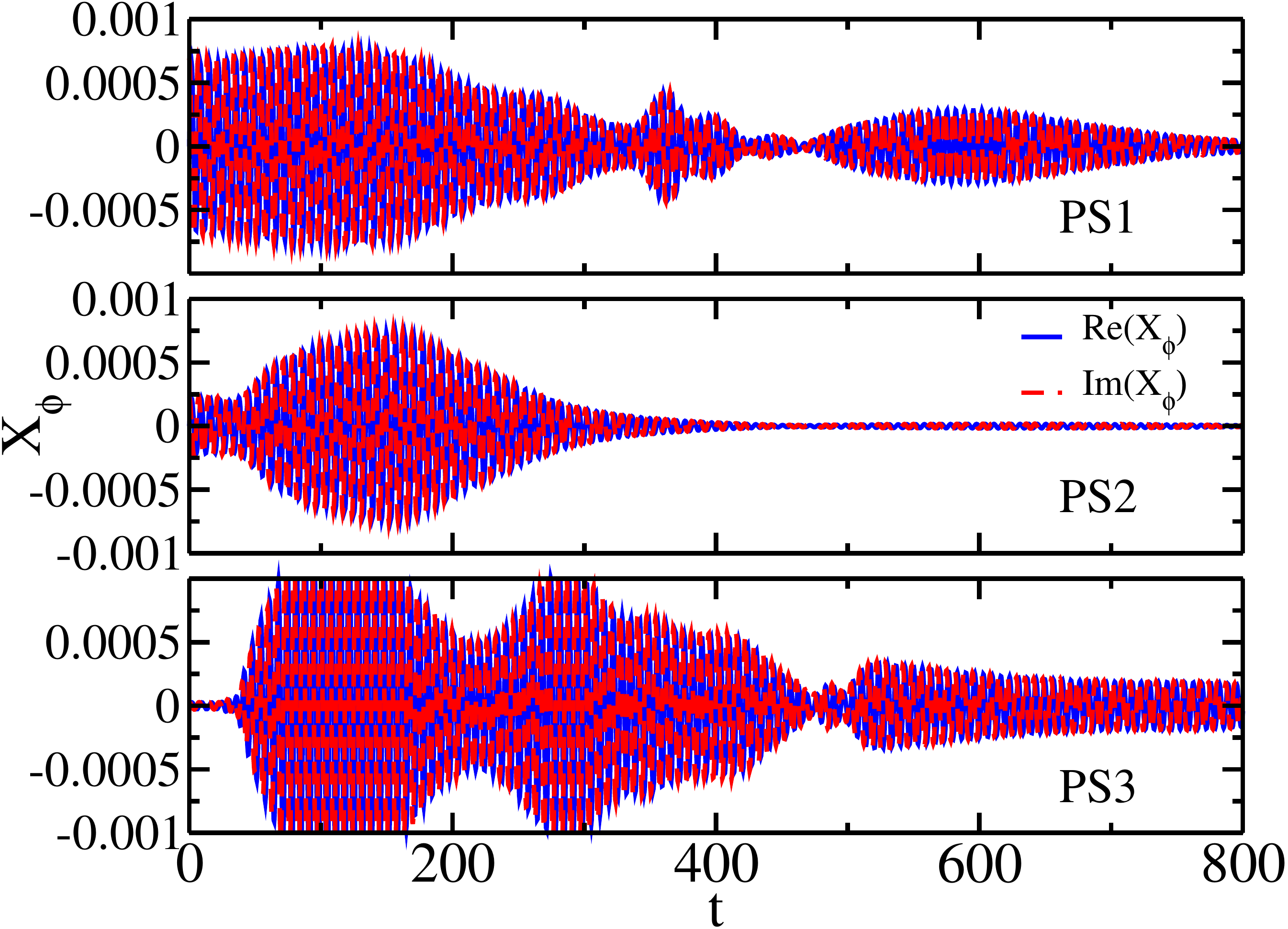} 
\caption{Time evolutions of the amplitude of the real and imaginary parts of the scalar potential extracted at $r=12$ for models PS1-PS1 (top panel), PS2-PS2 (middle panel) and PS3-PS3 (bottom panel).
}
\label{fig2c}
\end{figure}

Fig.~\ref{fig2b} displays the time evolution of the Proca field energy and of the irreducible mass of the AH for models PS1-PS1, PS2-PS2 and PS3-PS3. For these three models, the collapse of the solutions is triggered after the collision and an AH forms. After the collapse there is still a Proca field remnant outside the horizon, as Fig.~\ref{fig2c} shows. For PS1-PS1 and PS3-PS3 this remnant has a visible dynamics, reminiscent of a beating pattern, a signature of the presence of more than one quasi-bound state outside the AH. In the case of the PS3-PS3 model, the Proca remnant seems particularly long lived. We shall observe a possible impact of this feature in the gravitational waveform below.

 %%%%%%%%%%%%%%%%%%%%%%%%%%%%%%%%%%%%%%%%%%%%%%%%%%%%%%%%%%%%%
\subsubsection{Waveforms}
\label{waveforms}
 %%%%%%%%%%%%%%%%%%%%%%%%%%%%%%%%%%%%%%%%%%%%%%%%%%%%%%%%%%%%%

\begin{figure}[t!]
\centering
\includegraphics[height=2.5in]{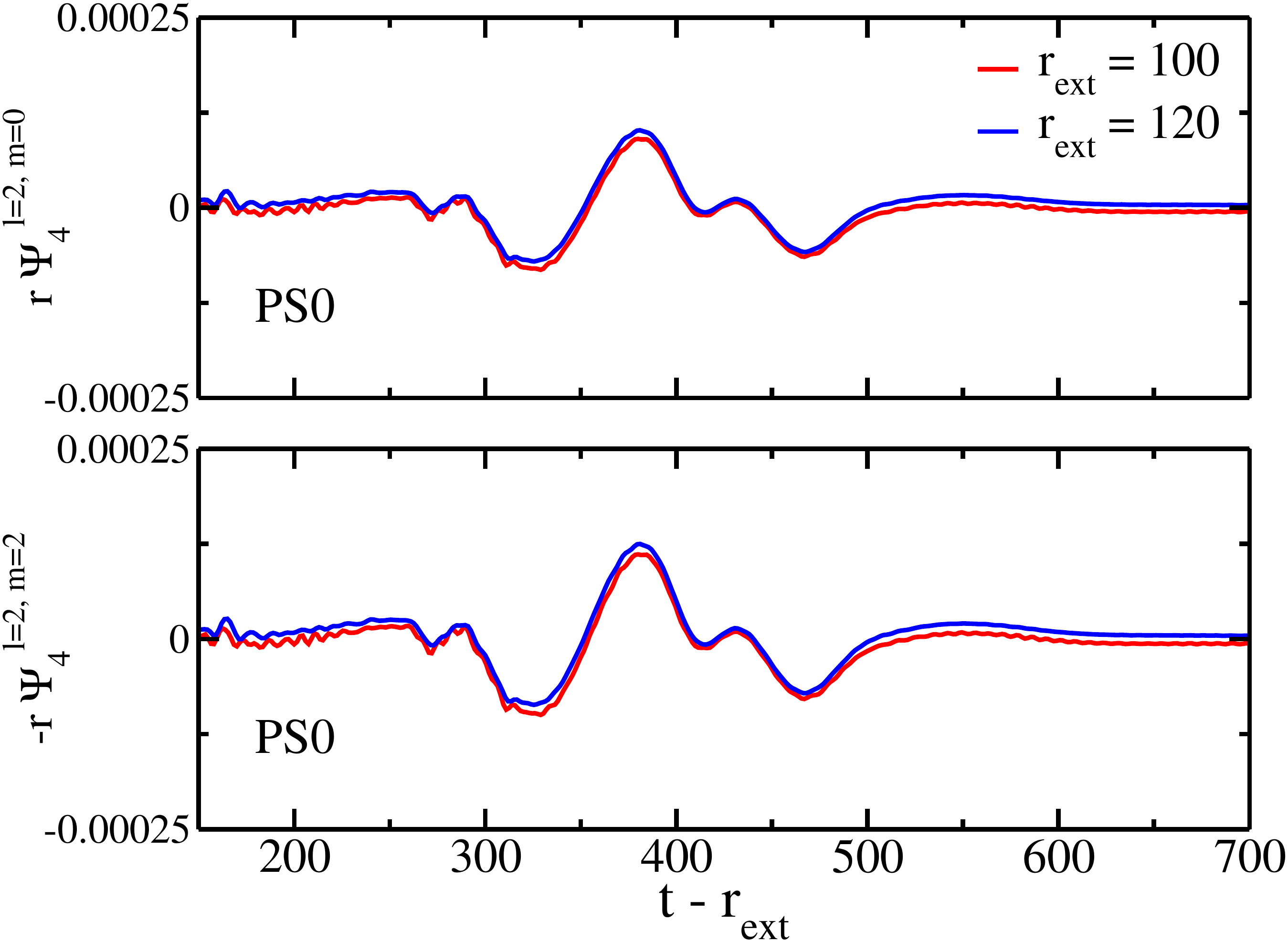} 
\includegraphics[height=2.5in]{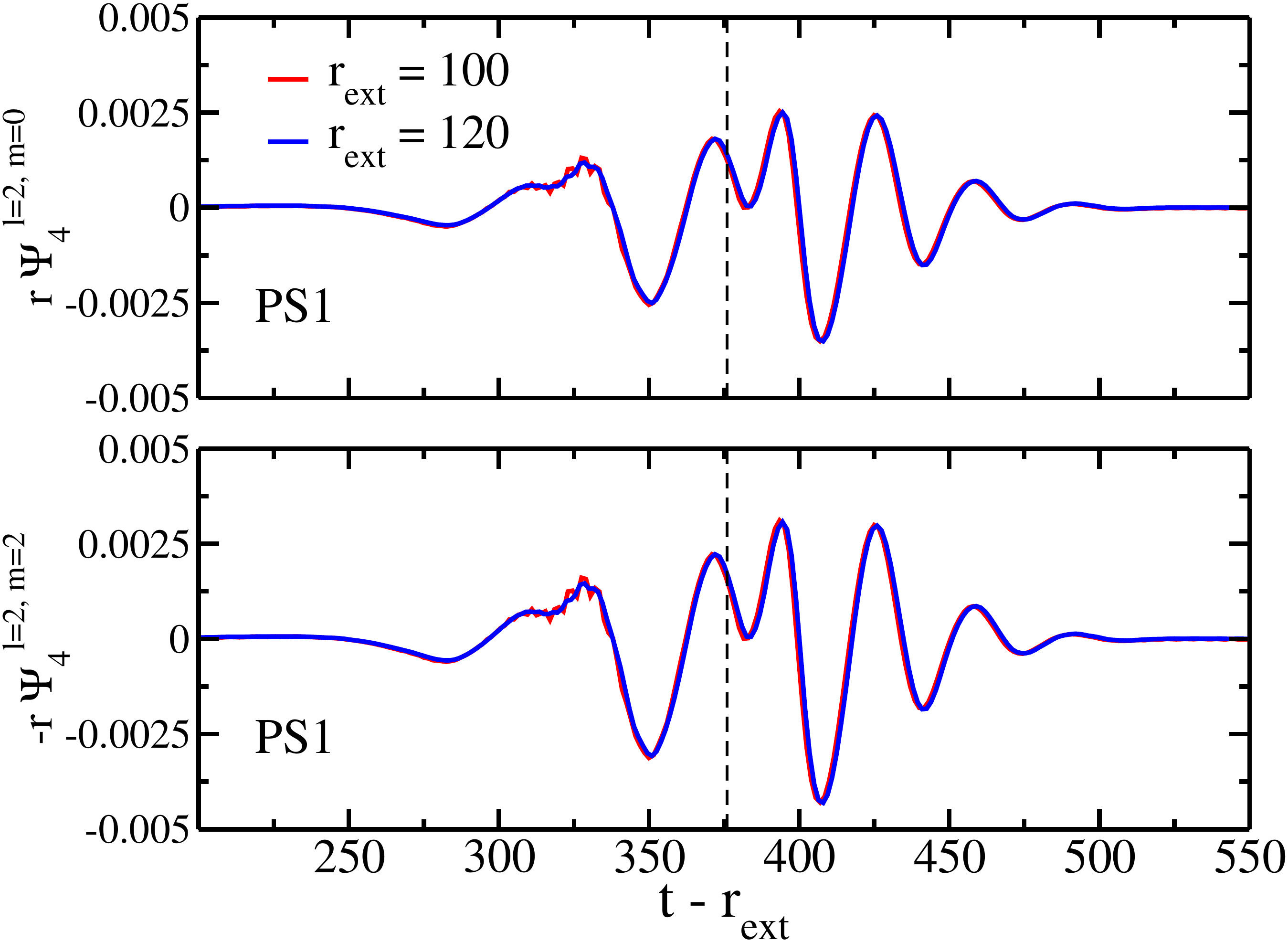}  
\includegraphics[height=2.5in]{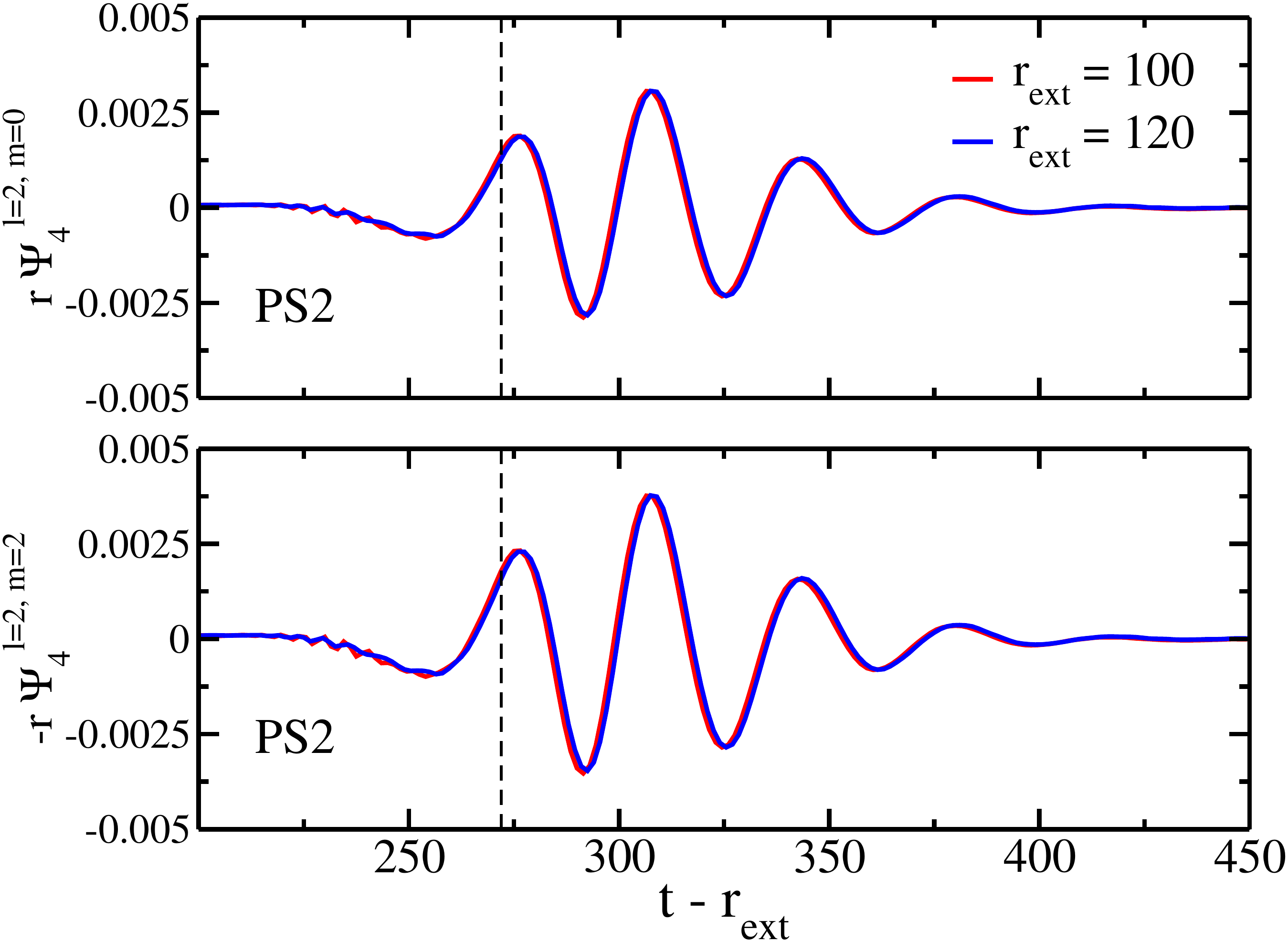}
\caption{Real part of $r\Psi_4^{l=2,m=0}$ and $r\Psi_4^{l=2,m=+2}$ for PS0-PS0 (top panels), PS1-PS1 (middle panels) and PS2-PS2 (bottom panels).}
\label{fig3}
\end{figure}

\begin{figure}[t!]
\centering
 \includegraphics[height=2.5in]{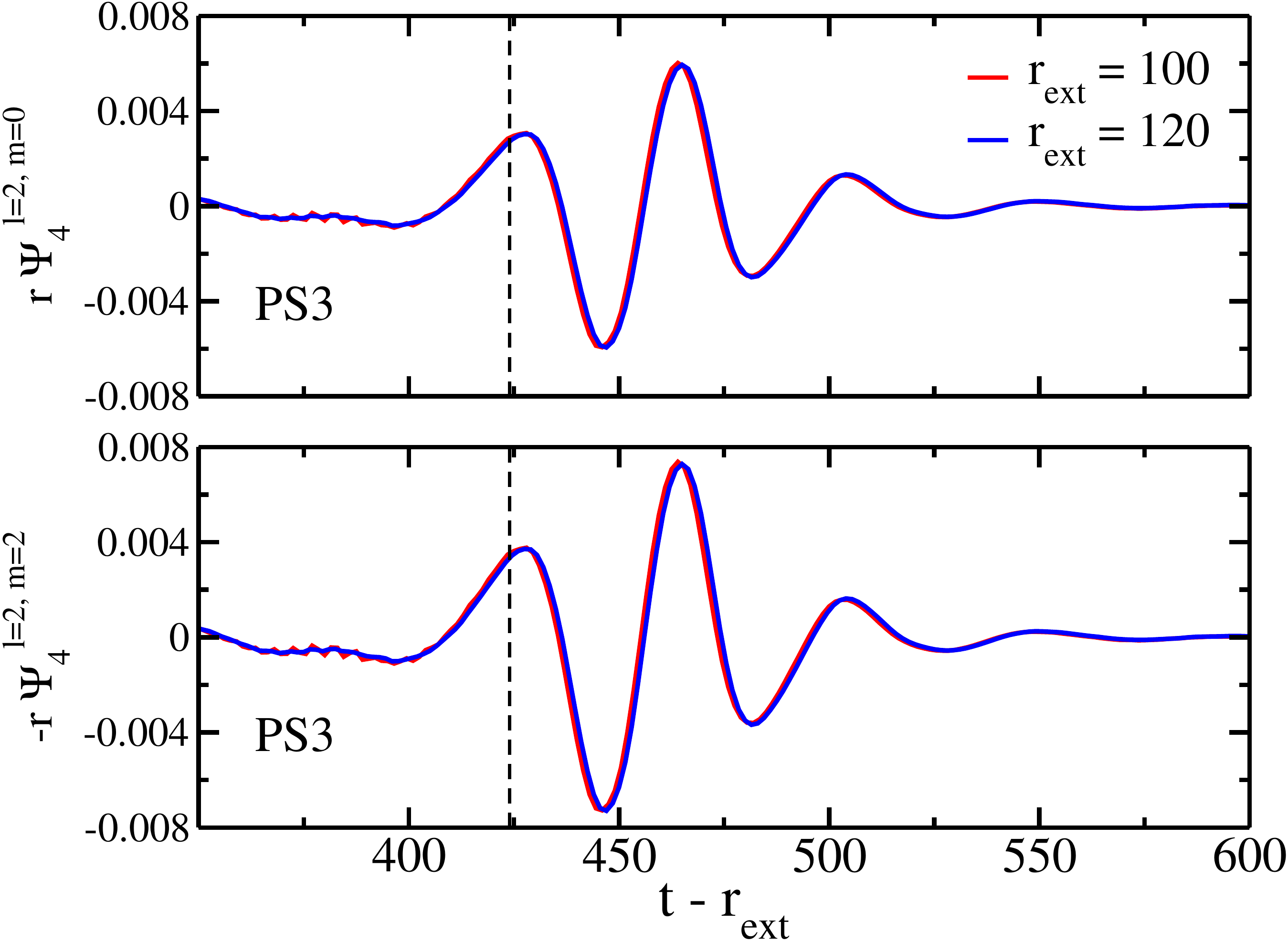}
 \includegraphics[height=2.5in]{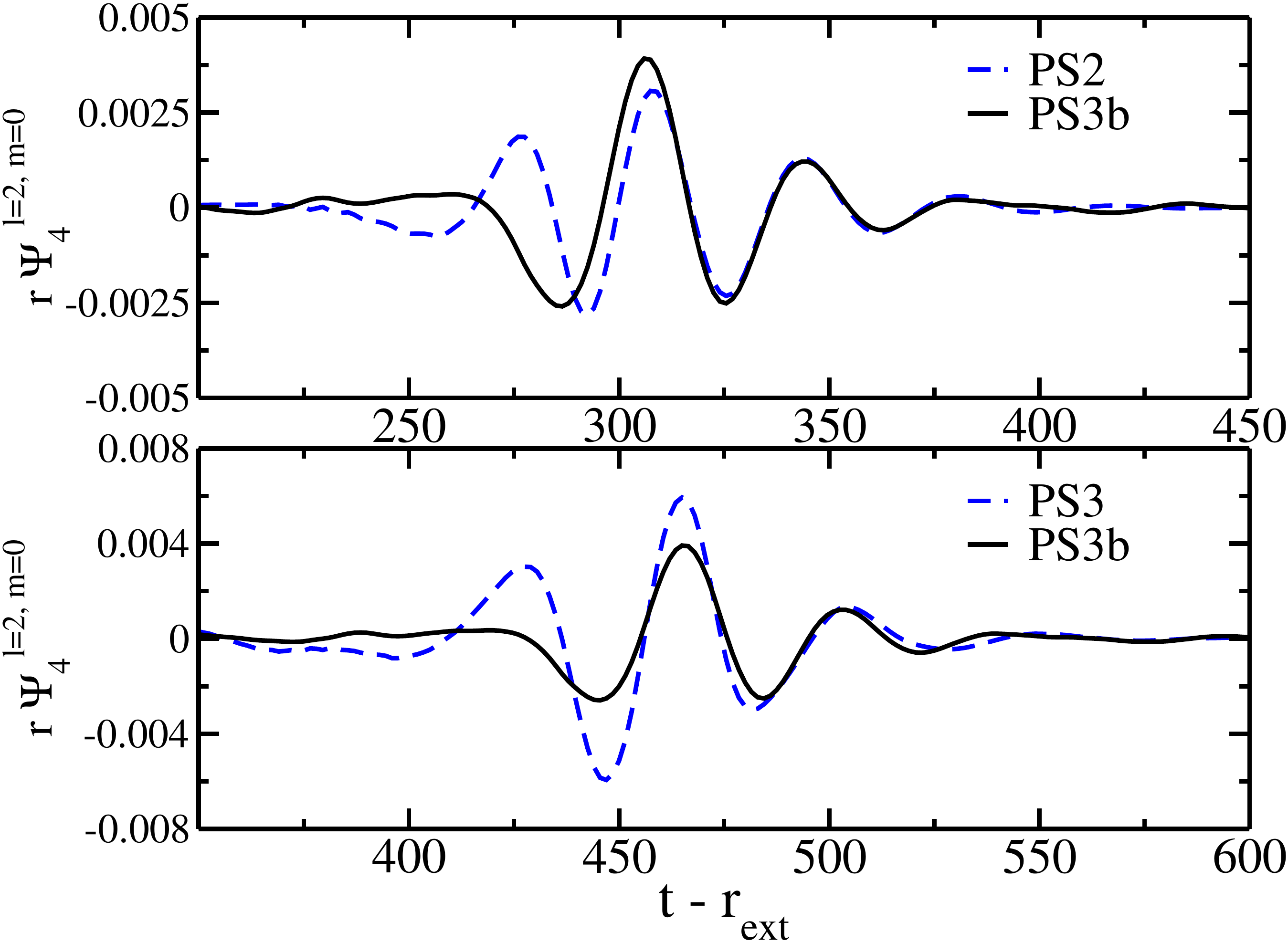} 
\caption{Real part of $r\Psi_4^{l=2,m=0}$ and $r\Psi_4^{l=2,m=+2}$ for PS3-PS3 (top panels). Comparison of the real part of $r\Psi_4^{l=2,m=0}$ between PS2-PS2, PS3-PS3 and PS3b-PS3b computed at $r_{\rm ext}=120$. The latter is conveniently time shifted (bottom panels).
}
\label{fig4}
\end{figure}

In Figs.~\ref{fig3} and \ref{fig4} we plot the resulting gravitational waveforms, showing the Newman-Penrose scalar $r\Psi_{4}^{l=2, m}$, for the modes $l=2$, $m=\lbrace0,+2\rbrace$, extracted at radii $r_{\rm ext}=\lbrace100,120\rbrace$. The waves, conveniently shifted and rescaled, overlap, as expected in the wave zone. Non-axisymmetric modes $l=2$, $m=\pm1$ are consistent with zero. In agreement with the results for head-on collisions of spherical boson stars in~\cite{palenzuela2007head}, we find that the coefficients $\mathcal{C}_{2,m}$ of the different modes are related in the following way:
\begin{eqnarray}
\text{Re}(\mathcal{C}_{2,+2}) &=&\text{Re}(\mathcal{C}_{2,-2}),\\
\text{Re}(\mathcal{C}_{2,+2}) &=&- \sqrt{3/2}\,\text{Re}(\mathcal{C}_{2,0}).\label{rescalingmodes}
\end{eqnarray}

For our sample of models, the amplitude of the gravitational waves increases monotonically with decreasing vector-field frequency. The largest amplitude is achieved for model PS3-PS3 which is more than an order of magnitude larger than for model PS0-PS0. All waveforms are of the burst-type, of which PS2-PS2 and PS3-PS3 are clear examples. Morphological differences are apparent in model PS0-PS0 and, in particular in model PS1-PS1.
For the latter, plotted in the middle panels of Fig.~\ref{fig3}, there is a time delay $\Delta t \sim 100$  between the first negative peak of the waveform (at around $t\sim 275$) and the collapse (at around $t\sim 375$). The waveform can be regarded as composed by two contributions: the first part, from $t\sim 250$ to $t\sim 375$, would correspond to the collision of the stars, forming a ``hypermassive" PS, a remnant that oscillates and eventually collapses to a BH at $t\sim 375$, triggering the second part of the wave. The collapse nearly coincides with a peak of the waveform and the corresponding time is highlighted by a vertical dashed line in the figure. For PS2-PS2 and PS3-PS3, these dynamics are absent, the AHs form promptly and their times of formation are very close to the first positive peak in the waveforms (vertical dashed lines).

\begin{figure}%[h!]
\centering
\includegraphics[height=2.5in]{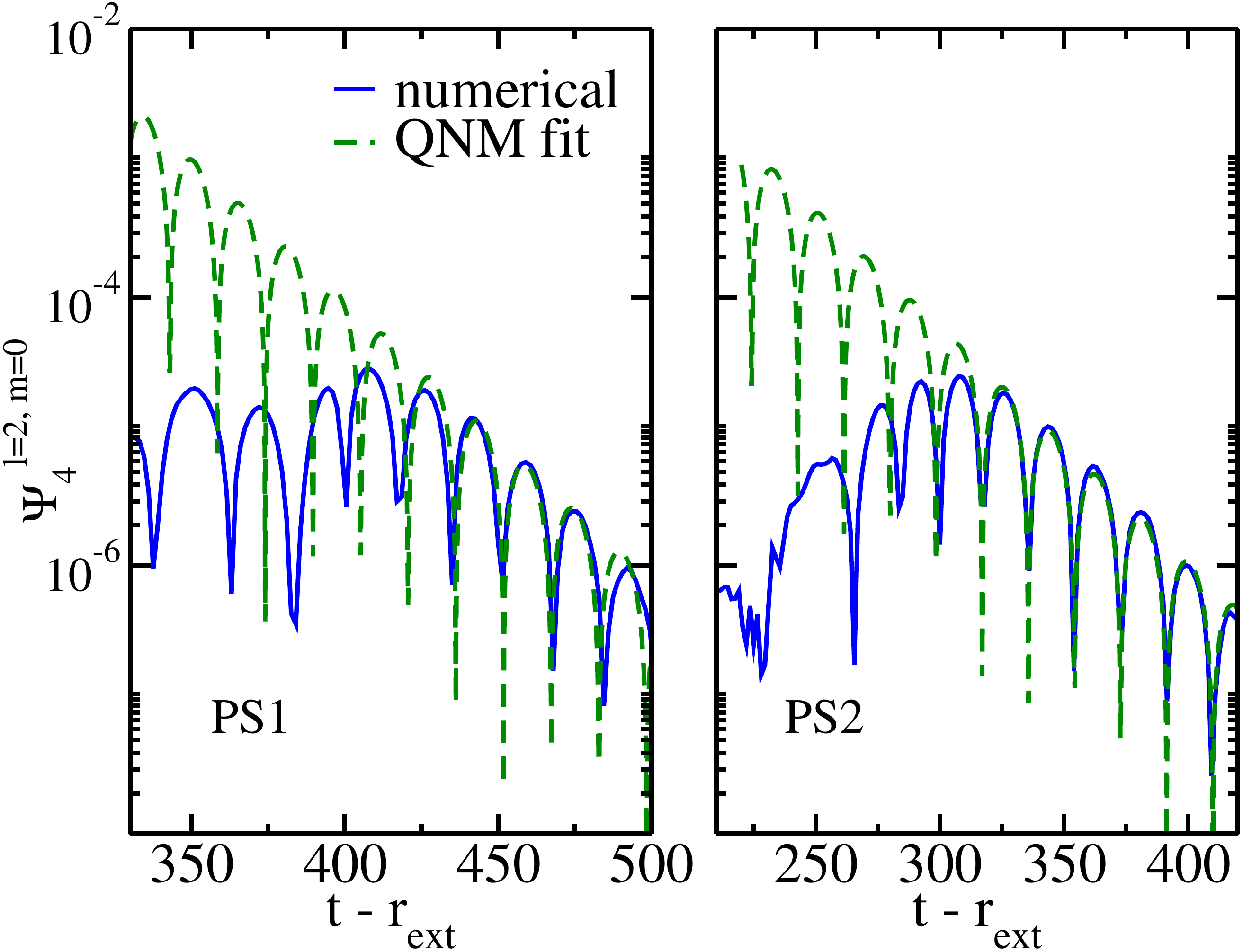} 
\includegraphics[height=2.5in]{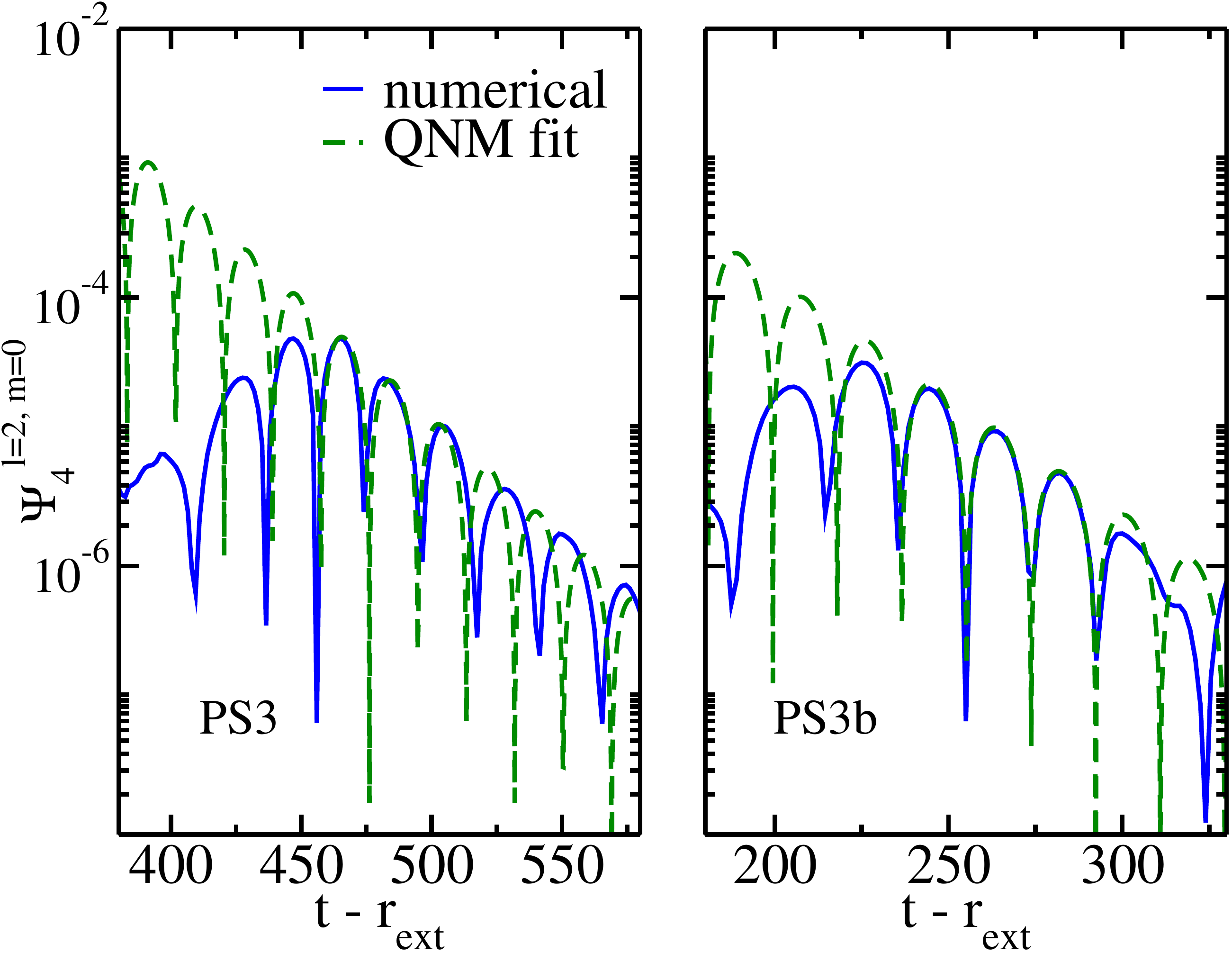}  
\caption{Black hole ringing and quasinormal modes (QNMs) for models PS1-PS1 (top left panel), PS2-PS2 (top right panel), PS3-PS3 (bottom left panel), and PS3b-PS3b (bottom right panel). The QNM fit shown for PS2-PS2, PS3-PS3 and PS3b-PS3b corresponds to the same BH mass. All waveforms are computed at $r_{\rm ext}=120$.
}
\label{fig5}
\end{figure}

The bottom panels of Fig.~\ref{fig4} show the gravitational waveform for the PS3b-PS3b model together with those of models PS2-PS2 and PS3-PS3. Recall that PS2 and PS3 have almost the same mass (less than 1\% difference), therefore the BH formed will have a similar mass. Thus, we can compare the quasinormal modes (QNMs) of the three cases. For PS2-PS2, the QNMs are in good agreement in at least four of the peaks, while for PS3-PS3 there are some differences in the frequency. This comparison becomes more  clear in the QNM ringdown plots shown in 
 Fig.~\ref{fig5}, where we plot the waveforms in logarithmic scale and we fit them to the QNMs of a Schwarzschild BH, following the results of~\cite{Leaver:1985ax}. PS2-PS2, PS3-PS3 and PS3b-PS3b have almost the same total mass, therefore the fit is the same. The agreement for PS2-PS2 and PS3b-PS3b is very good. The masses that we obtain from the fits have an error of $7\%$ for PS1-PS1 and $4\%$ for PS2-PS2, PS3-PS3 and PS3b-PS3b, with respect to the masses computed with the \texttt{AHFinder} algorithm of the \texttt{Einstein Toolkit}. However, for the PS3-PS3 waveform the frequency does not match the PS3b-PS3b nor the QNM fit. A possible interpretation is that this is due to the fact that there is still a rich Proca field environment around the newly form BH, $cf.$ Figs.~\ref{fig2b} and~\ref{fig2c}.
 
\begin{figure}[h!]
\centering
 \includegraphics[height=2.5in]{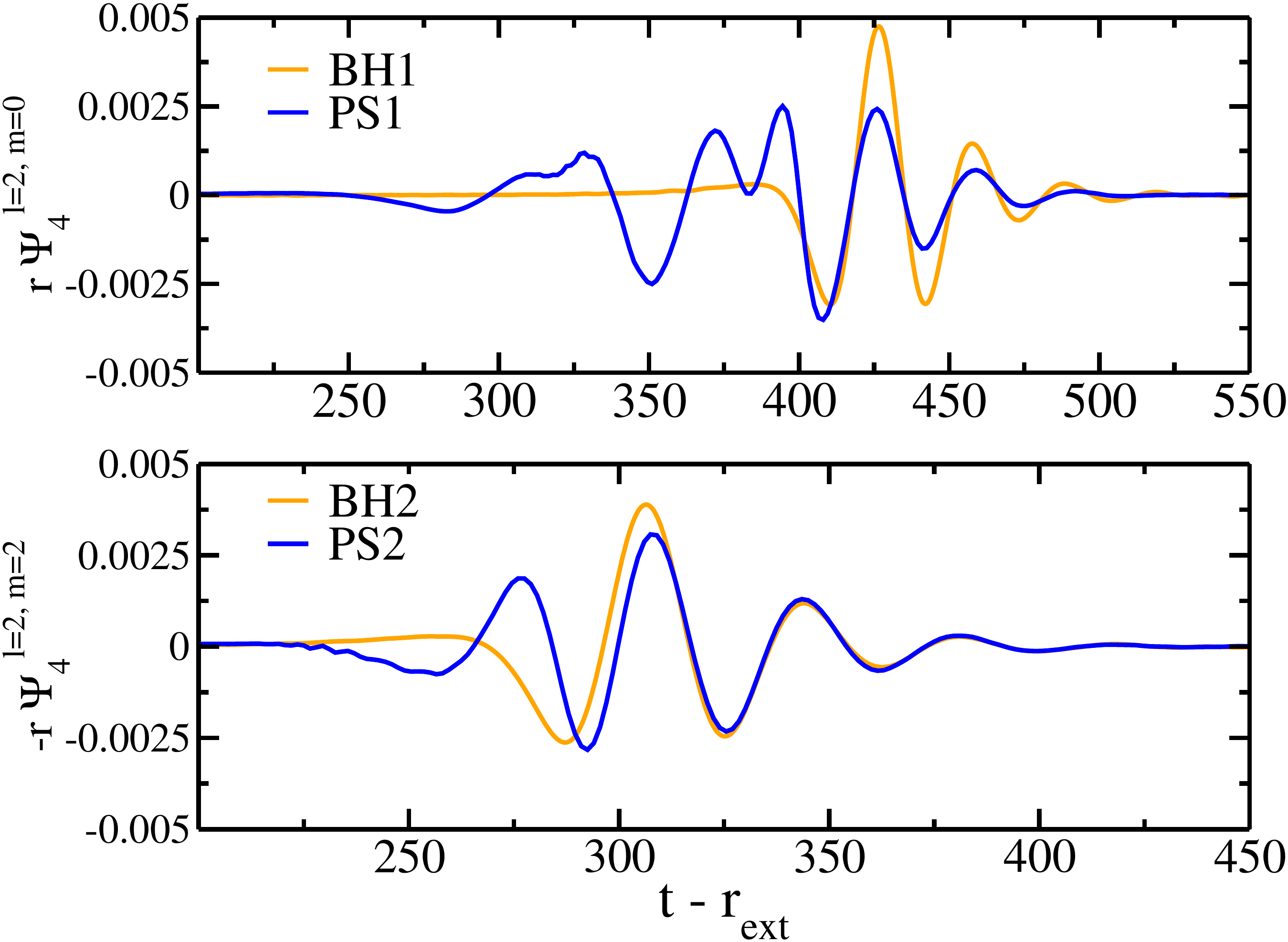} \includegraphics[height=2.5in]{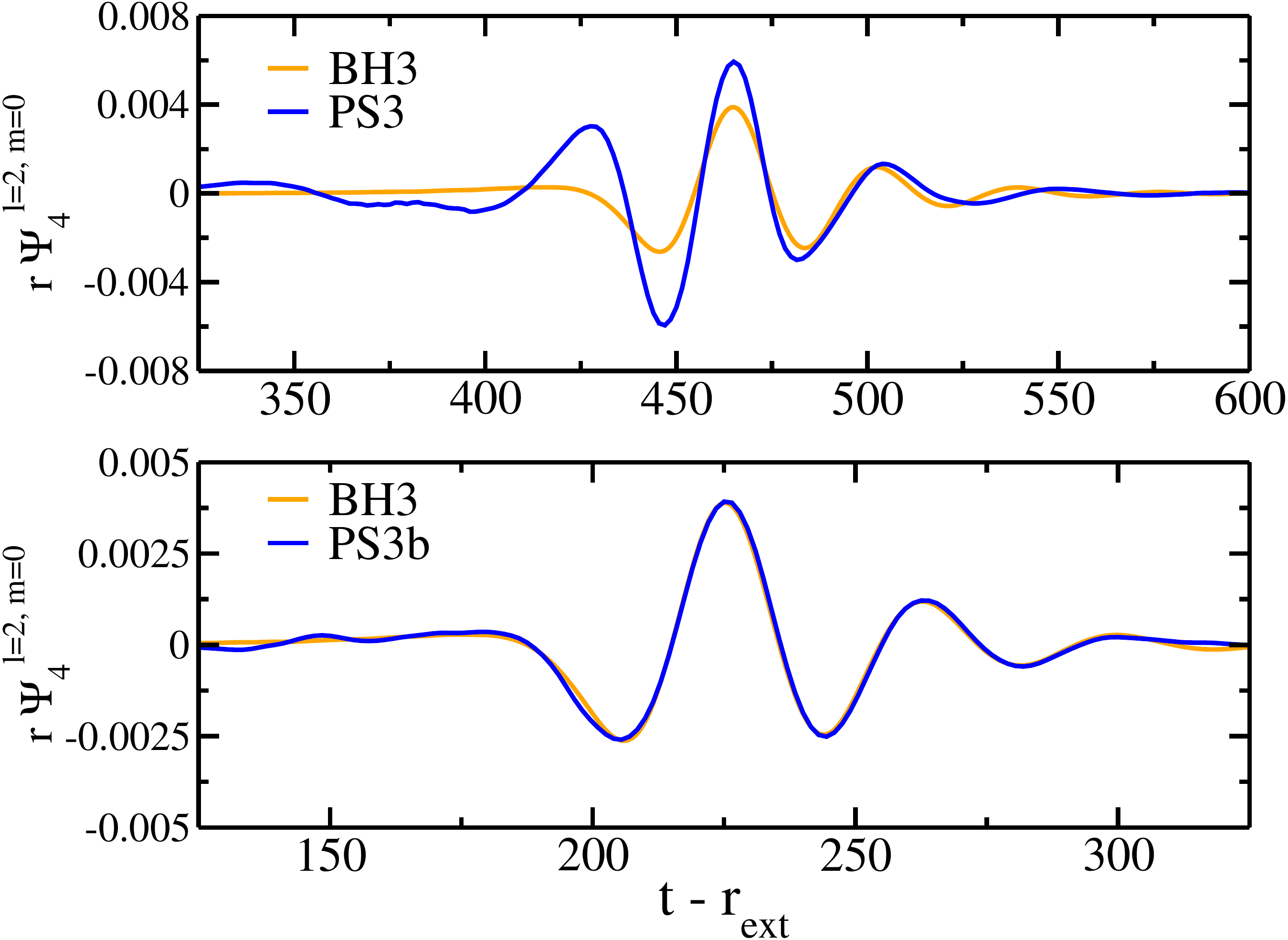} 
\caption{Comparison of the real part of $r\Psi_4^{l=2,m=0}$ computed at $r_{\rm ext}=120$ from head-on collisions of PSs (blue lines) and Schwarzschild BHs of the same mass (orange lines).
}
\label{fig5b}
\end{figure}

\begin{figure}[h!]
\centering
\includegraphics[height=2.5in]{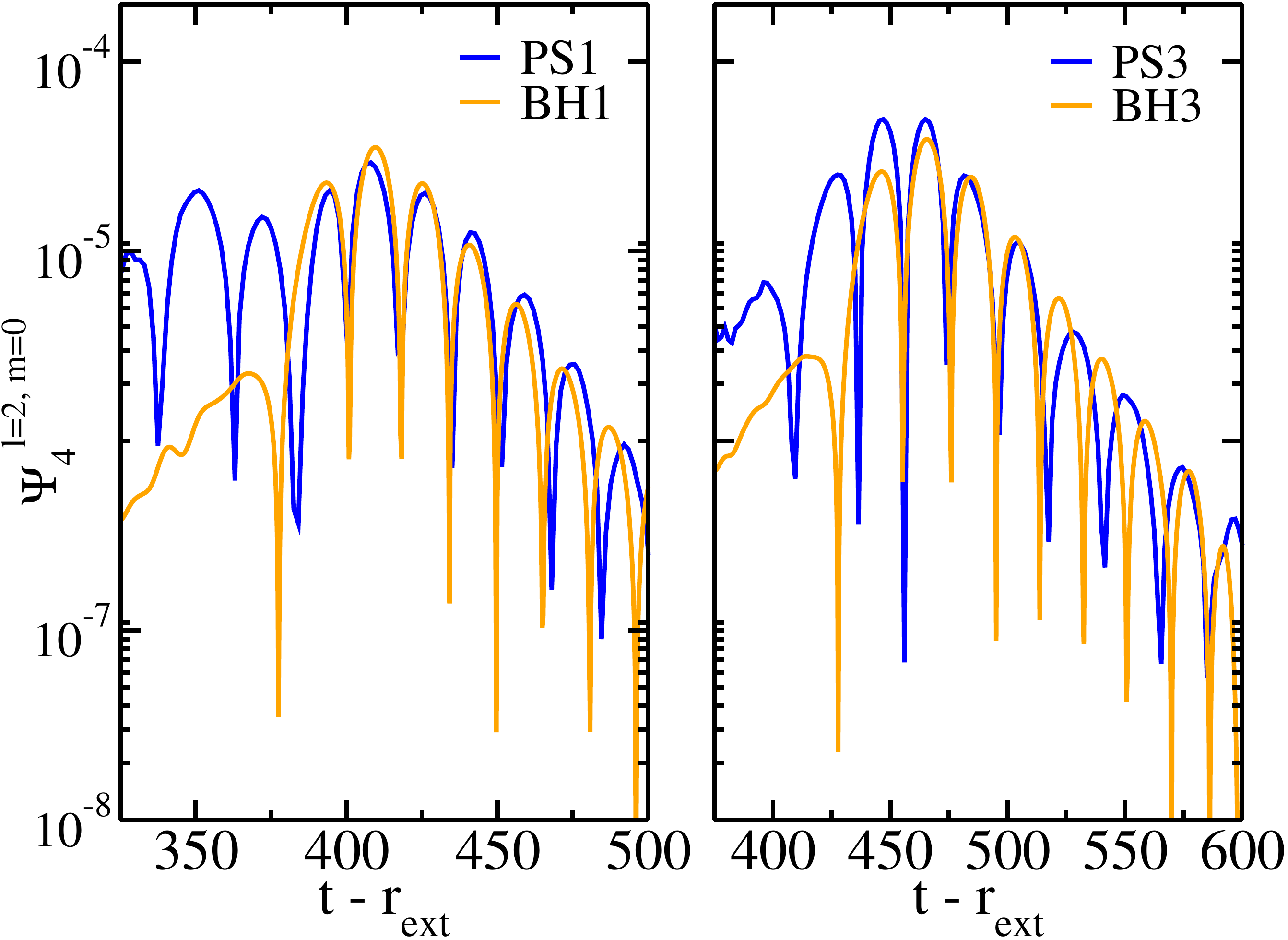}
\caption{Black hole ringing and quasinormal modes (QNMs) for models PS1-PS1 and the corresponding BH binary with the same mass (left panel). Same for PS3-PS3 (right panel).
}
\label{fig5c}
\end{figure}

%\begin{figure}[h!]
%\centering
%\includegraphics[height=3.5in]{Phi1-Model1.pdf}  \includegraphics[height=3.5in]{Phi1-Model2-Model3.pdf} 
%\caption{(Top panels) Time evolution of scalar potential $\Phi$ for PS1-PS1 extracted at $r=0$. Time evolution of scalar potential $\Phi$ for PS2-PS2 (middle panel) and PS3-PS3 (bottom panel) extracted at $r=0$.
%}
%\label{fig2}
%\end{figure}

%\begin{figure}[h!]
%\centering
%\includegraphics[height=2.5in]{movie0250.pdf}
%\includegraphics[height=2.5in]{movie0250.pdf}
%\includegraphics[height=2.5in]{movie0250.pdf}
%\caption{Snapshots for PS2-PS2.
%}
%\label{fig6}
%\end{figure}
\begin{figure*}%[h!]
\centering
\includegraphics[height=1.3in]{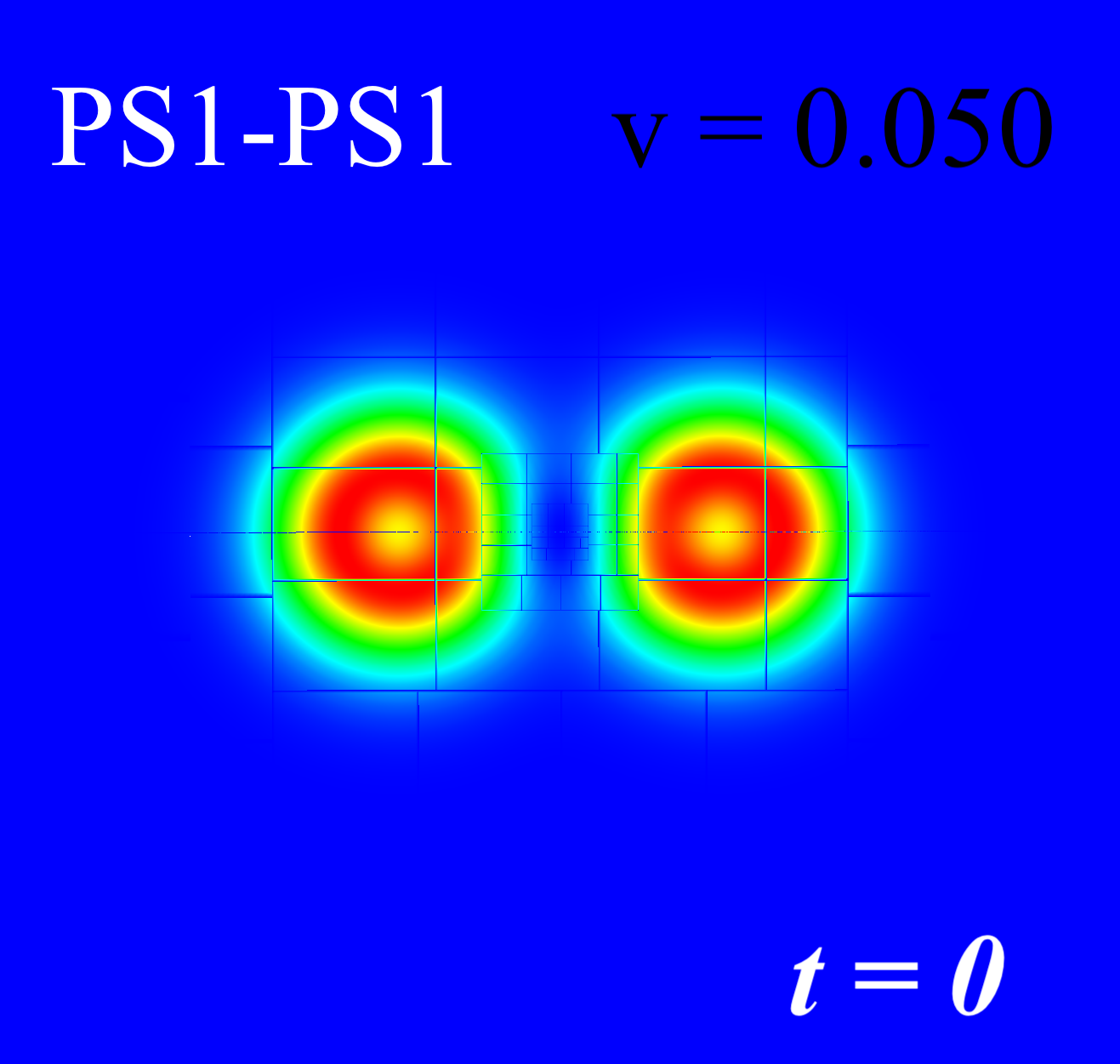} \includegraphics[height=1.3in]{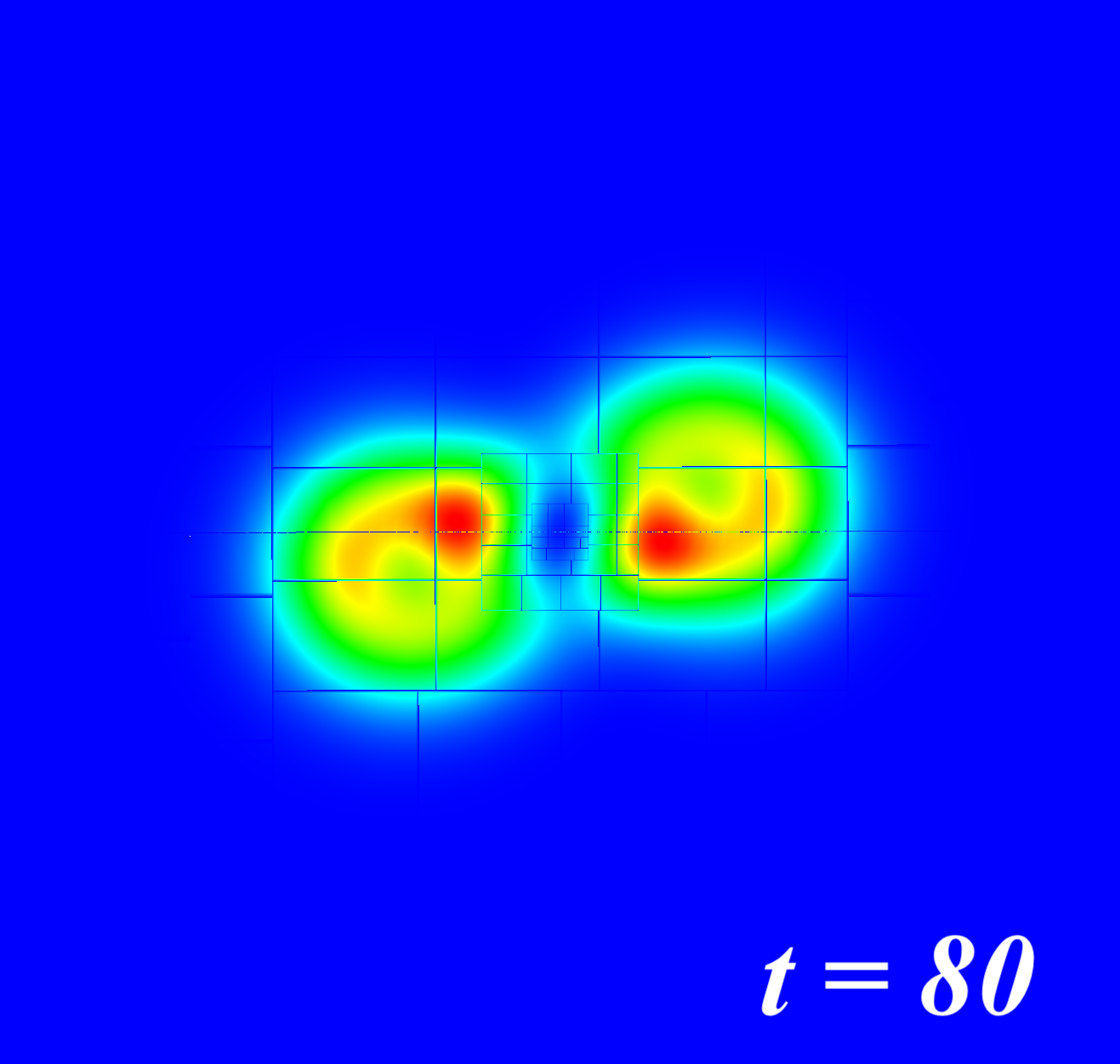} \includegraphics[height=1.3in]{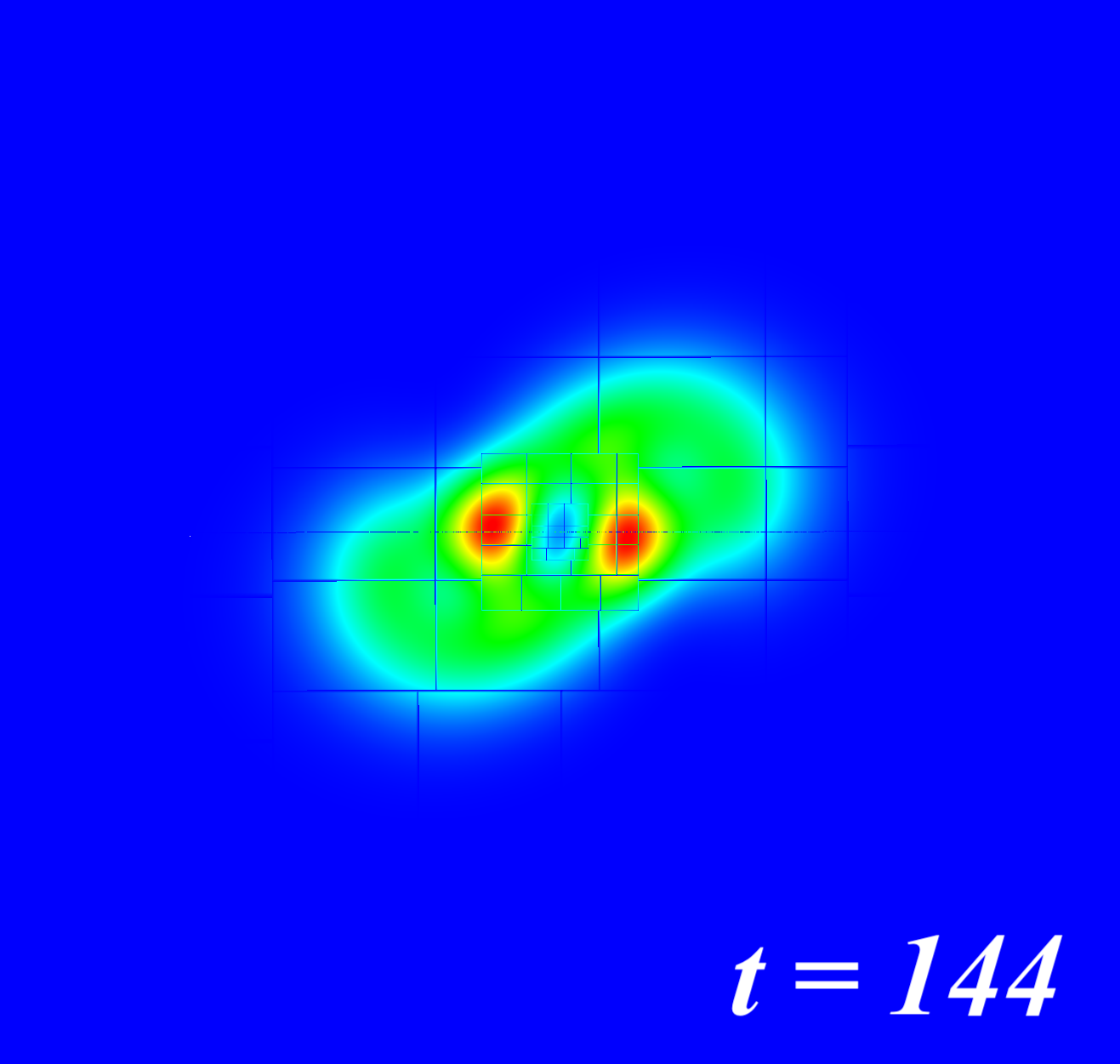} \includegraphics[height=1.3in]{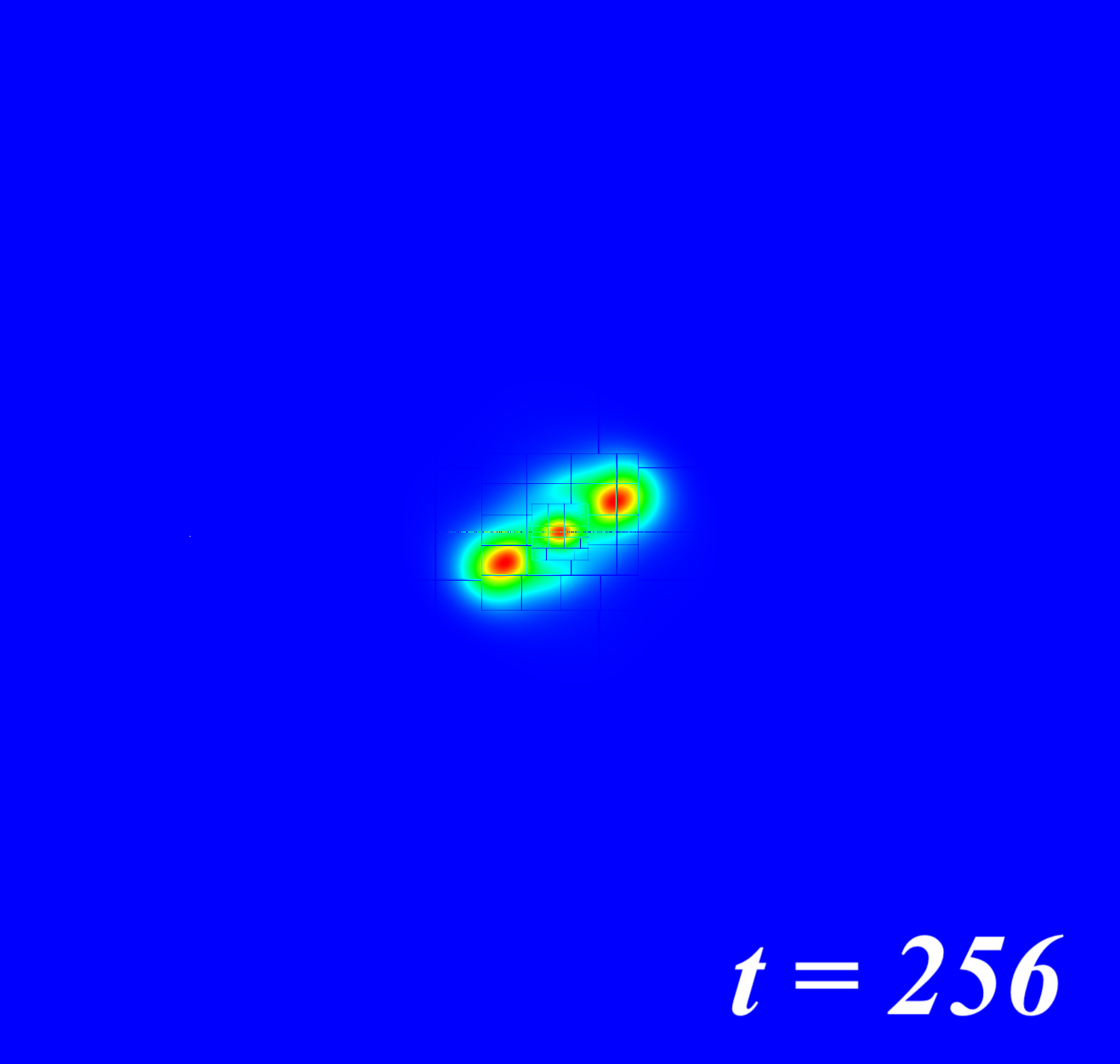} \includegraphics[height=1.3in]{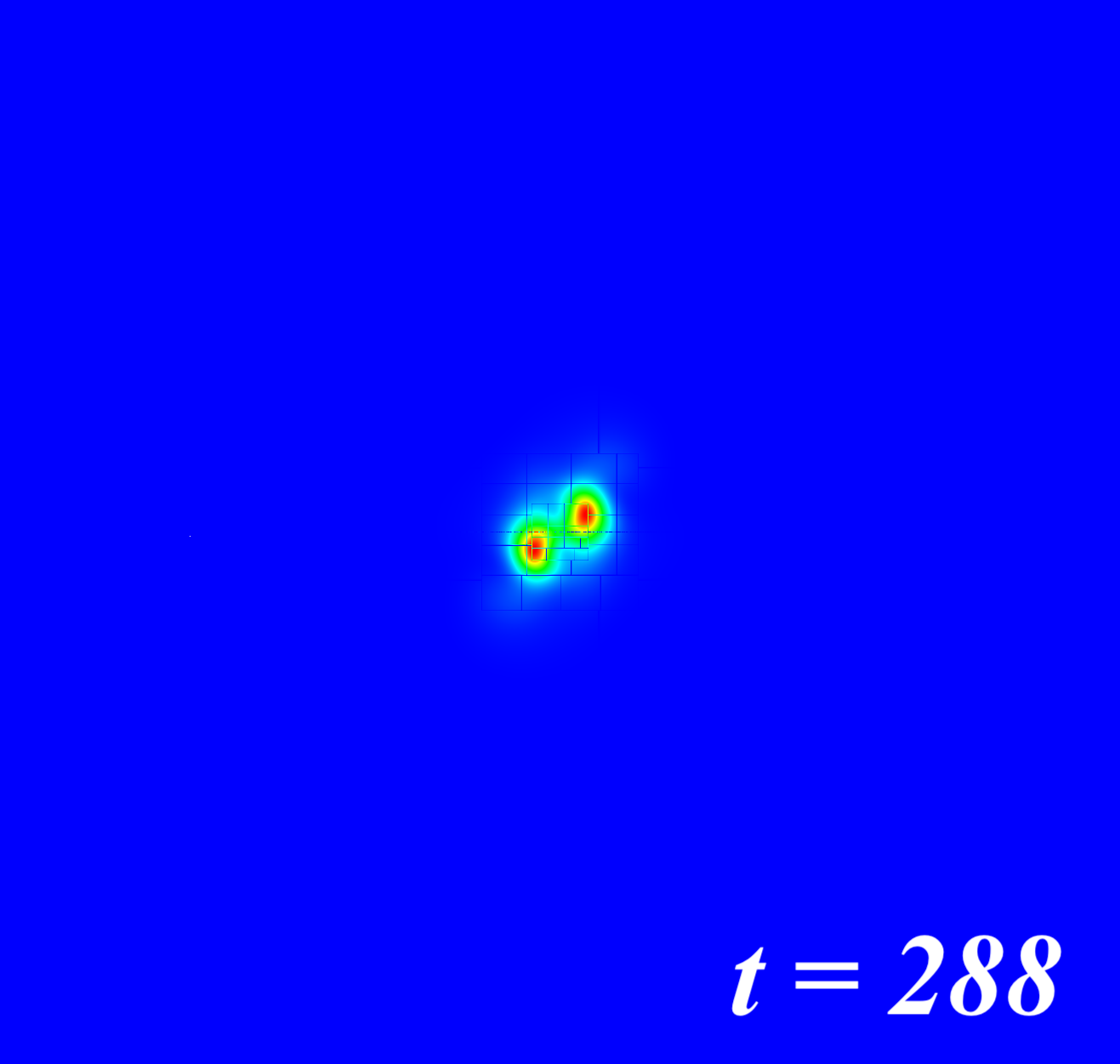}\\
\smallskip
\includegraphics[height=1.3in]{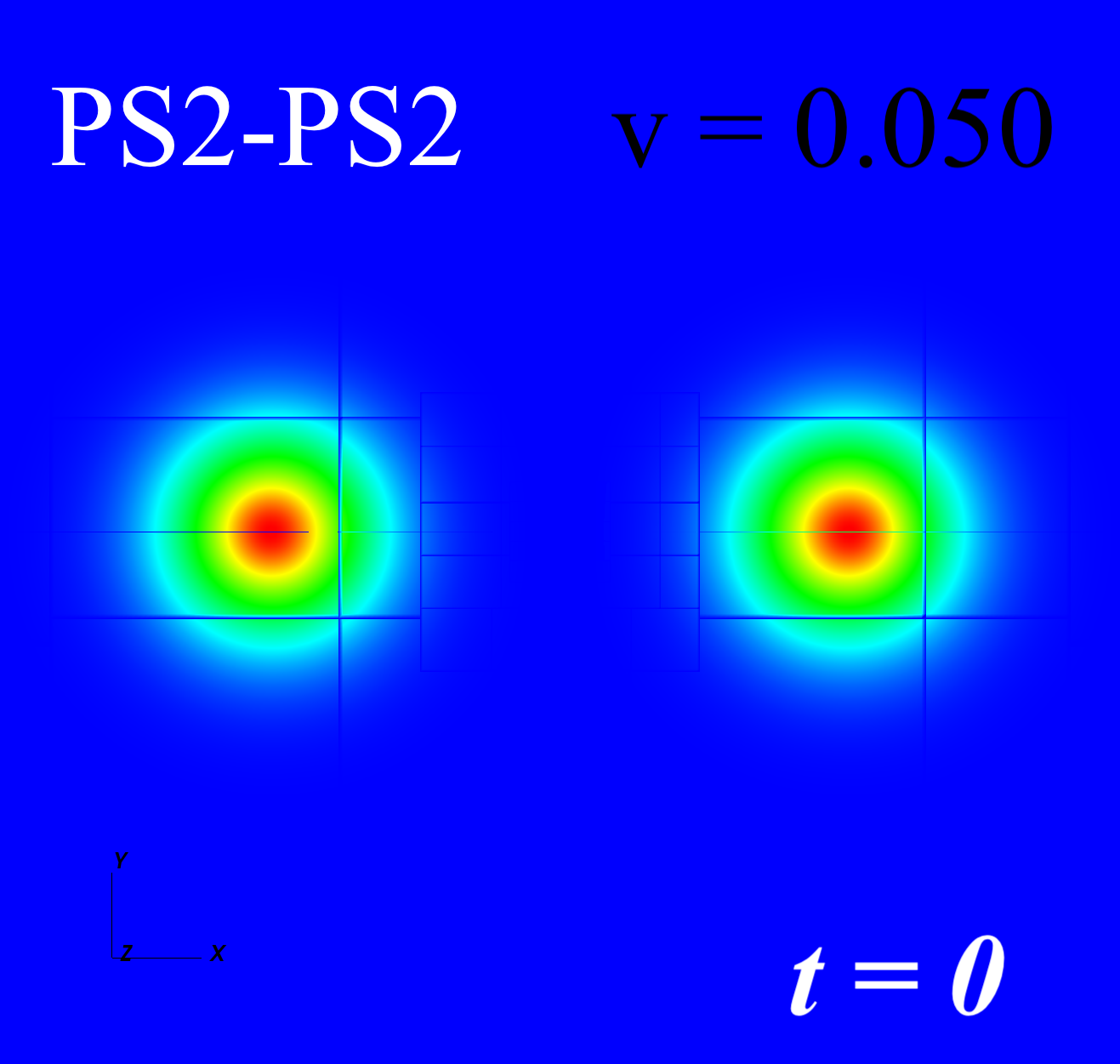} \includegraphics[height=1.3in]{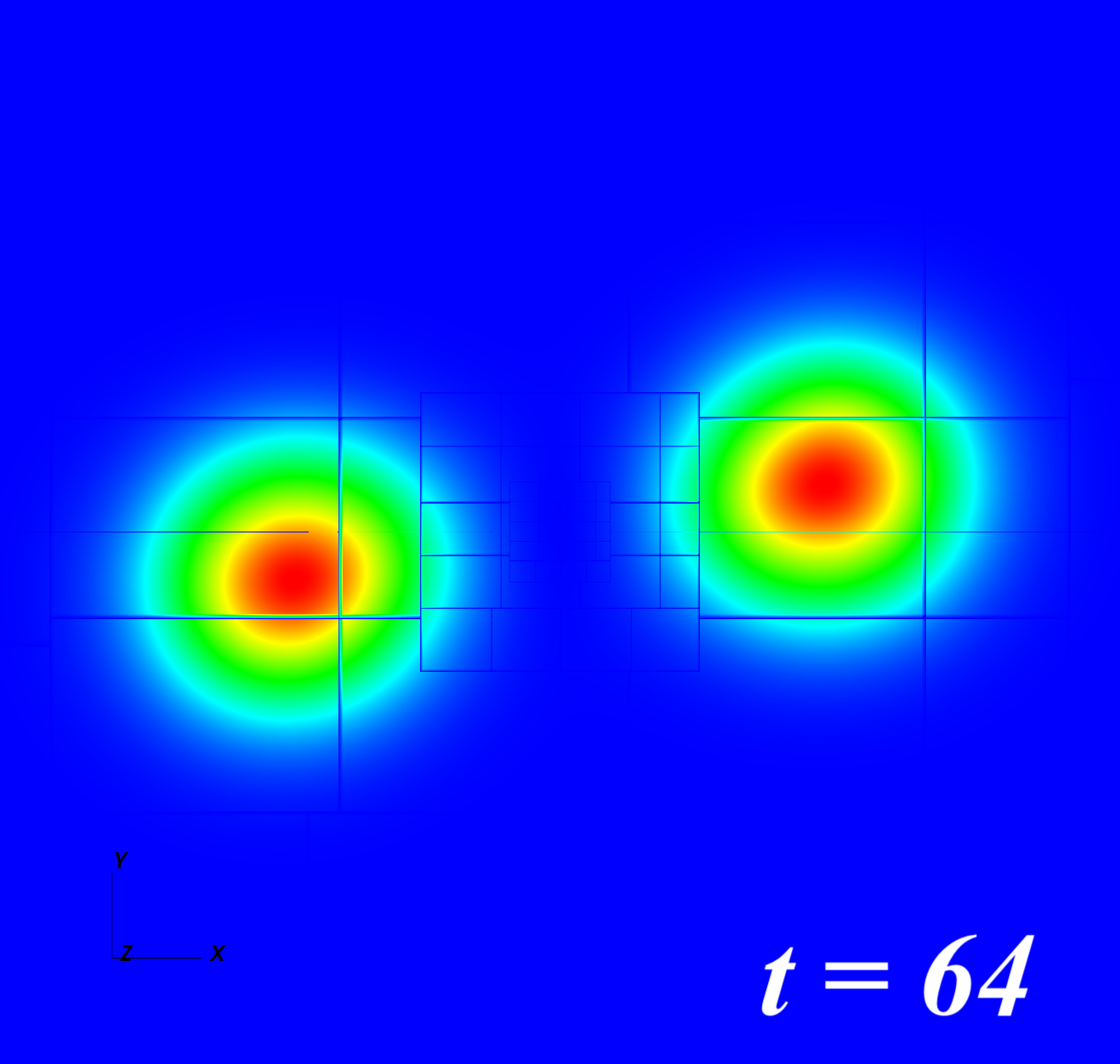} \includegraphics[height=1.3in]{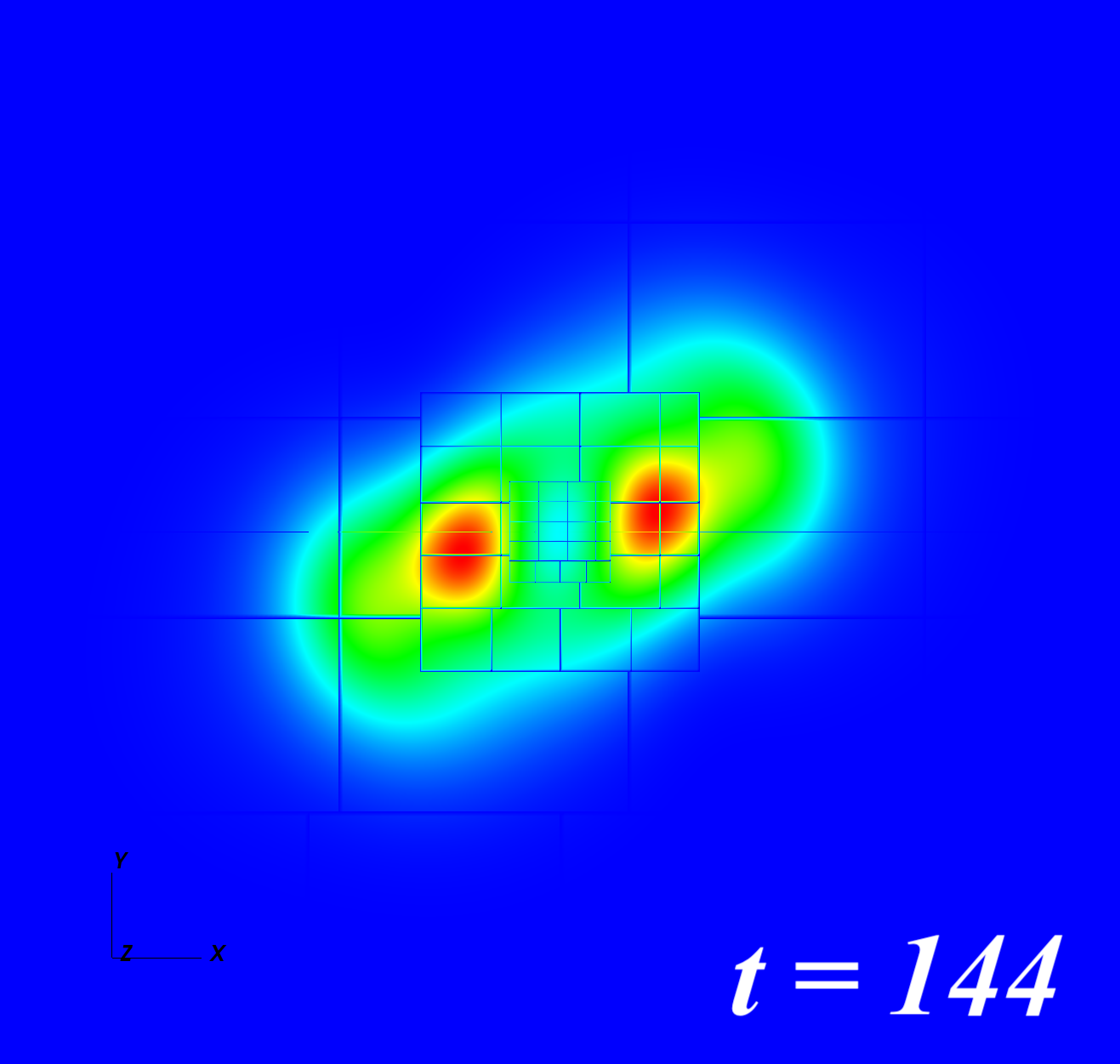} \includegraphics[height=1.3in]{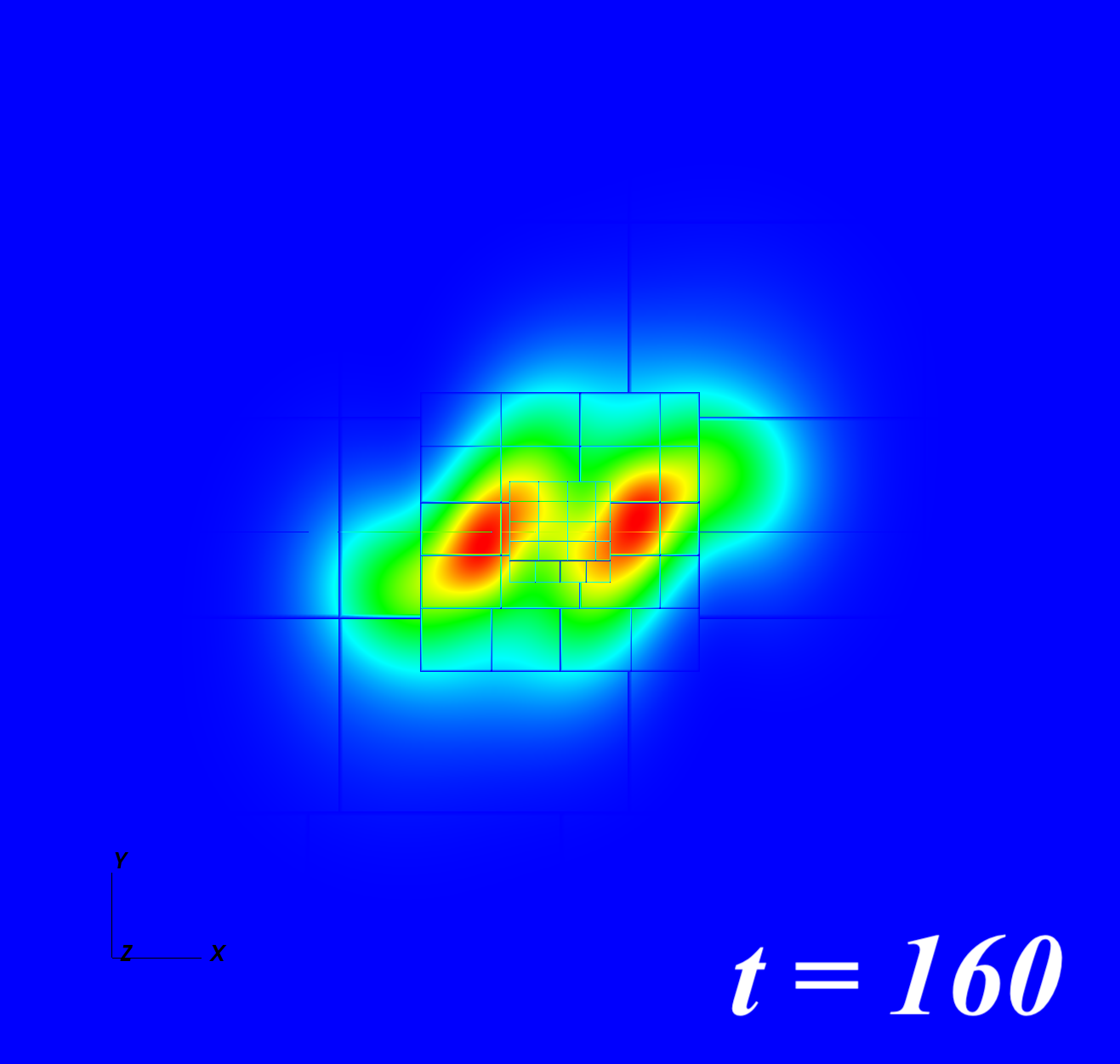} \includegraphics[height=1.3in]{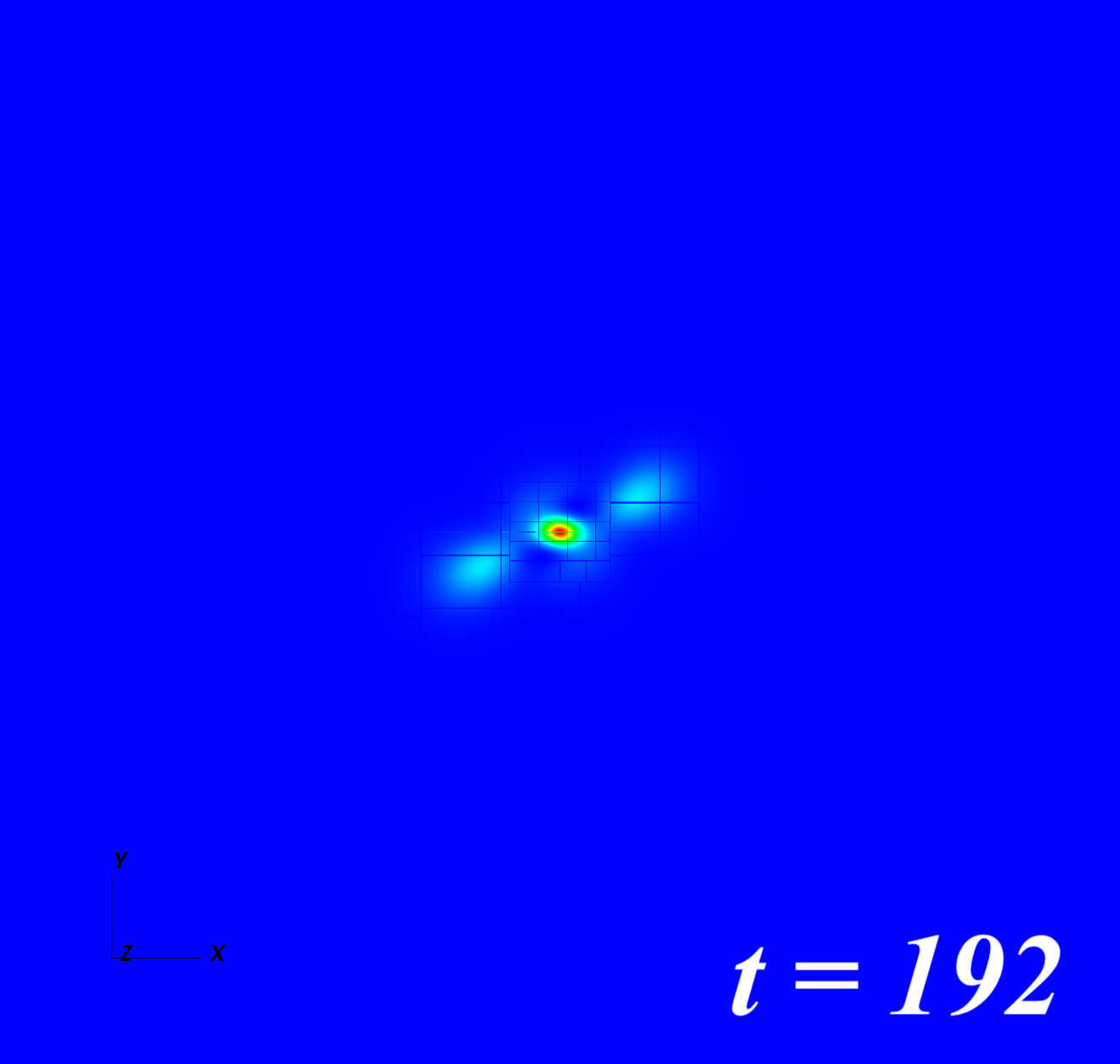}\\
\smallskip
\includegraphics[height=1.3in]{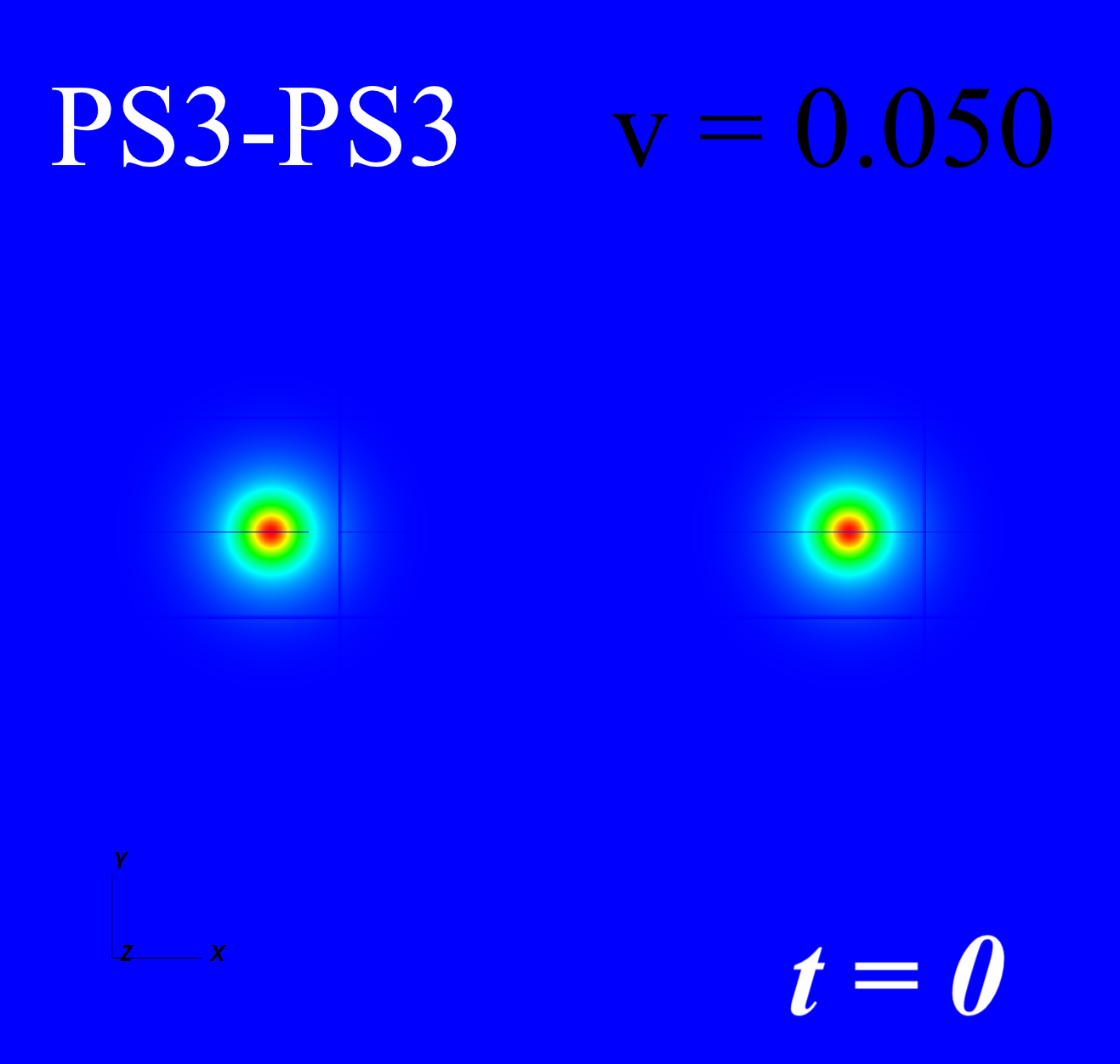} \includegraphics[height=1.3in]{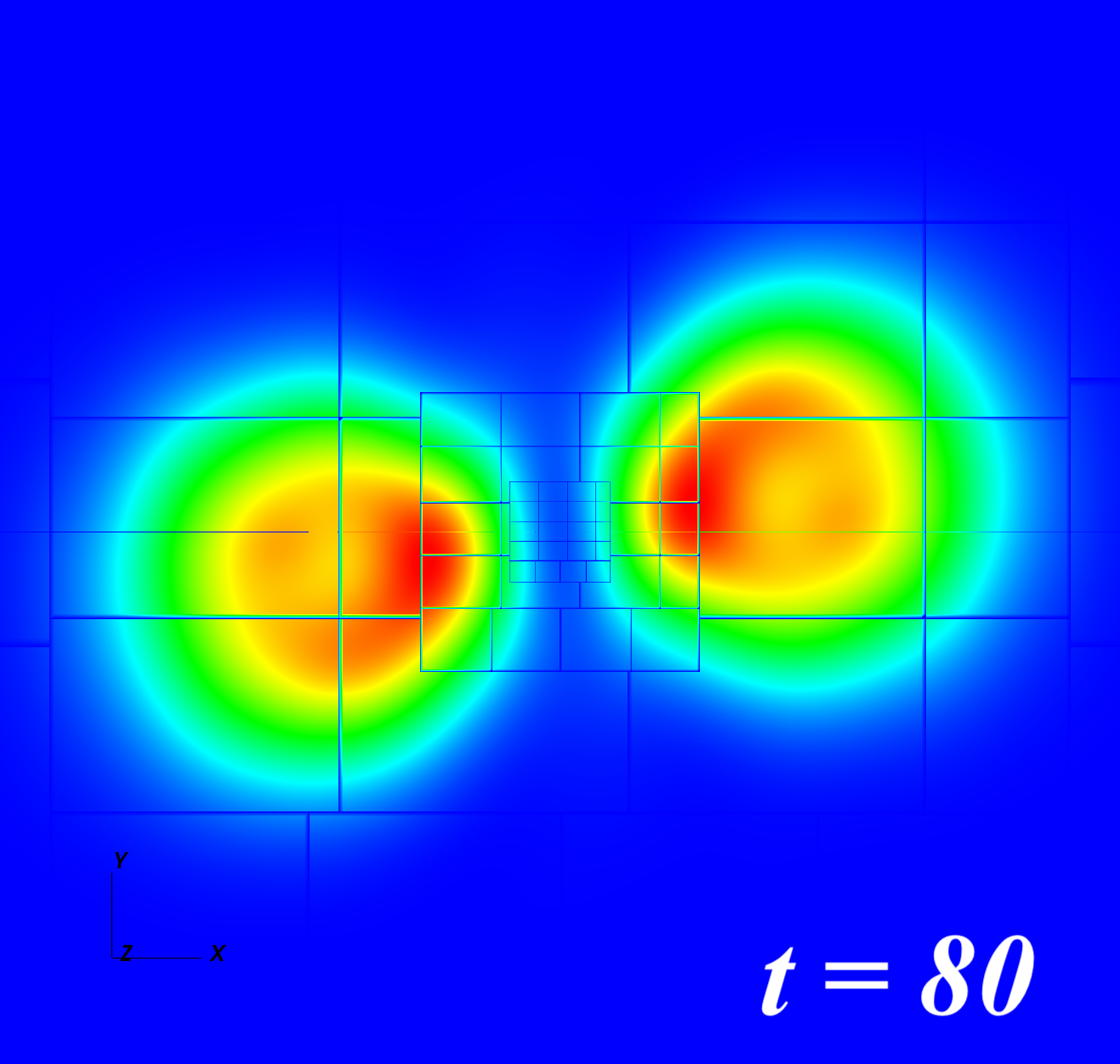} \includegraphics[height=1.3in]{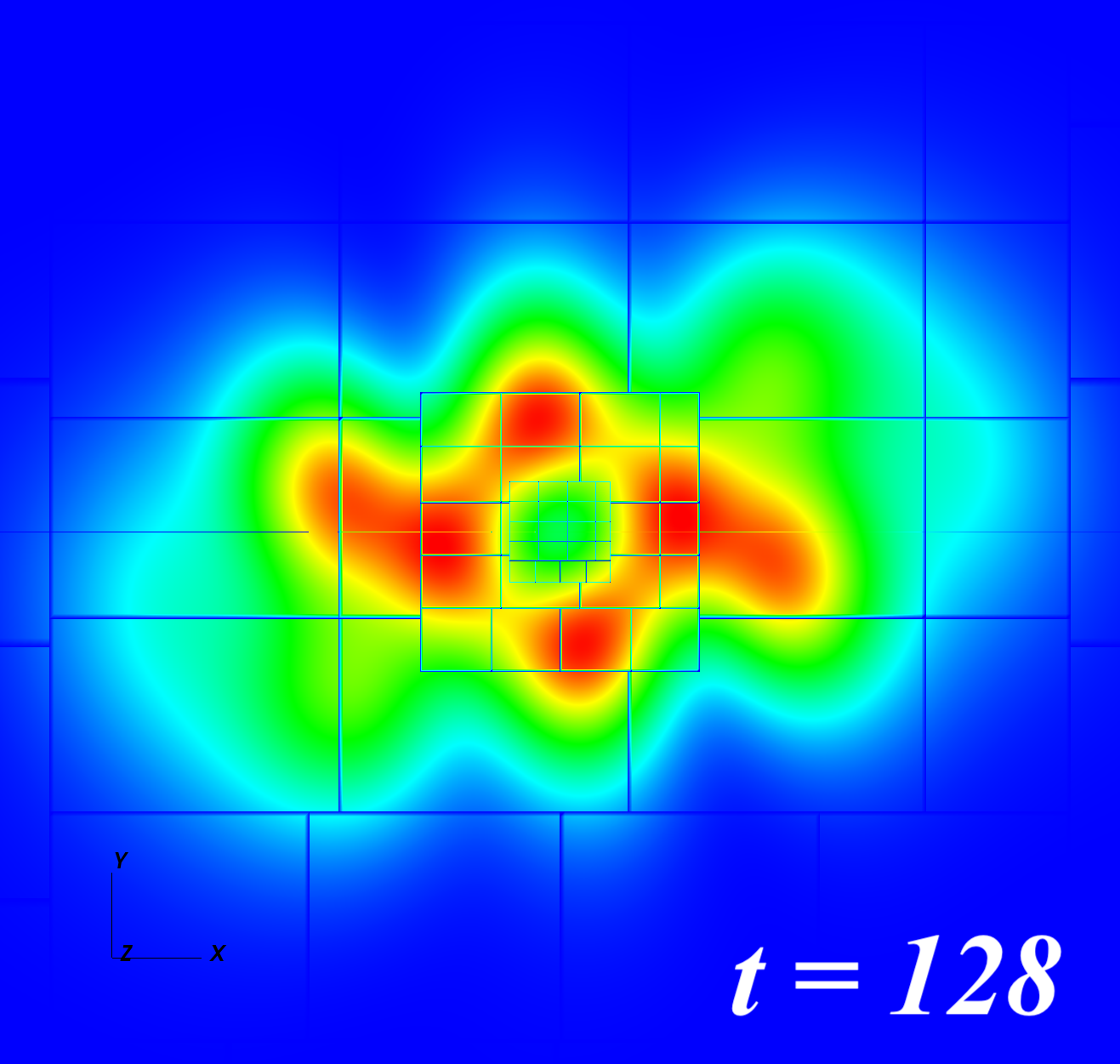} \includegraphics[height=1.3in]{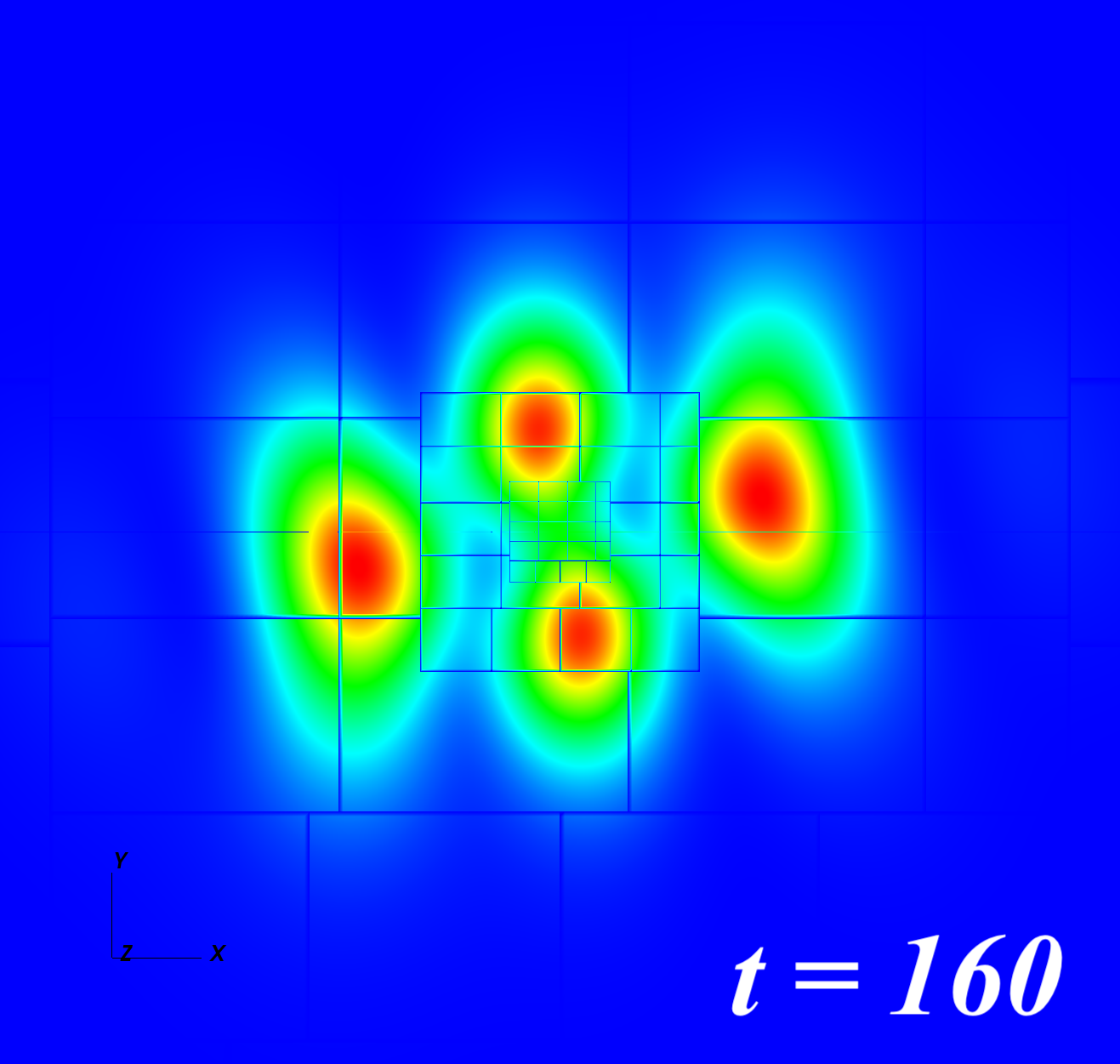} \includegraphics[height=1.3in]{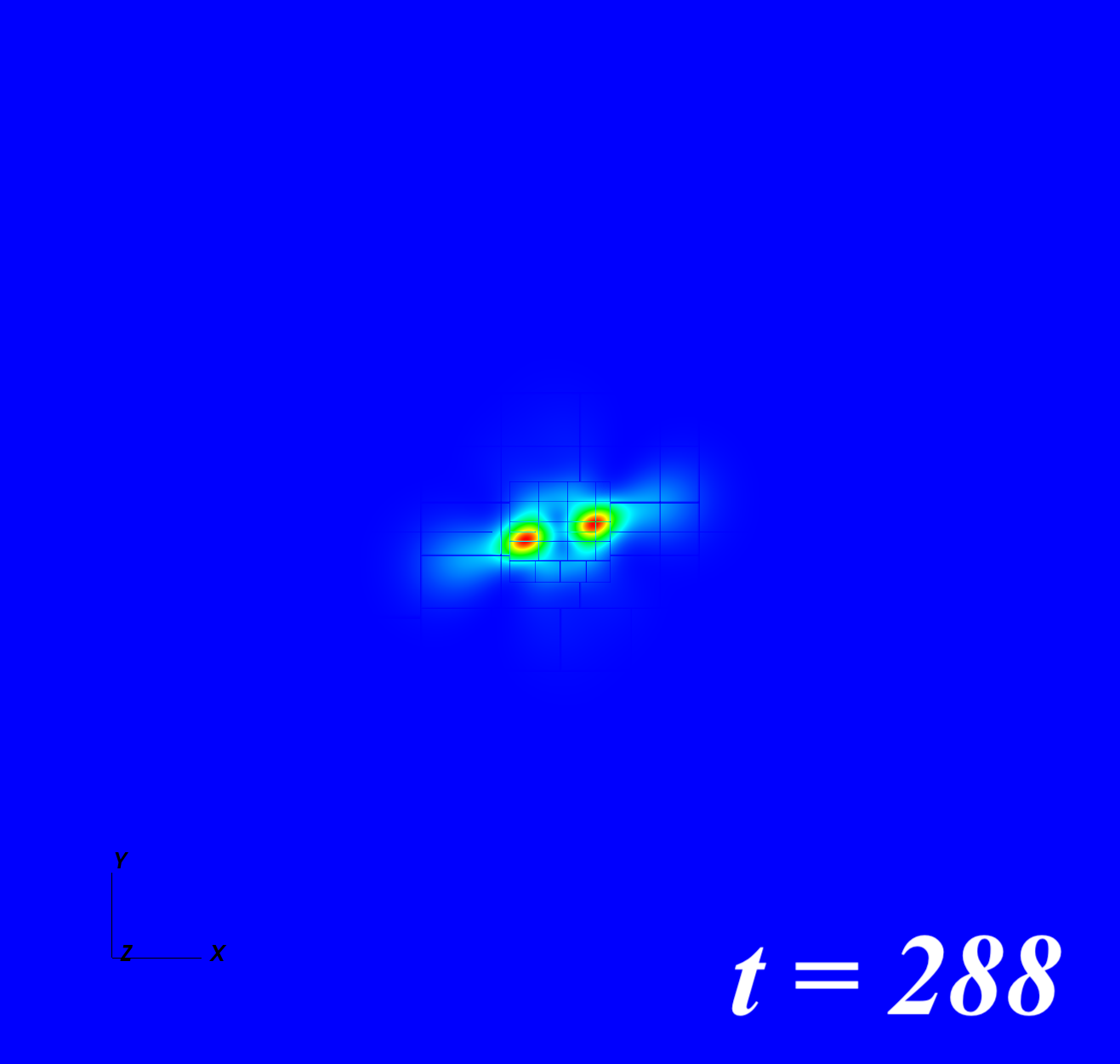}\\

\caption{Snapshots of the energy density for the orbital merger of PS1-PS1, PS2-PS2 and PS3-PS3 in the equatorial plane. The vertical axis correspond to the $y$ direction and the horizontal to the $x$ direction.
}
\label{fig:merger0}
\end{figure*}

 %%%%%%%%%%%%%%%%%%%%%%%%%%%%%%%%%%%%%%%%%%%%%%%%%%%%%%%%%%%%%
\subsubsection{Comparison with BH collisions}
 %%%%%%%%%%%%%%%%%%%%%%%%%%%%%%%%%%%%%%%%%%%%%%%%%%%%%%%%%%%%%
 
 We also perform head-on collisions of Schwarzschild BHs with the same mass as the PSs in each model. This allows to compare the resulting gravitational waveforms with those from the PS binaries. The real part of $r\Psi_4^{l=2,\,m=0}$ for both cases is shown in Fig.~\ref{fig5b}. Observe that for model PS3b-PS3b, we obtain essentially the same gravitational waveform. This is the expected result; in this case, the PSs collapse to BHs before the actual collision due to the initial perturbation we have imposed, thus leading to a head-on collision of BHs. In the case of PS2-PS2, the ringdown phase overlaps almost perfectly that same phase for the BH collision in agreement with the QNMs analysis in Fig.~\ref{fig5}. The first part of the wave, however, is different: the frequency is higher but the amplitude is slightly smaller at the peak of the emission. 

On the other hand, as already shown in the previous subsection, for PS1-PS1 and PS3-PS3, the comparison indicates that the waveforms depart from those of the Schwarzschild BH collisions. The amplitudes are different for both models. Some of the QNM oscillations can be fitted with the Schwarzschild BH-BH waveforms but not all, as the frequency changes, see~Fig.~\ref{fig5c}.  The deviation in the frequency could be a signature of the presence of quasi-bound states around the BH and might be in particular related to the compactness of the remaining Proca field around the BH. The gravitational radiation induced by accreting shells of matter evolving in fixed BH backgrounds was first studied in~\cite{papadopoulos1999matter} by numerically solving the linearized curvature perturbation equations (see also~\cite{PF2} for a dynamical spacetime study). It was found that  the excitation of the BH QNM ringing strongly depends on the shell thickness, becoming increasingly clear with progressively more compact shells (see Fig.~10 in~\cite{papadopoulos1999matter}). In the infinitesimally thin limit, the gravitational energy asymptotes to a finite value, about a third of the point particle upper limit. Those findings confirmed earlier ideas 
about the QNM excitation mechanism made by~\cite{sun-price-1990}, namely that the {\it strong} excitation is induced by curvature profiles that have
spatial  wavelengths comparable to the width of the BH potential. 

\begin{figure*}%[t!]
\centering
\includegraphics[height=1.25in]{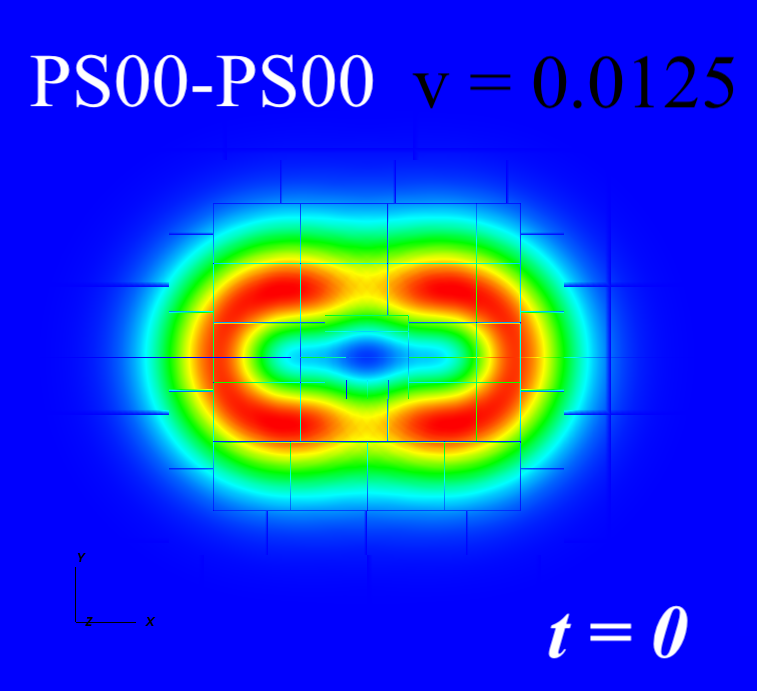} \includegraphics[height=1.25in]{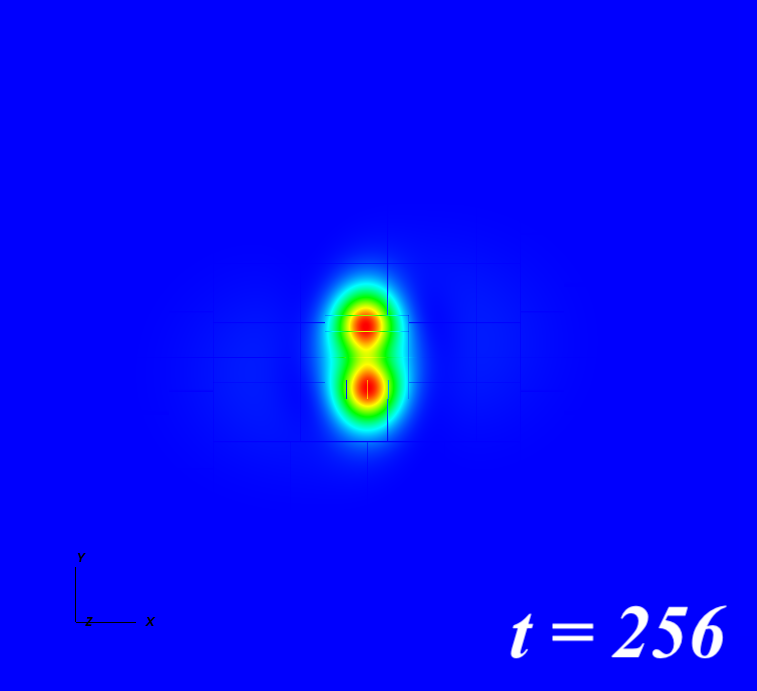} \includegraphics[height=1.25in]{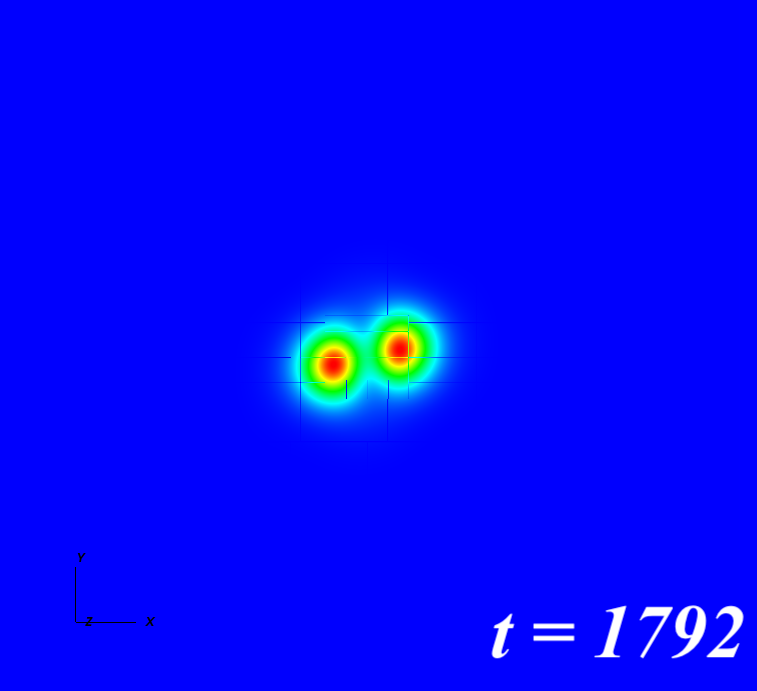} \includegraphics[height=1.25in]{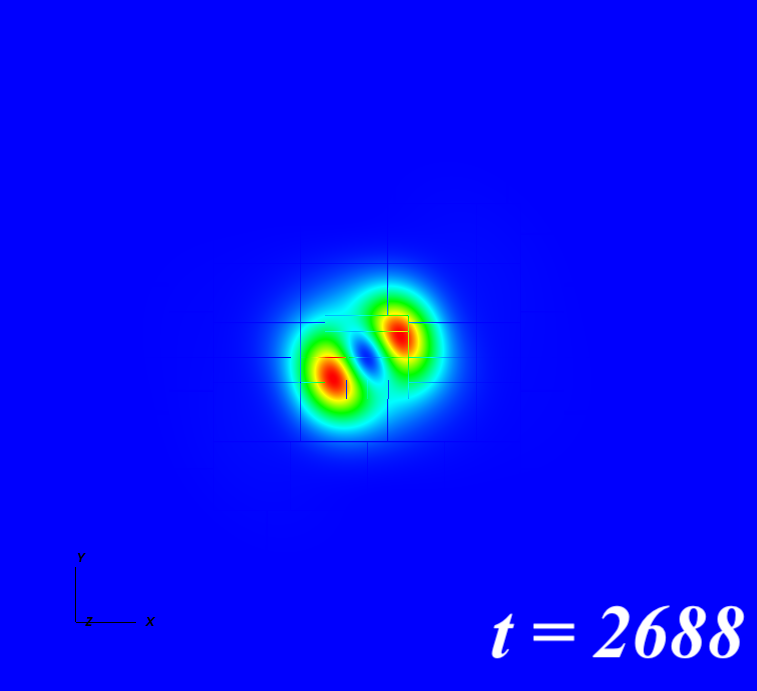} \includegraphics[height=1.25in]{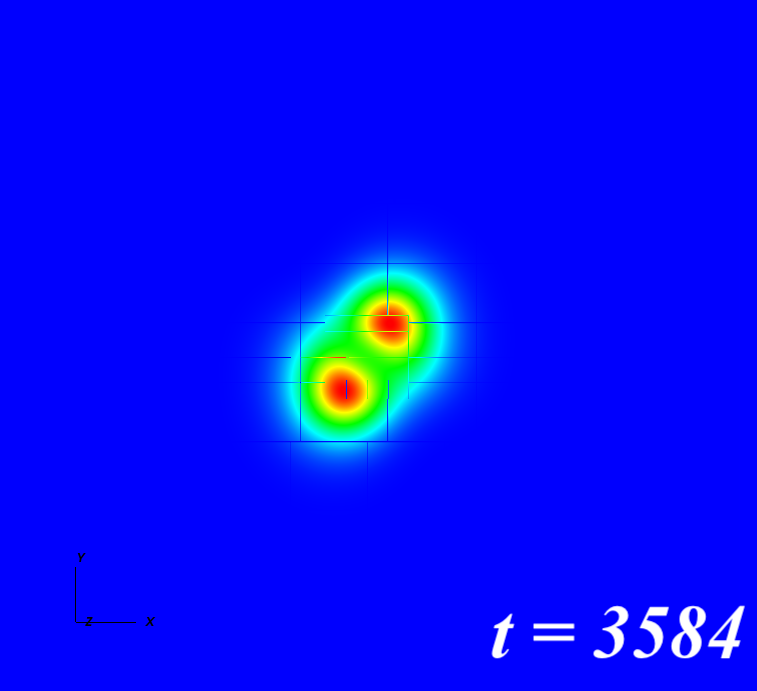}\\
\smallskip
\smallskip
\includegraphics[height=1.25in]{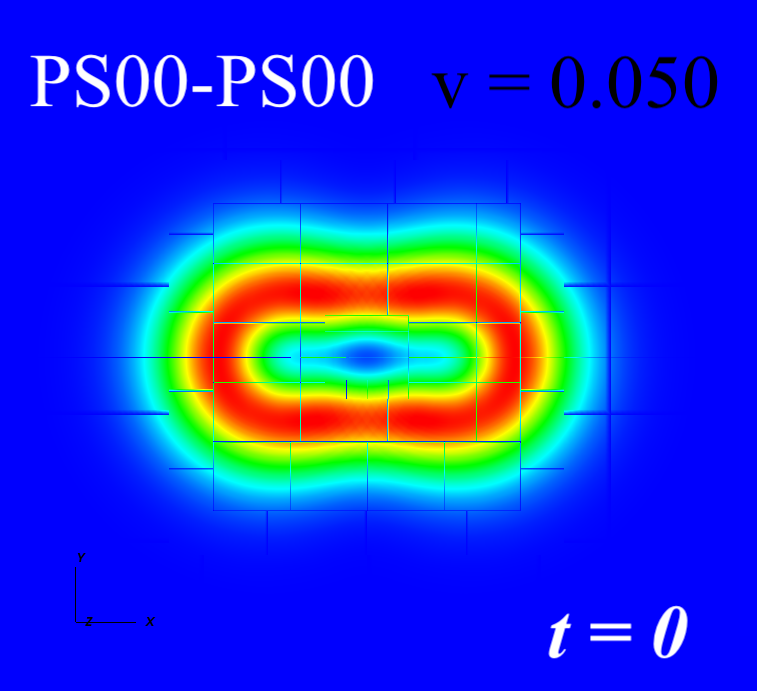} \includegraphics[height=1.25in]{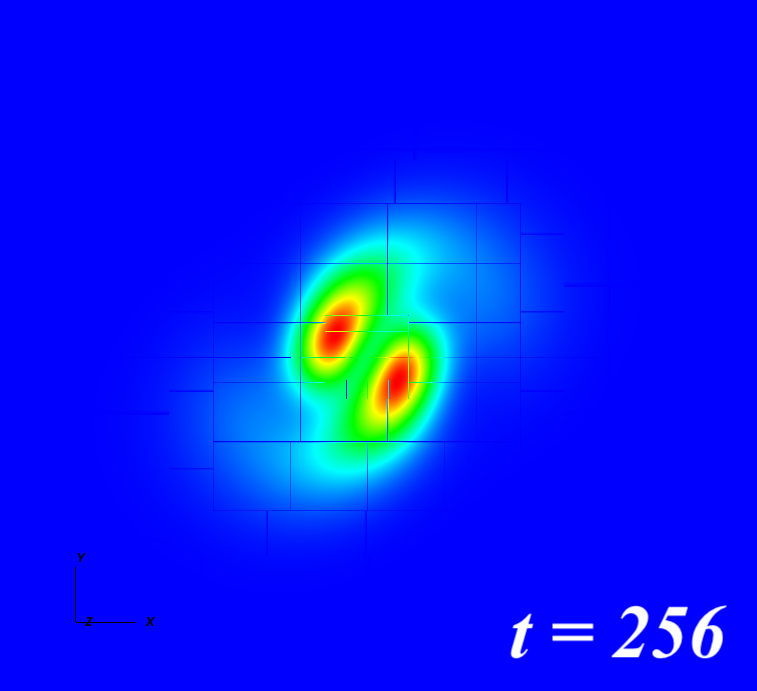} \includegraphics[height=1.25in]{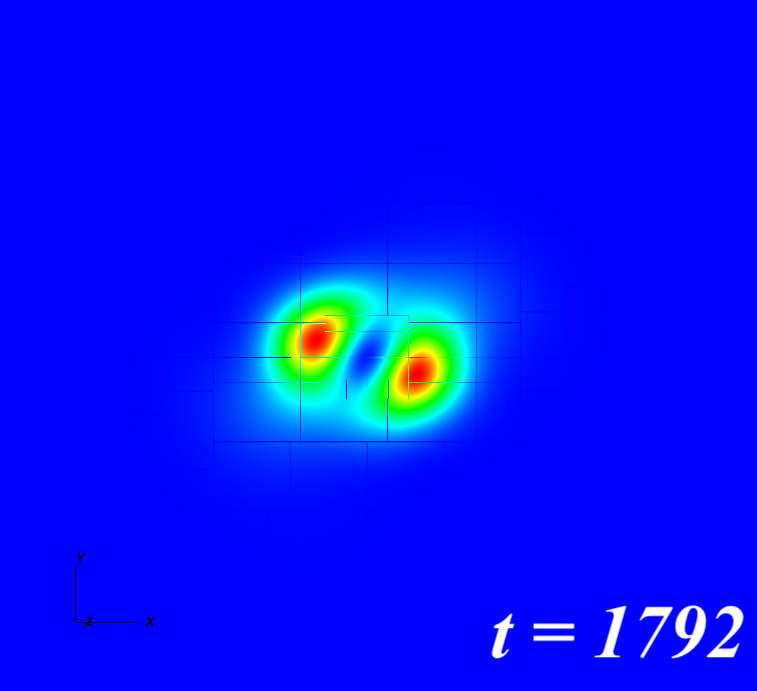} \includegraphics[height=1.25in]{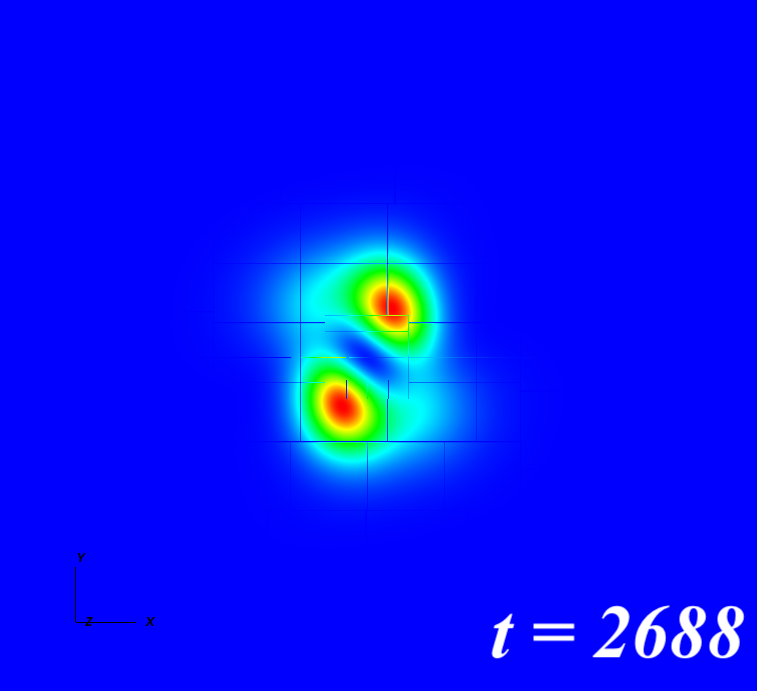} \includegraphics[height=1.25in]{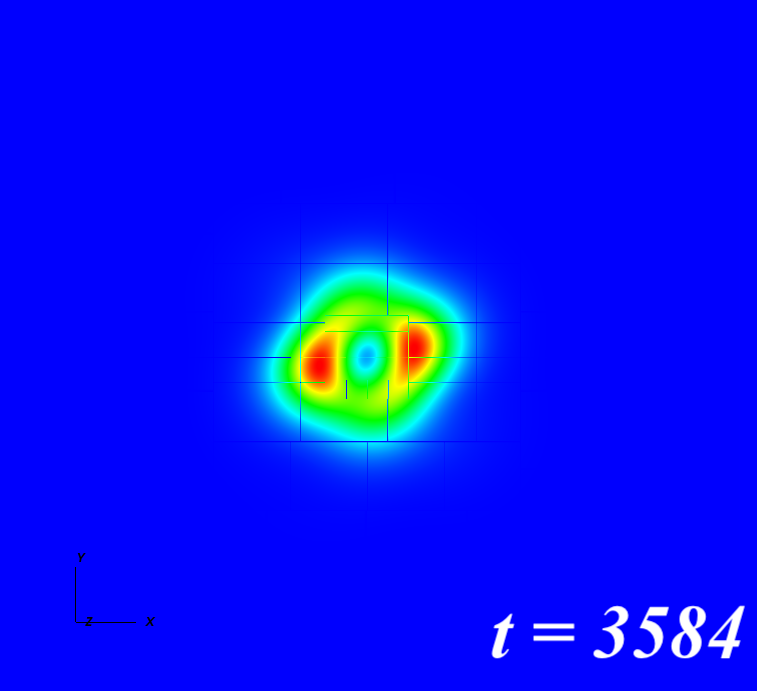}\\
\smallskip
\smallskip
\includegraphics[height=1.25in]{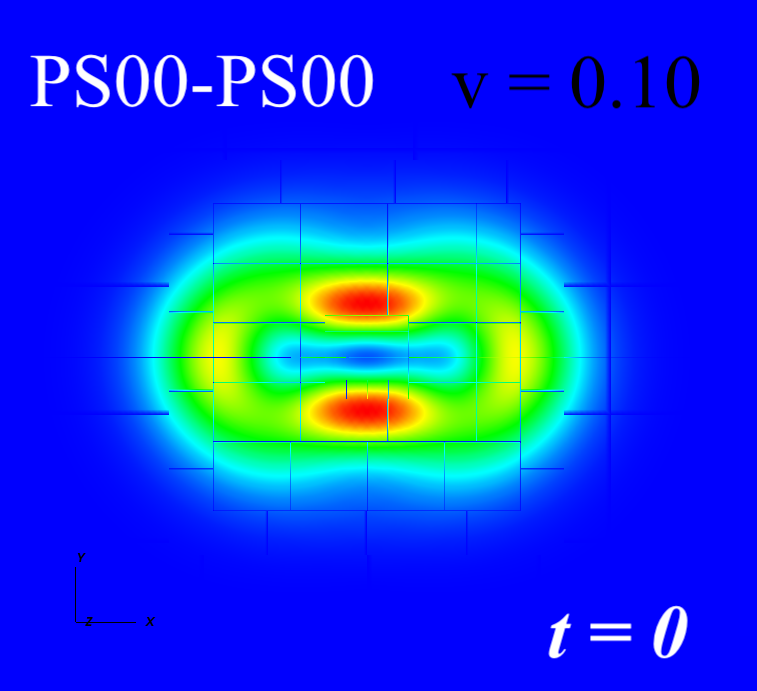} \includegraphics[height=1.25in]{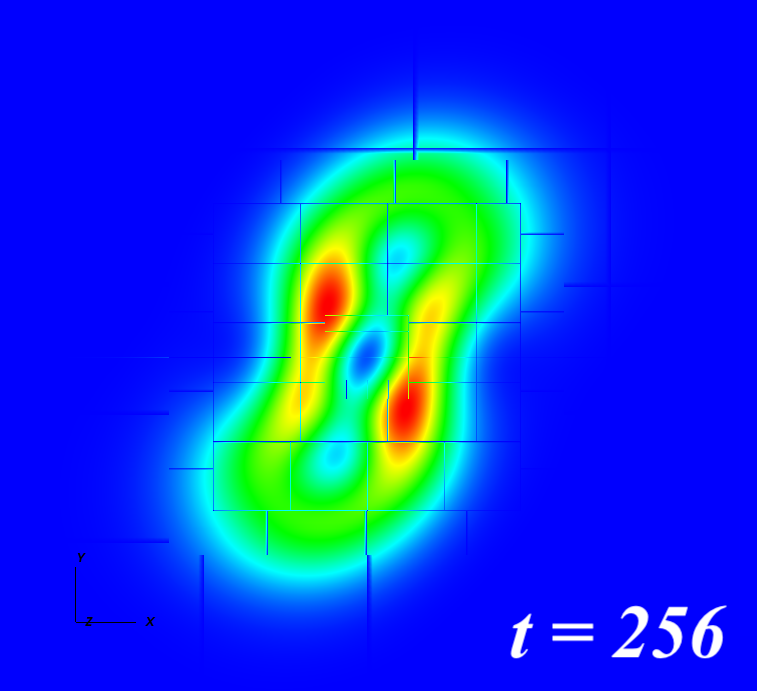} \includegraphics[height=1.25in]{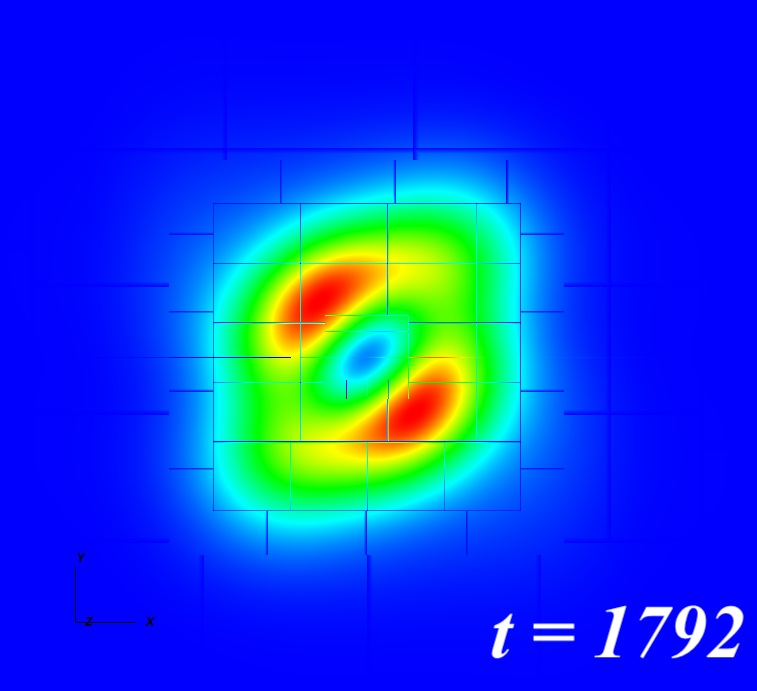} \includegraphics[height=1.25in]{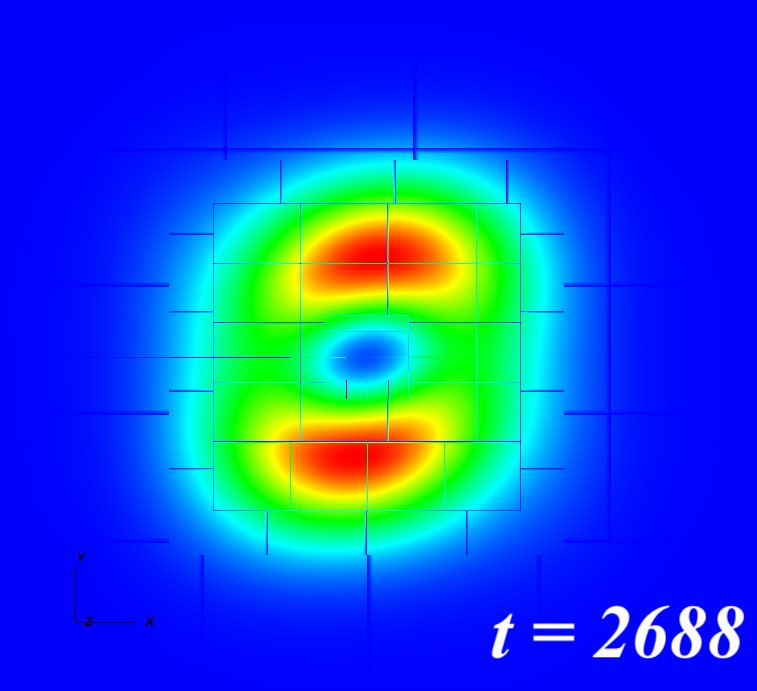} \includegraphics[height=1.25in]{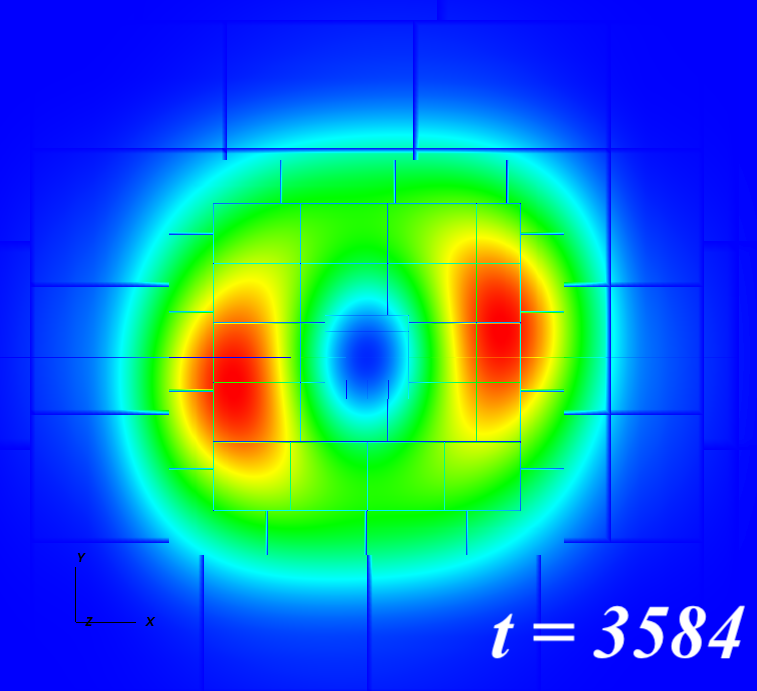}\\

\caption{Snapshots of the energy density for the orbital merger of PS00-PS00 with five initial velocites, namely $v_{y}=\lbrace{0.0125, 0.05, 0.10\rbrace}$ in the equatorial plane. The vertical axis correspond to the $y$ direction and the horizontal to the $x$ direction.
}
\label{fig:merger00}
\end{figure*}

%%%%%%%%%%%%%%%%%%%%%%%%%%%%%%%%%%%%%%%%%%%%%%%%%%%%%%%%%%%%%
\subsection{Orbital binary simulations}
%%%%%%%%%%%%%%%%%%%%%%%%%%%%%%%%%%%%%%%%%%%%%%%%%%%%%%%%%%%%%

We turn now to describe the orbital binary merger simulations of models PS00-PS00, PS1-PS1, PS2-PS2, and PS3-PS3. The stars are boosted in the $y$ direction. In this case, the objects are initially separated by $\Delta x = 30$.

 %%%%%%%%%%%%%%%%%%%%%%%%%%%%%%%%%%%%%%%%%%%%%%%%%%%%%%%%%%%%%
\subsubsection{Visualisation of the binary mergers}
 %%%%%%%%%%%%%%%%%%%%%%%%%%%%%%%%%%%%%%%%%%%%%%%%%%%%%%%%%%%%%
 
The results are similar to the head-on collision case. The stars do not complete a full orbit; nonetheless, the final object has non-zero angular momentum. Fig.~\ref{fig:merger0} shows snapshots of the evolution of the energy density of models PS1-PS1, PS2-PS2 and PS3-PS3 in the equatorial plane for an initial velocity $v_{y}=0.050$. The collision happens at $t\sim128$ for models PS1-PS1 and PS3-PS3, and at $t\sim144$ for PS2-PS2. The merged object has still some angular momentum left. An AH appears at $t\sim350$ for model PS1-PS1, at $t\sim200$ for model PS2-PS2 and at $t\sim300$ for PS3-PS3.

In Fig.~\ref{fig:merger00} we exhibit the evolution of the energy density for the model PS00-PS00, with three different initial velocities, $v_{y}=0.0125$, $v_{y}=0.050$ and $v_{y}=0.10$. For PS00-PS00, the result of the merger does not lead to BH formation; instead the final object is a Proca star with angular momentum. The larger the initial velocity, the larger the fraction of the initial Proca mass and angular momentum that is ejected during and after the merger. With the largest initial velocity, $v_{y}=0.10$, almost all the Proca field is dispersed away at the end of the simulation.

%\begin{figure*}[t!]
%\centering
%\includegraphics[height=1.3in]{ibc01_movie_0000.png} \includegraphics[height=1.3in]{ibc01_movie_0005.png} \includegraphics[height=1.3in]{ibc01_movie_0009.png} \includegraphics[height=1.3in]{ibc01_movie_0016.png} \includegraphics[height=1.3in]{ibc01_movie_0018.png}\\
%\smallskip
%\includegraphics[height=1.3in]{ibc02_movie_0000.png} \includegraphics[height=1.3in]{ibc02_movie_0004.png} \includegraphics[height=1.3in]{ibc02_movie_0009.png} \includegraphics[height=1.3in]{ibc02_movie_0010.png} \includegraphics[height=1.3in]{ibc02_movie_0012.png}\\
%\smallskip
%\includegraphics[height=1.3in]{ibc03_movie_0000.png} \includegraphics[height=1.3in]{ibc03_movie_0005.png} \includegraphics[height=1.3in]{ibc03_movie_0008.png} \includegraphics[height=1.3in]{ibc03_movie_0010.png} \includegraphics[height=1.3in]{ibc03_movie_0018.png}\\
%
%\caption{Snapshots of the energy density for PS0-PS0, PS1-PS1, PS2-PS2 and PS3-PS3 in the equatorial plane. The vertical axis correspond to the $y$ direction and the horizontal to the $x$ direction.
%}
%\label{fig6}
%\end{figure*}

\begin{figure}[t!]
\centering
\includegraphics[height=2.7in]{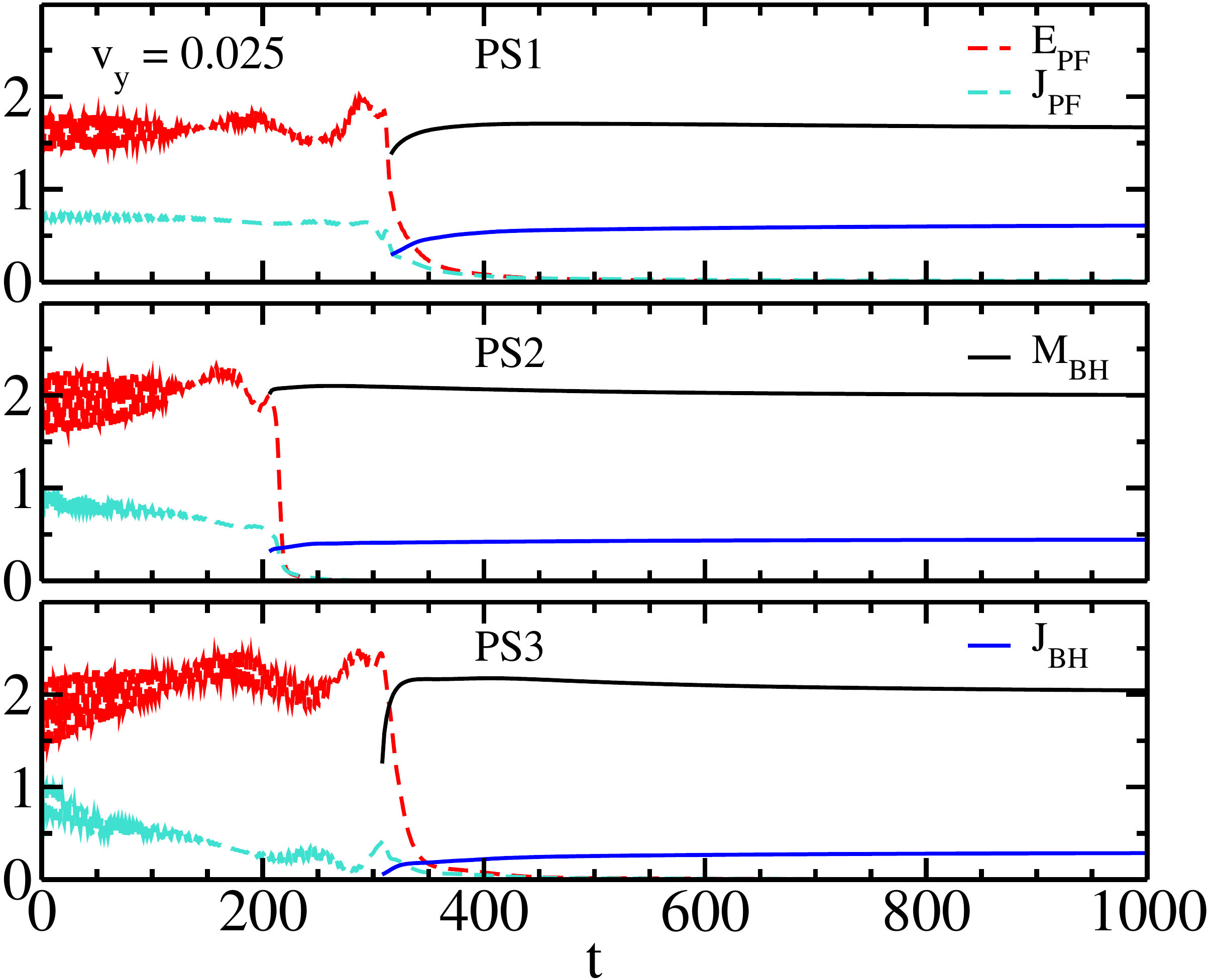}
\includegraphics[height=2.7in]{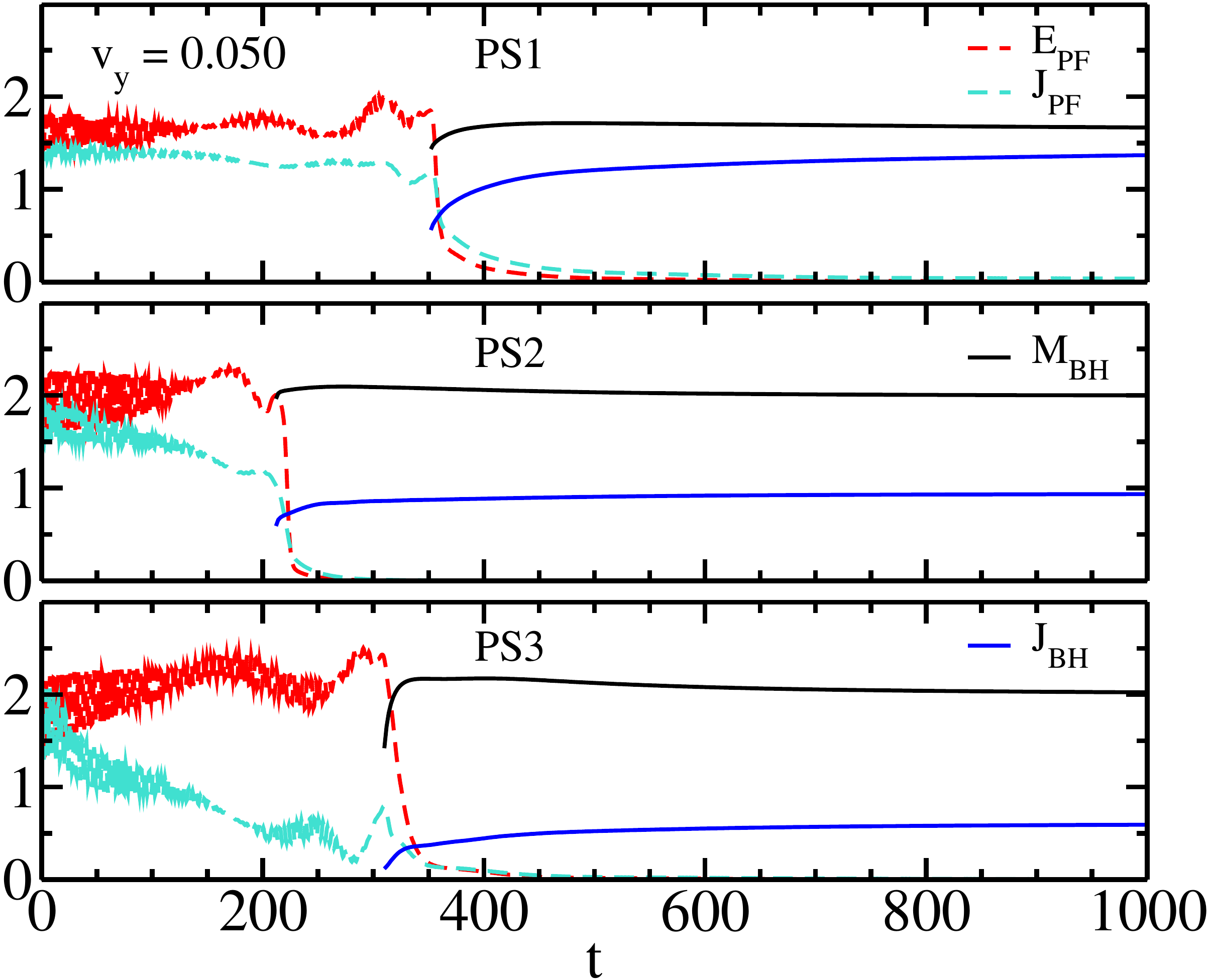}
\caption{ Time evolution of the Proca field energy and angular momentum, together with the BH mass and spin for models PS1-PS1, PS2-PS2 and PS3-PS3, for two initial velocities $v_{y}=0.025$ (top panel) and $v_{y}=0.050$ (bottom panel).
}
\label{fig:merger1a}
\end{figure}

\begin{figure}[t!]
\centering
\includegraphics[height=2.5in]{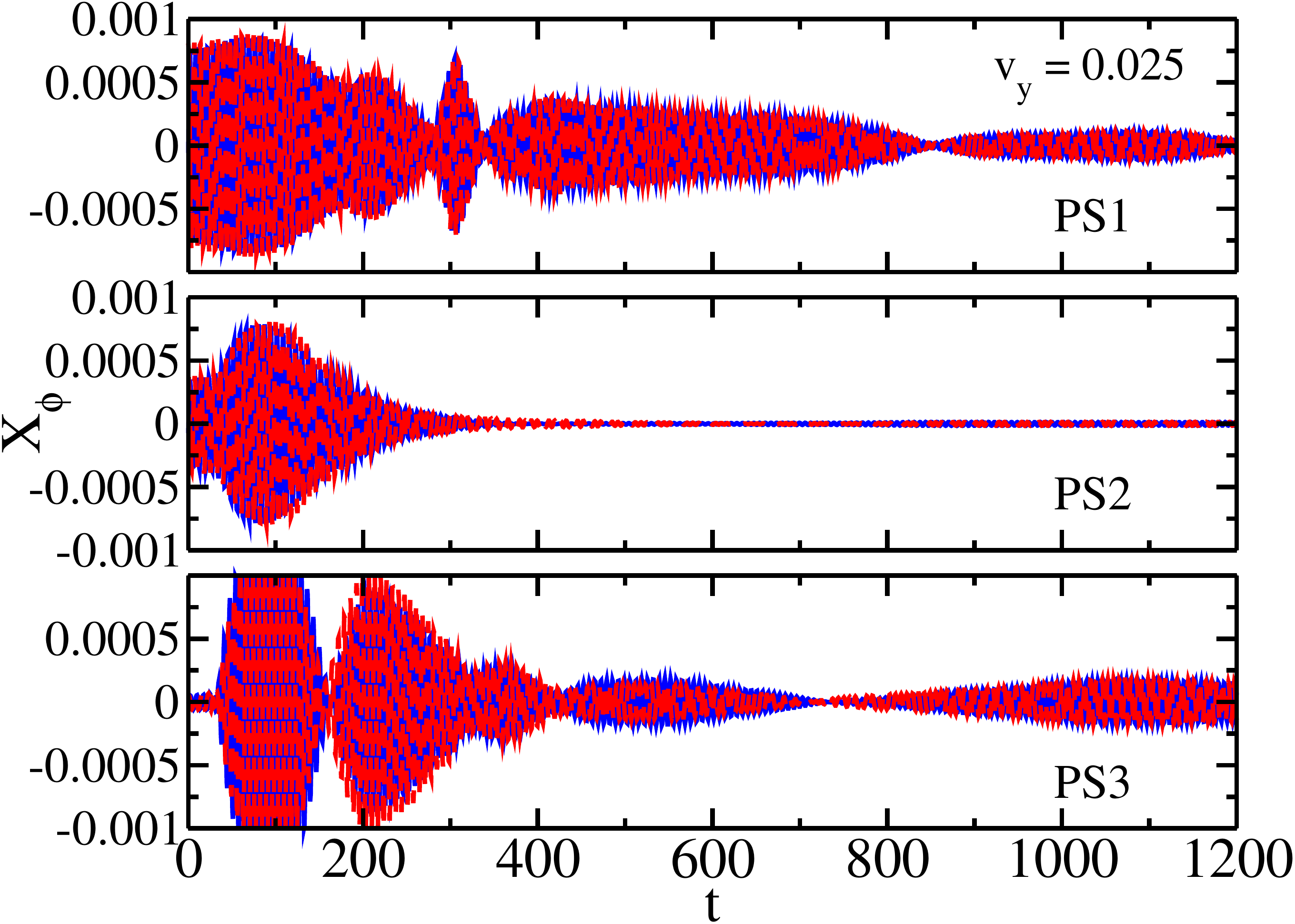}  
\includegraphics[height=2.5in]{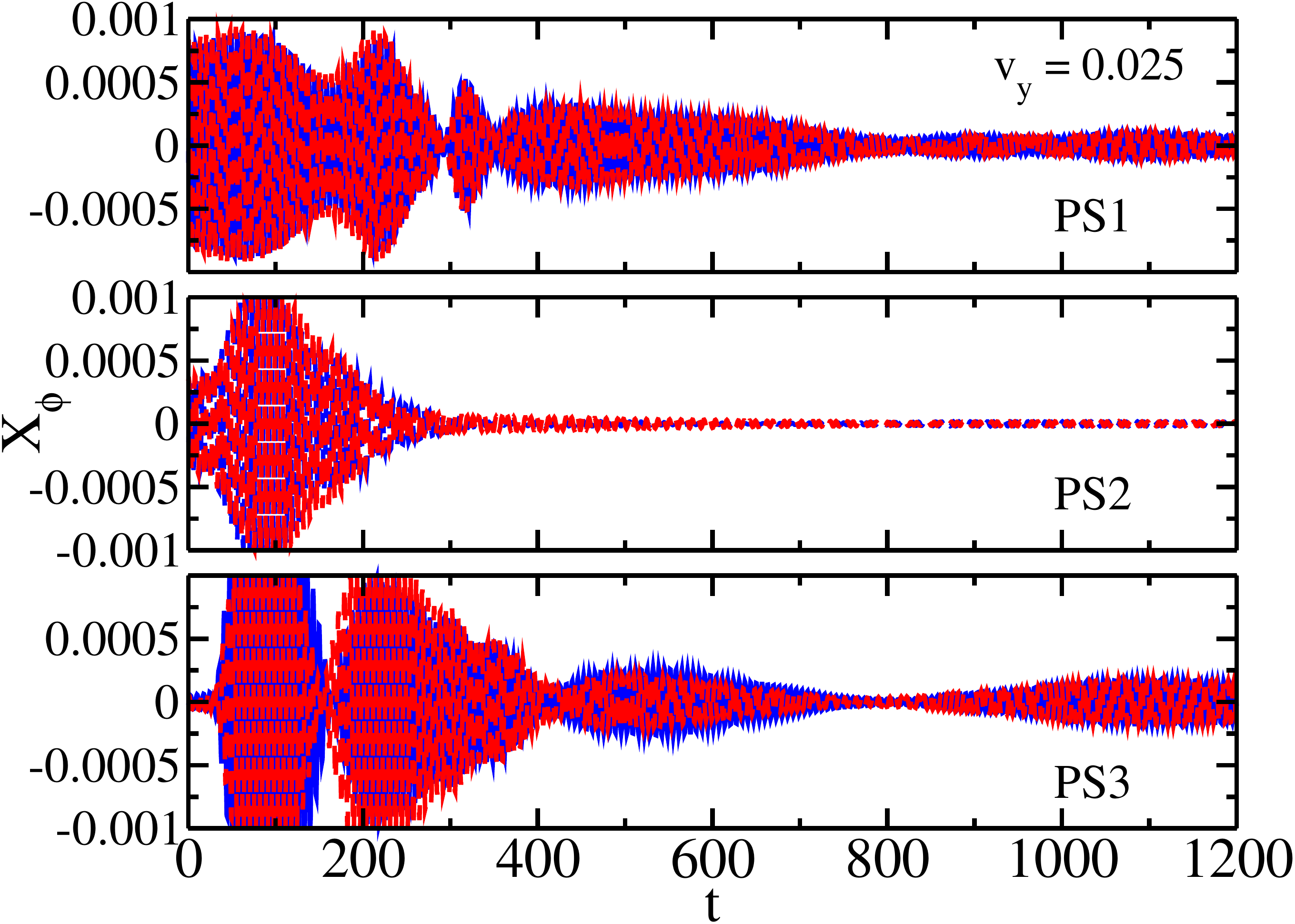}  
\caption{ Time evolution of the real and imaginary parts of the scalar potential amplitude for models PS1-PS1, PS2-PS2 and PS3-PS3 for two initial velocities $v_{y}=0.025$ (top panel) and $v_{y}=0.050$ (bottom panel).
}
\label{fig:merger1b}
\end{figure}

 %%%%%%%%%%%%%%%%%%%%%%%%%%%%%%%%%%%%%%%%%%%%%%%%%%%%%%%%%%%%%
\subsubsection{Proca energy and horizon formation}
 %%%%%%%%%%%%%%%%%%%%%%%%%%%%%%%%%%%%%%%%%%%%%%%%%%%%%%%%%%%%%

In Fig.~\ref{fig:merger1a} we plot the time evolution of the Proca field energy, Eq.~(\ref{energy}), and the Proca field angular momentum
\begin{equation}
J_{\rm PF}=\int_{\Sigma}drd\theta 
d\varphi\,T^{t}_{\varphi}\alpha\sqrt{\gamma} \ ,
\label{angmom}
\end{equation}   
together with the BH mass and the BH spin $J$ for models PS1-PS1, PS2-PS2 and PS3-PS3 (bottom panels). In all these cases there is AH formation. These models collapse after the merger and form a Kerr BH with angular momentum. There is a remnant Proca field outside the rotating horizon, forming a quasi-bound state. We do not see evidence of infinitely long-lived Proca remnants (hair) forming around the BH.

The binary takes longer to collapse for a larger initial velocity $v_{y}$, but for the models PS2-PS2 and PS3-PS3 the difference is small for the velocities we have chosen. A large part of the total angular momentum is lost before the collapse for these models. On the other hand, for model PS1-PS1 the difference in collapse time is more noticeable: it takes about 50 times longer to collapse for the largest velocity and the final BH stores almost all of the initial angular momentum. 

Fig.~\ref{fig:merger1b} shows that a quasi-bound state appears outside the horizon after the collapse and formation of a BH. As in the head-on collision case, the final amount of Proca field remaining is smaller for model PS2-PS2 than for models PS1-PS1 and PS3-PS3.

\begin{figure}[t!]
\centering
\includegraphics[height=2.5in]{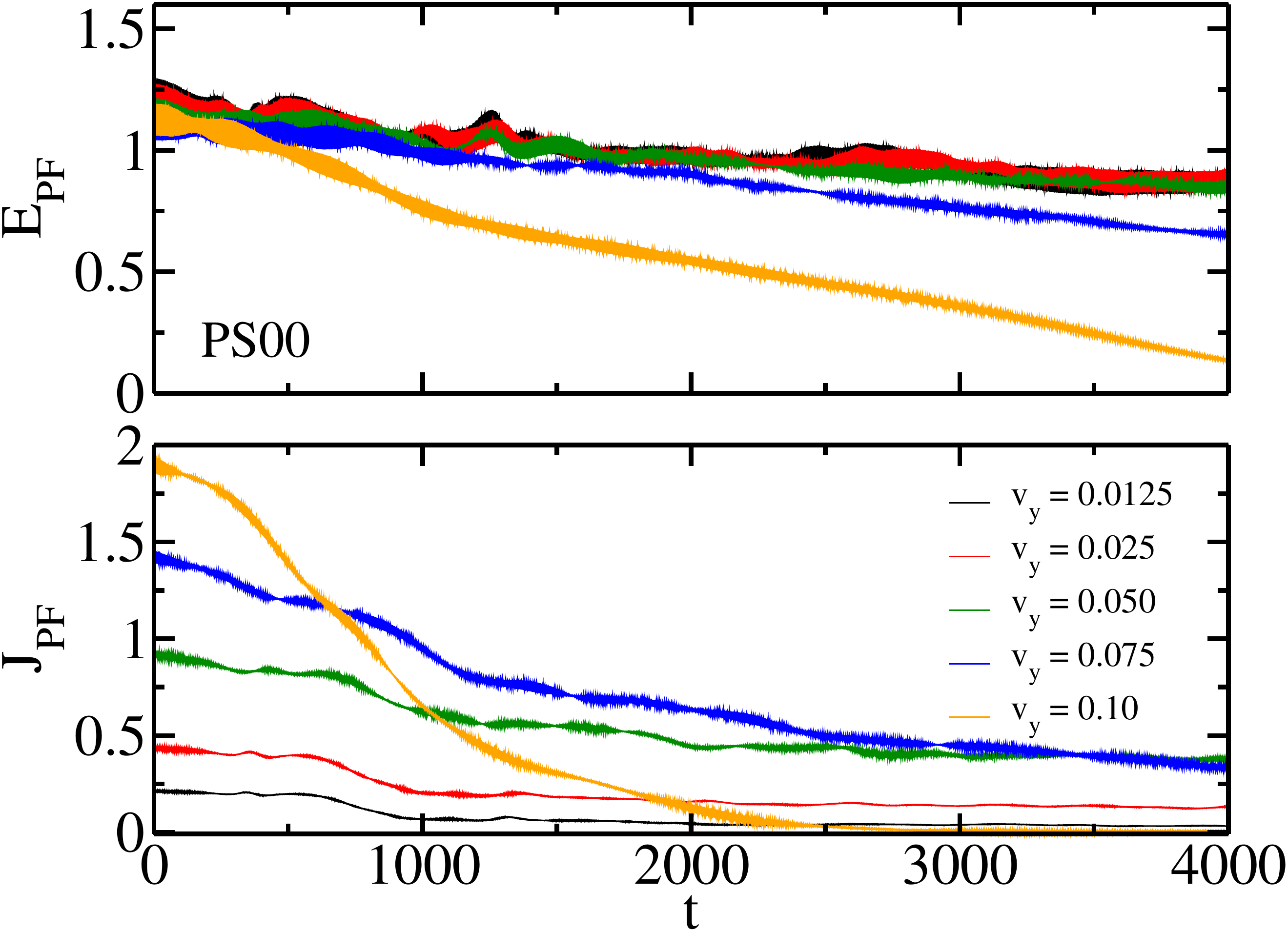}
\includegraphics[height=2.5in]{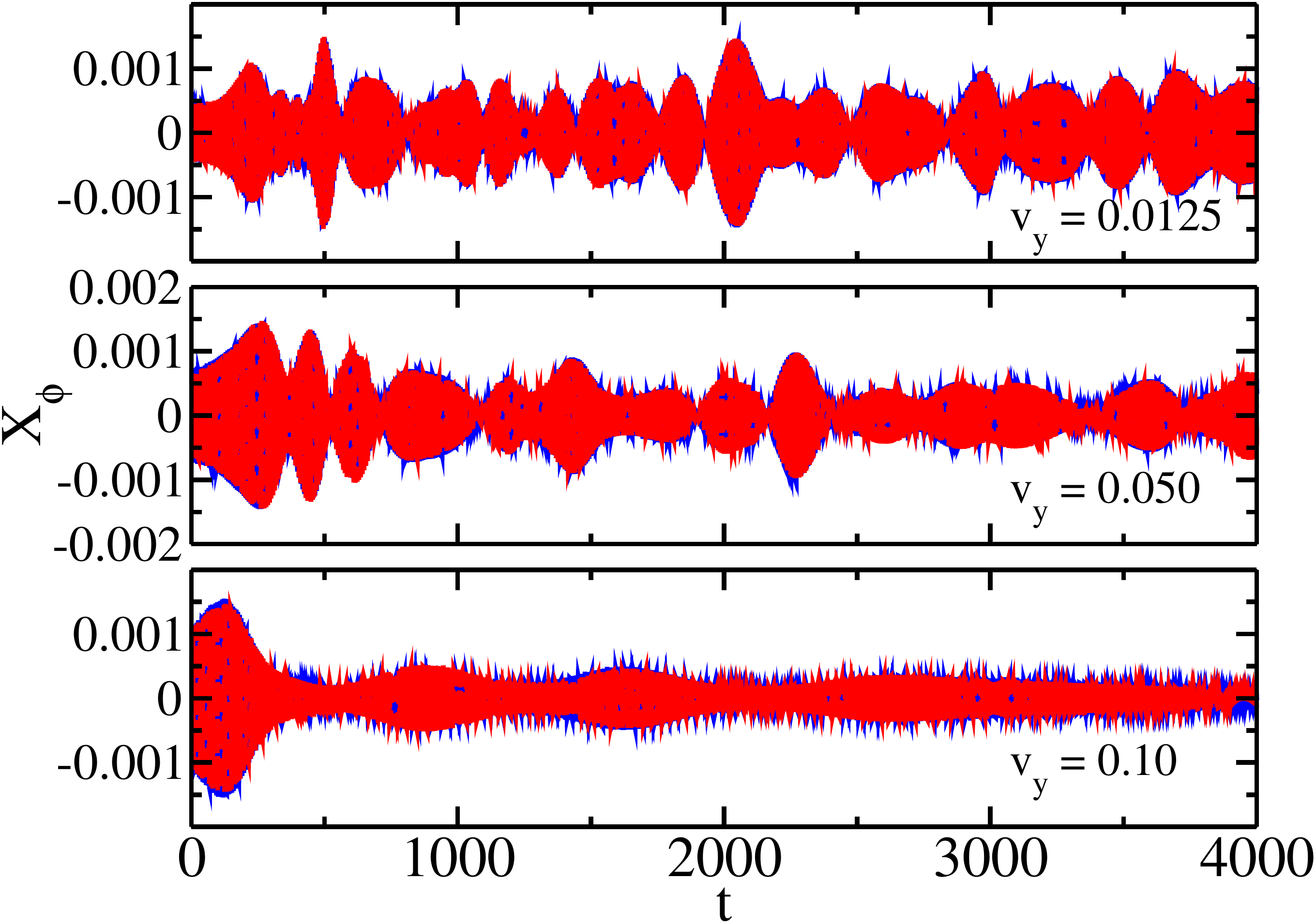}  
\caption{Time evolution of the Proca field energy and angular momentum (top panels) and of the amplitude of the scalar potential for model PS00-PS00 (bottom panels).
}
\label{fig:merger2}
\end{figure}

For PS00-PS00 there is no BH formation. In Fig.~\ref{fig:merger2}, we show the time evolution of the energy and angular momentum of this model that we evolve with five different initial boost velocities, therefore increasing the initial angular momentum, namely $v_{y}=\lbrace0.0125, 0.025, 0.050, 0.075, 0.10\rbrace$. The PS formed after the collision has non-zero residual angular momentum. The total Proca energy is larger than the maximum mass of a rotating PS with m=1 ($M\sim1.124$, see Fig. 6 in~\cite{Herdeiro:2017phl}). To prevent the collapse, the star can lose energy through two mechanisms: gravitational cooling (ejecting Proca particles) and gravitational-wave emission (which also carries angular momentum). One may ask if there is formation of a rotating Proca star after the merger; we see that if we increase the initial velocity the angular momentum is rapidly lost and goes below the energy of the Proca field. This result  could be due to the constraint-violating initial data we use; thus, it is interesting to revisit this problem once constraint-satisfying initial data becomes available, as it would provide a more reliable answer. For very large initial velocities, the final star can be completely dispersed away. For the largest velocity, we see that the amplitude of the scalar potential is decreasing with time.

%\begin{figure}[h!]
%\centering
%\includegraphics[height=3.5in]{Phi1-Model1.pdf}  \includegraphics[height=3.5in]{Phi1-Model2-Model3.pdf} 
%\caption{(Top panels) Time evolution of scalar potential $\Phi$ for PS1-PS1 extracted at $r=0$. Time evolution of scalar potential $\Phi$ for PS2-PS2 (middle panel) and PS3-PS3 (bottom panel) extracted at $r=0$.
%}
%\label{fig2}
%\end{figure}

 %%%%%%%%%%%%%%%%%%%%%%%%%%%%%%%%%%%%%%%%%%%%%%%%%%%%%%%%%%%%%
\subsubsection{Waveforms}
 %%%%%%%%%%%%%%%%%%%%%%%%%%%%%%%%%%%%%%%%%%%%%%%%%%%%%%%%%%%%%

\begin{figure}%[h!]
\centering
\includegraphics[height=2.5in]{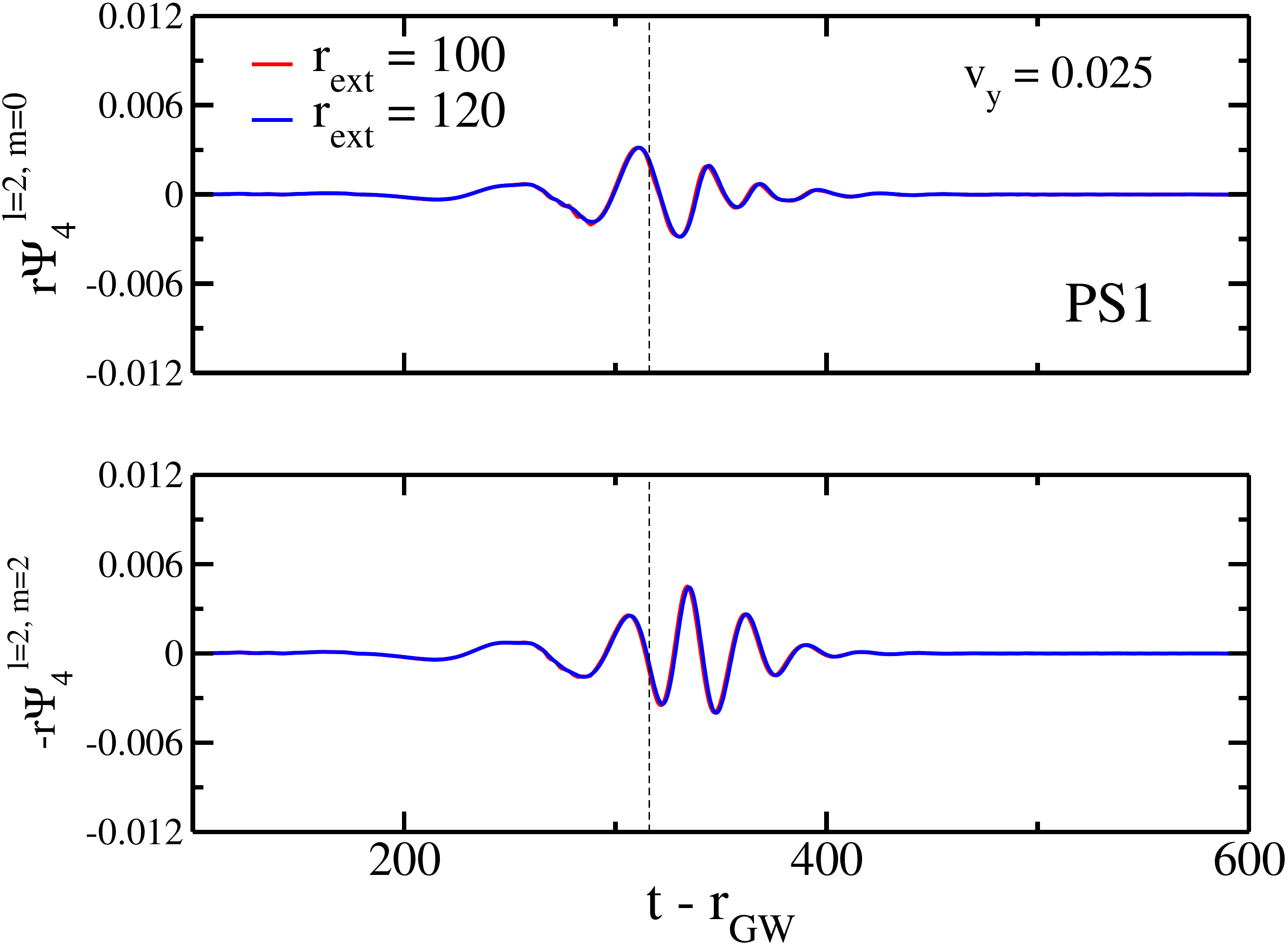}  
\includegraphics[height=2.5in]{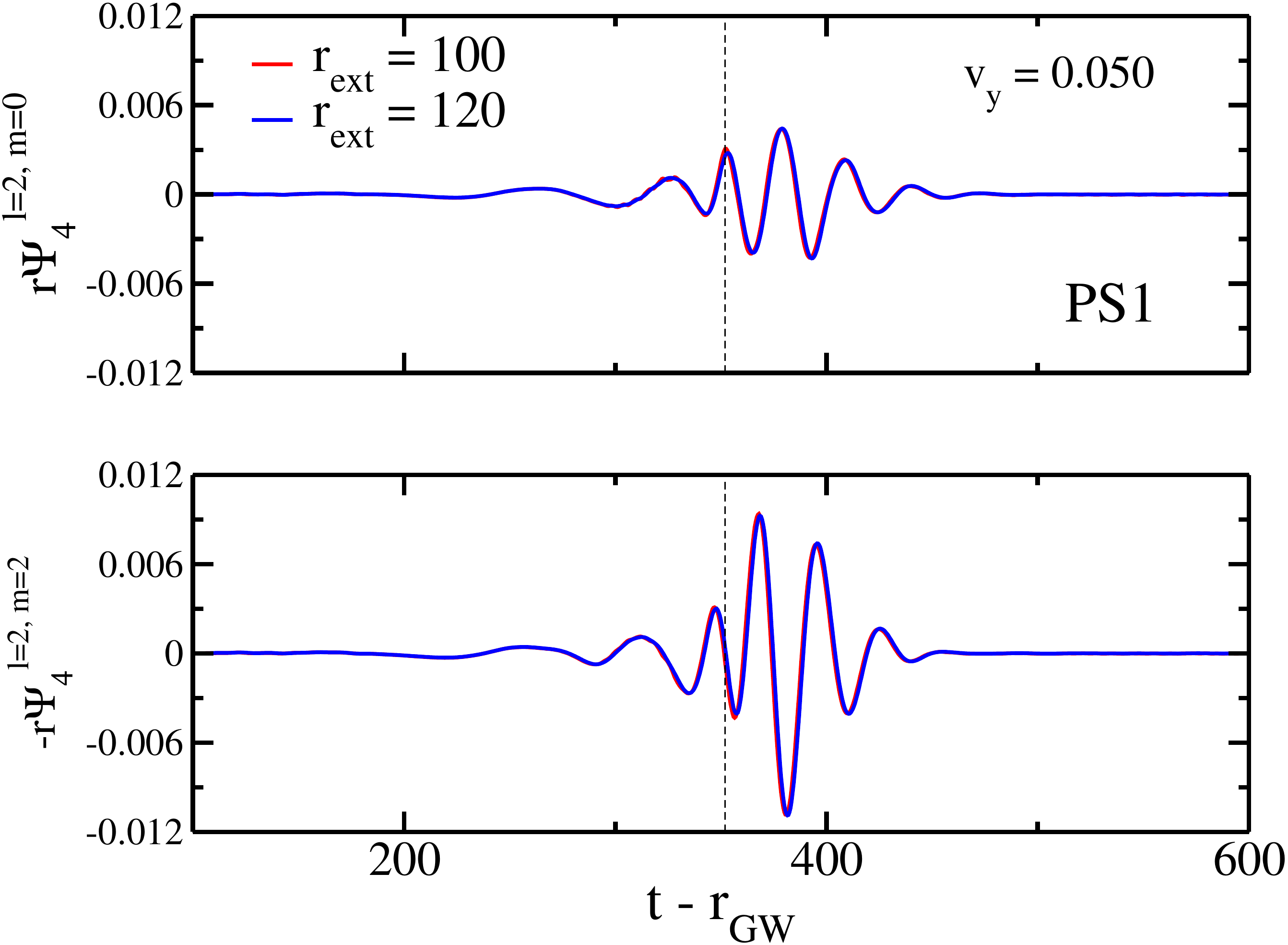}
\caption{Real part of $r\Psi_4^{l=2,m=0}$ and $r\Psi_4^{l=2,m=+2}$ for PS1-PS1 with initial boost velocity $v_{y}=0.025$ (top panels) and $v_{y}=0.050$ (bottom panels). 
}
\label{fig:merger3}
\end{figure}

\begin{figure}%[h!]
\centering
\includegraphics[height=2.5in]{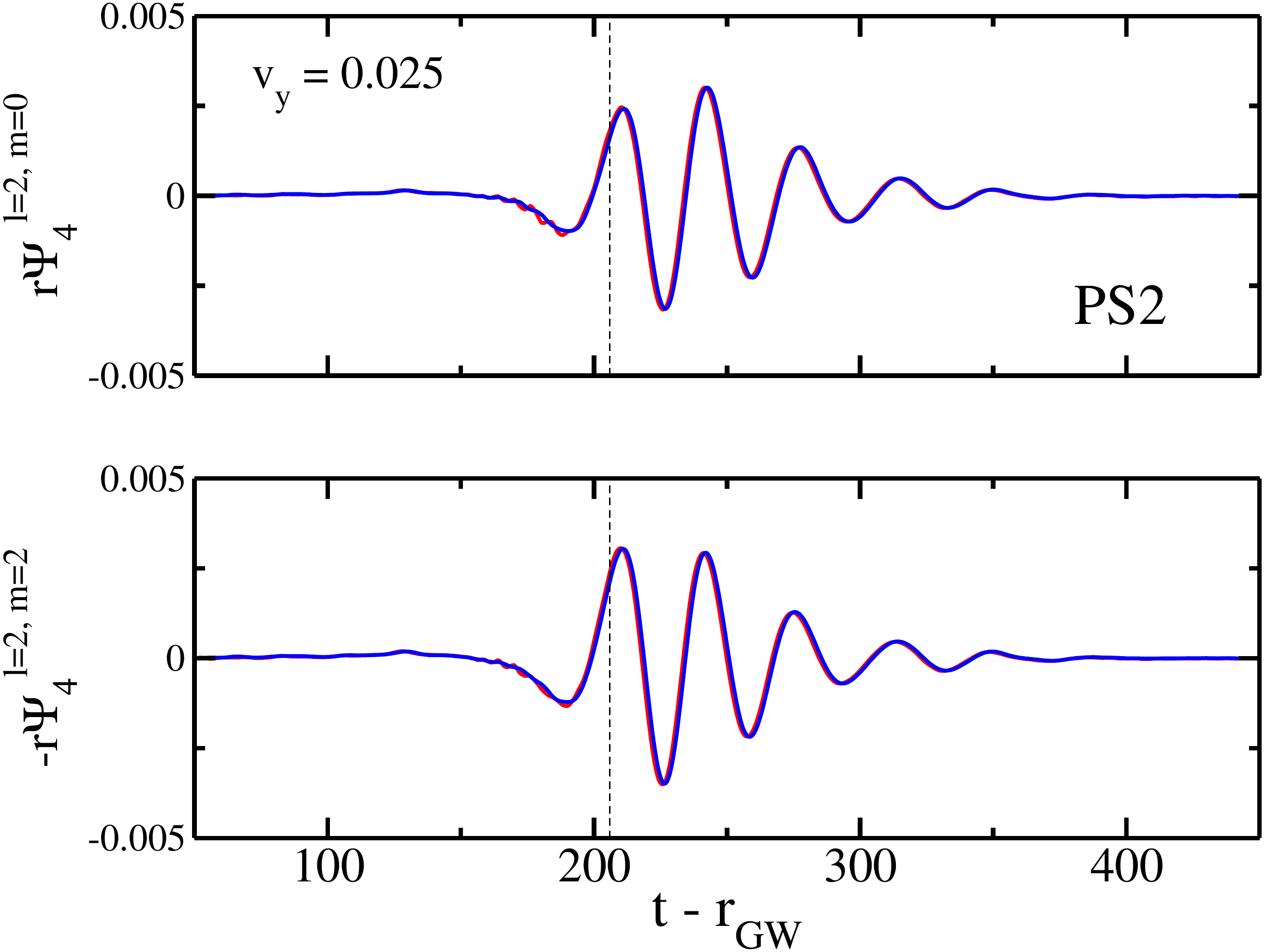}  
\includegraphics[height=2.5in]{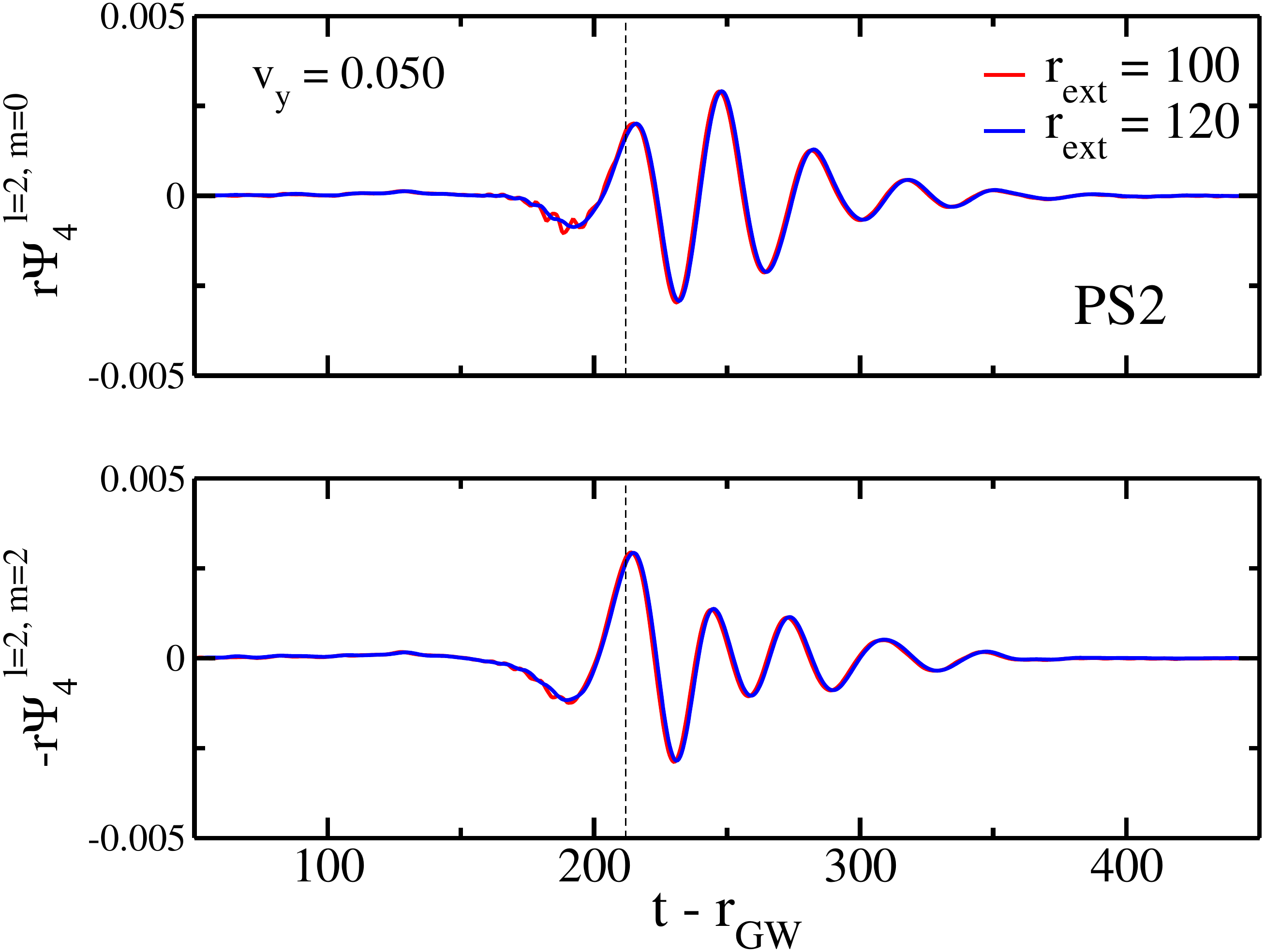}
\caption{Real part of $r\Psi_4^{l=2,m=0}$ and $r\Psi_4^{l=2,m=+2}$ for PS2-PS2 with initial boost velocity $v_{y}=0.025$ (top panels) and $v_{y}=0.050$ (bottom panels).
}
\label{fig:merger4}
\end{figure}

\begin{figure}%[h!]
\centering
\includegraphics[height=2.5in]{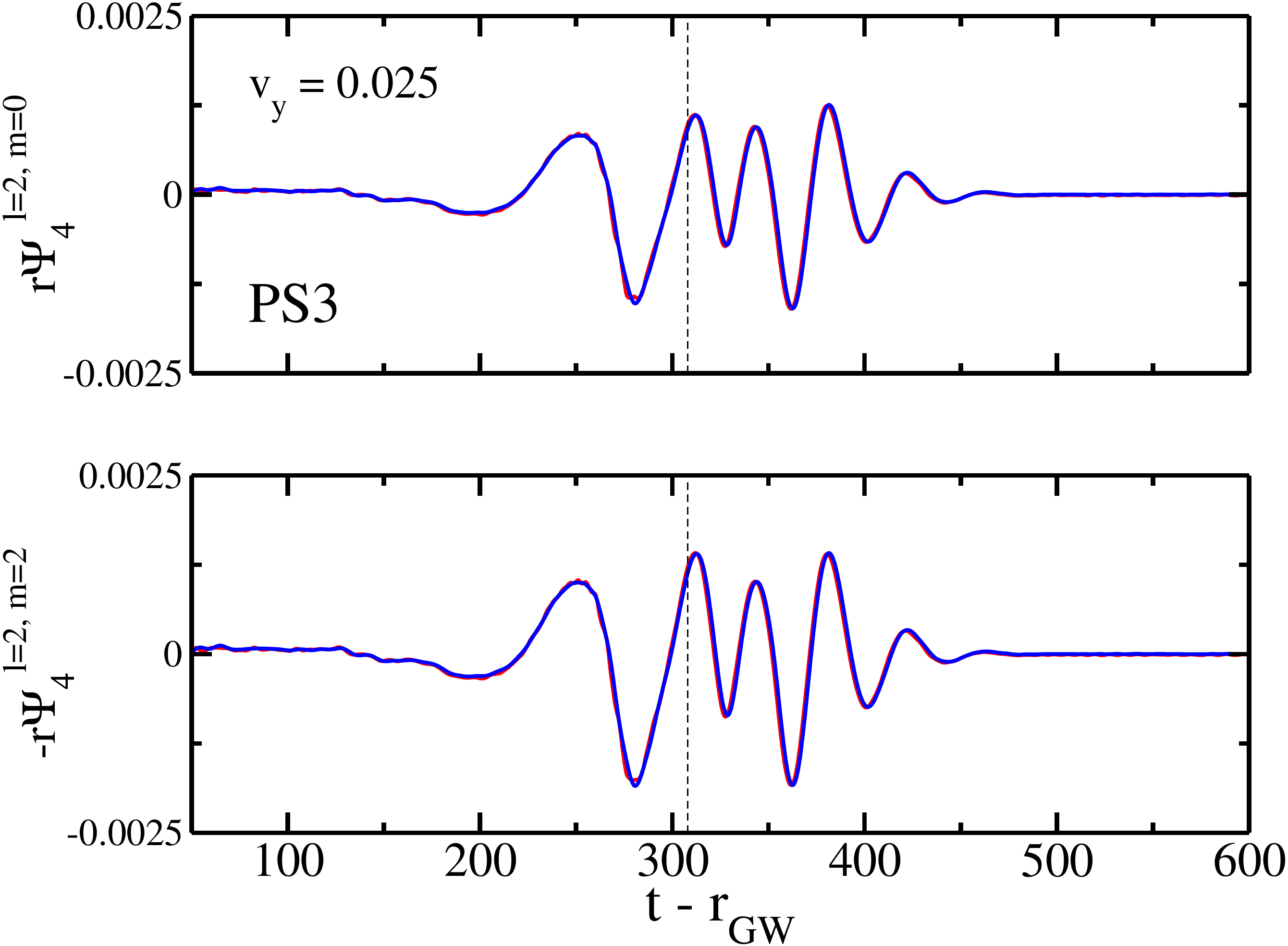}  
\includegraphics[height=2.5in]{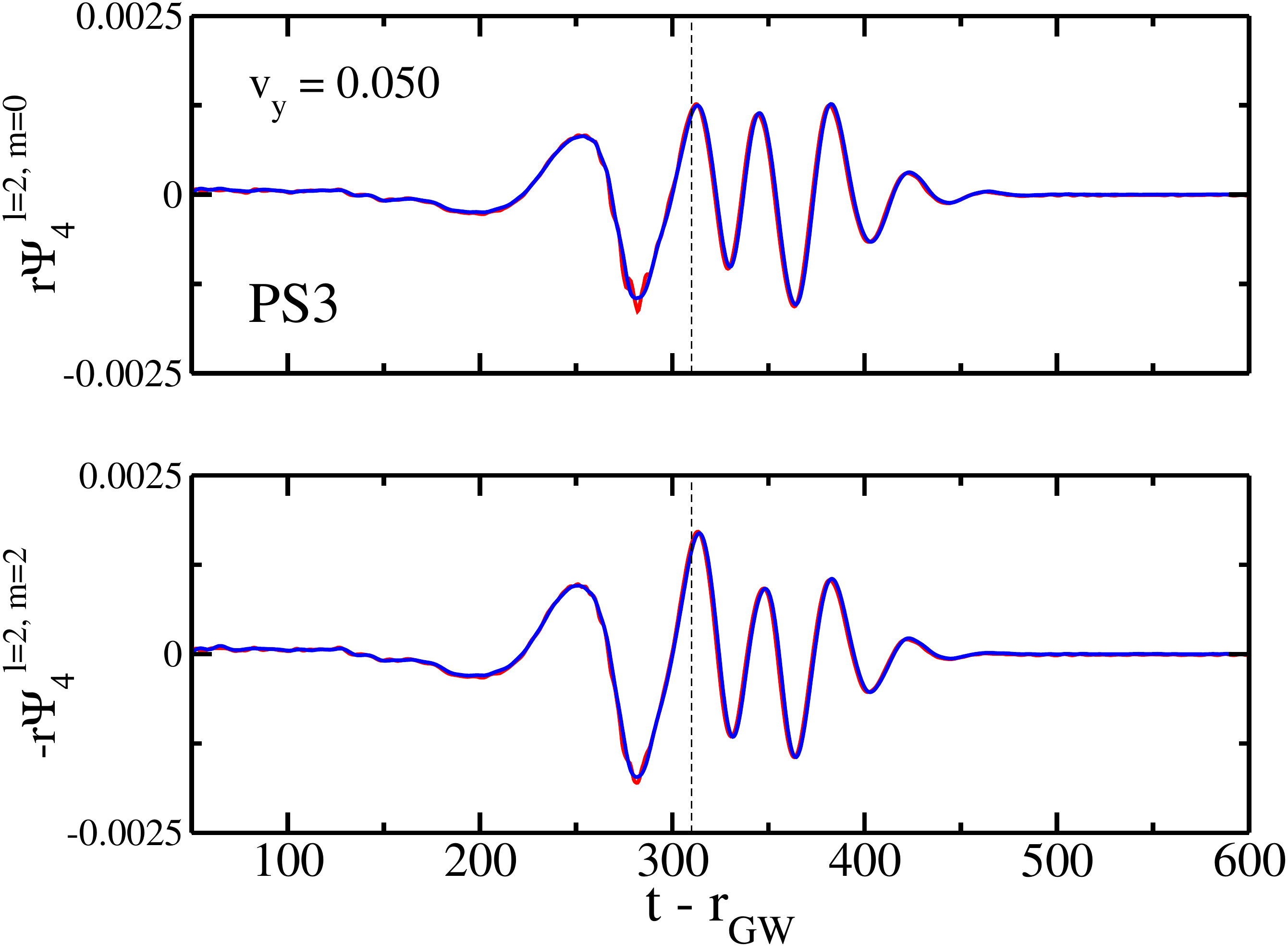}
\caption{Real part of $r\Psi_4^{l=2,m=0}$ and $r\Psi_4^{l=2,m=+2}$ for PS3-PS3 with initial boost velocity $v_{y}=0.025$ (top panels) and $v_{y}=0.050$ (bottom panels).
}
\label{fig:merger5}
\end{figure}

\begin{figure}%[h!]
\centering
\includegraphics[height=2.5in]{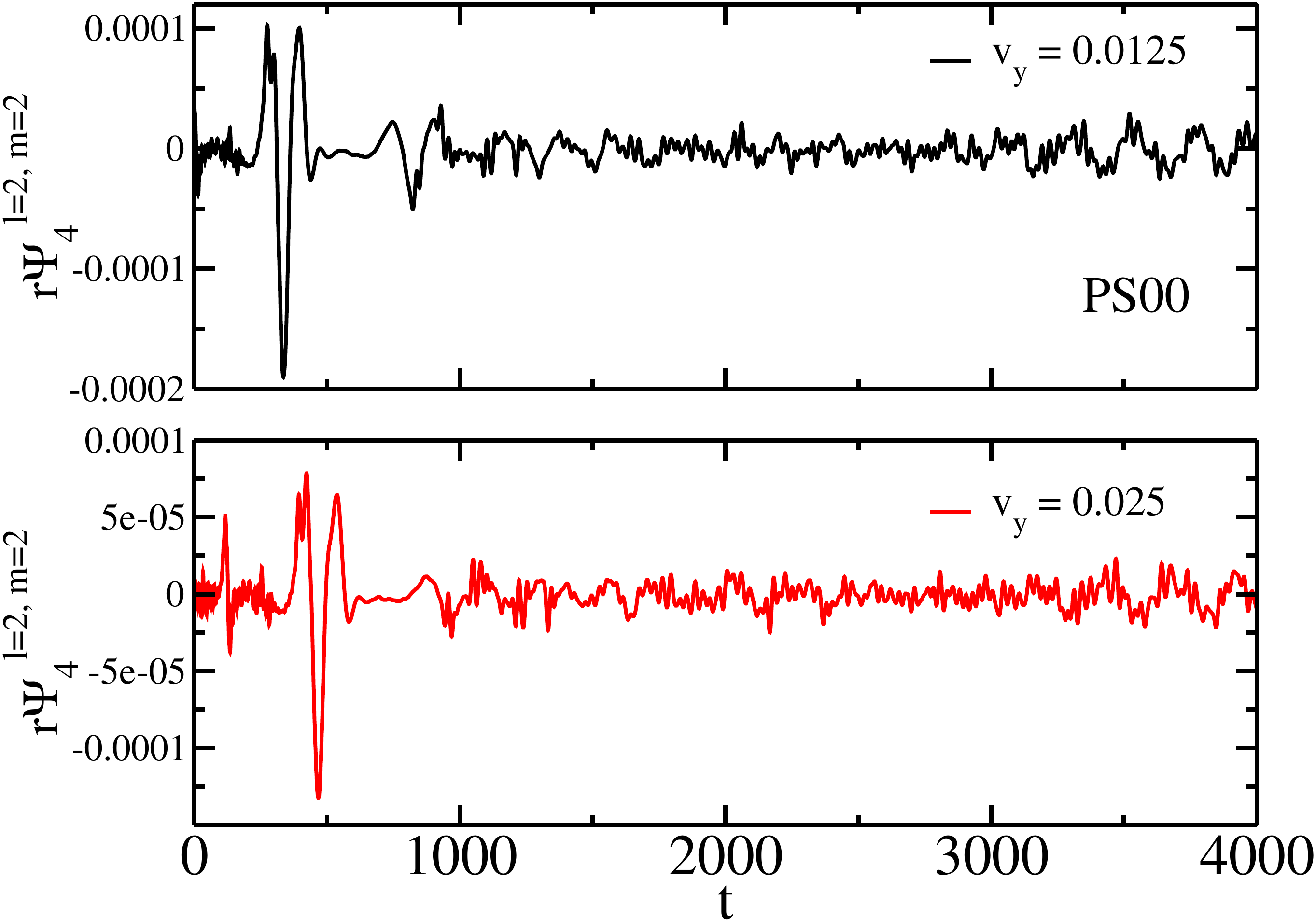}  
\caption{Real part of $r\Psi_4^{l=2,m=0}$ and $r\Psi_4^{l=2,m=+2}$ for PS00-PS00 extracted at $r_{\text{ext}}=120$ with initial boost velocity $v_{y}=0.0125$ (top panel), and $v_{y}=0.025$ (bottom panel).
}
\label{fig:merger6}
\end{figure}

In Figs.~\ref{fig:merger3} to \ref{fig:merger6} we exhibit the gravitational waveforms produced in the orbital mergers, showing again the  $l=2$, $m=\lbrace0,+2\rbrace$ modes, extracted at two radii, $r_{\rm ext}=\lbrace100,120\rbrace$. The waveforms, conveniently shifted and rescaled, overlap in the wave zone. For the PS1-PS1 merger one observes that, as one increases the initial velocity, the non-axisymmetric m=2 mode grows visibly more than the axisymmetric m=0 mode. In fact the same trend occurs for the PS2-PS2 and PS3-PS3 mergers, but it is less pronounced. Therefore, the rescaling presented in Eq.~(\ref{rescalingmodes}) is no longer true. Moreover, the two parts of the waveforms shown in Figs. (\ref{fig:merger3})-(\ref{fig:merger5}) seem to indicate that for models PS1-PS1 and PS3-PS3 a transient hypermassive PS forms before it collapses to a BH. Finally, for model PS00-PS00 (cf.~\ref{fig:merger6}) the gravitational-wave emission does not decay as the result of this merger is a highly perturbed PS. The waveform is filled with high frequency noise, probably coming from reflections with the outer boundary. The gravitational waveforms for this case are markedly similar to previous results in the scalar case~\cite{Bezares:2017mzk,Palenzuela:2017kcg,Bezares:2018qwa}.

%%%%%%%%%%%%%%%%%%%%%%%%%%%%%%%%%%%%%%%%%%%%%%%%%%%%%%%%%%%%%
\section{Conclusions}
\label{secconclusions}
%%%%%%%%%%%%%%%%%%%%%%%%%%%%%%%%%%%%%%%%%%%%%%%%%%%%%%%%%%%%%%

In this paper we have used numerical-relativity techniques to study both head-on collisions and orbital binary mergers of Proca stars of equal ADM mass and vector field frequency and we have extracted the gravitational waves produced in those collisions. This work continues our numerical exploration of the dynamics of PSs initiated in~\cite{sanchis2017numerical,di2018dynamical}. PSs are macroscopic, self-gravitating, Bose-Einstein condensates built out of a massive, complex, vector field~\cite{Brito:2015pxa}. Since they can achieve a compactness comparable to that of BHs, they are a type of BH mimicker with appealing dynamical properties, namely they are stable against perturbations~\cite{Brito:2015pxa,sanchis2017numerical} and they can form dynamically through a gravitational cooling mechanism~\cite{di2018dynamical}. 

Our investigation shows that the head-on collision of these spherically symmetric solutions may lead either to the formation of a more massive PS, which we dub ``hypermassive" PS or, if the initial PSs are sufficiently massive/compact, to the formation of a Schwarzschild BH. Horizon formation, however, only occurs after an intermediate phase, which leaves an imprint in the waveform, making it distinct from that of a head-on collision of Schwarzschild BHs. After horizon formation the BH QNMs match those of a Schwarzschild BH. However, we have found that for those cases where the final BH is surrounded by a sufficiently extended and long-lived quasi-stationary Proca cloud, differences with the BH ringdown are noticeable. 

In the orbital binary case, we have also observed two fates: $(i)$ the formation of a Proca star remnant, initially out of equilibrium and with angular momentum, but loosing such angular momentum as it approaches equilibrium; $(ii)$ the formation of a Kerr BH, also initially surrounded by a cloud of quasi-bound states of the Proca field, which also tends to be absorbed/scattered during the evolution. We have not found evidence for the formation of either rotating Proca stars or infinitely long-lived Proca hair around a rotating horizon. In the case of the rotating Proca stars, since these are thought to be perturbatively stable in some region of parameter space (see~\cite{Herdeiro:2015gia} for a discussion in the analogue case of rotating boson stars), it may be that the island of initial conditions leading to their formation has not yet been scanned by our simulations. In the near future we plan to scan the space of initial data, in particular initial velocities. For larger velocities,  obtaining reliable results requires constraint-preserving initial data. Obtaining such data and using it for performing further numerical evolutions is work underway.

%%%%%%%%%%%%%%%%%%%%%%%%%%%%%%%%%%%%%%%%%%%%%%%
\section*{Acknowledgements}
%%%%%%%%%%%%%%%%%%%%%%%%%%%%%%%%%%%%%%%%%%%%%%%

NSG thanks Miguel Zilh\~ao and Vassilios Mewes for useful discussions. This work has been supported by the Spanish MINECO (grant AYA2015-66899-C2-1-P), by the Generalitat Valenciana (ACIF/2015/216), by the FCT (Portugal) IF programme, by the FCT grant PTDC/FIS-OUT/28407/2017, by  CIDMA (FCT) strategic project UID/MAT/04106/2013, by CENTRA (FCT) strategic project UID/FIS/00099/2013 and by  the  European  Union's  Horizon  2020  research  and  innovation  (RISE) programmes H2020-MSCA-RISE-2015 Grant No.~StronGrHEP-690904 and H2020-MSCA-RISE-2017 Grant No.~FunFiCO-777740. The authors would like to acknowledge networking support by the COST Action CA16104.  Computations have been performed at the Servei d'Inform\`atica de la Universitat de Val\`encia.

\bigskip

\begin{appendix}
\section{Code assessment}\label{appendix}
\begin{figure}[t!]
\centering
\includegraphics[height=2.45in]{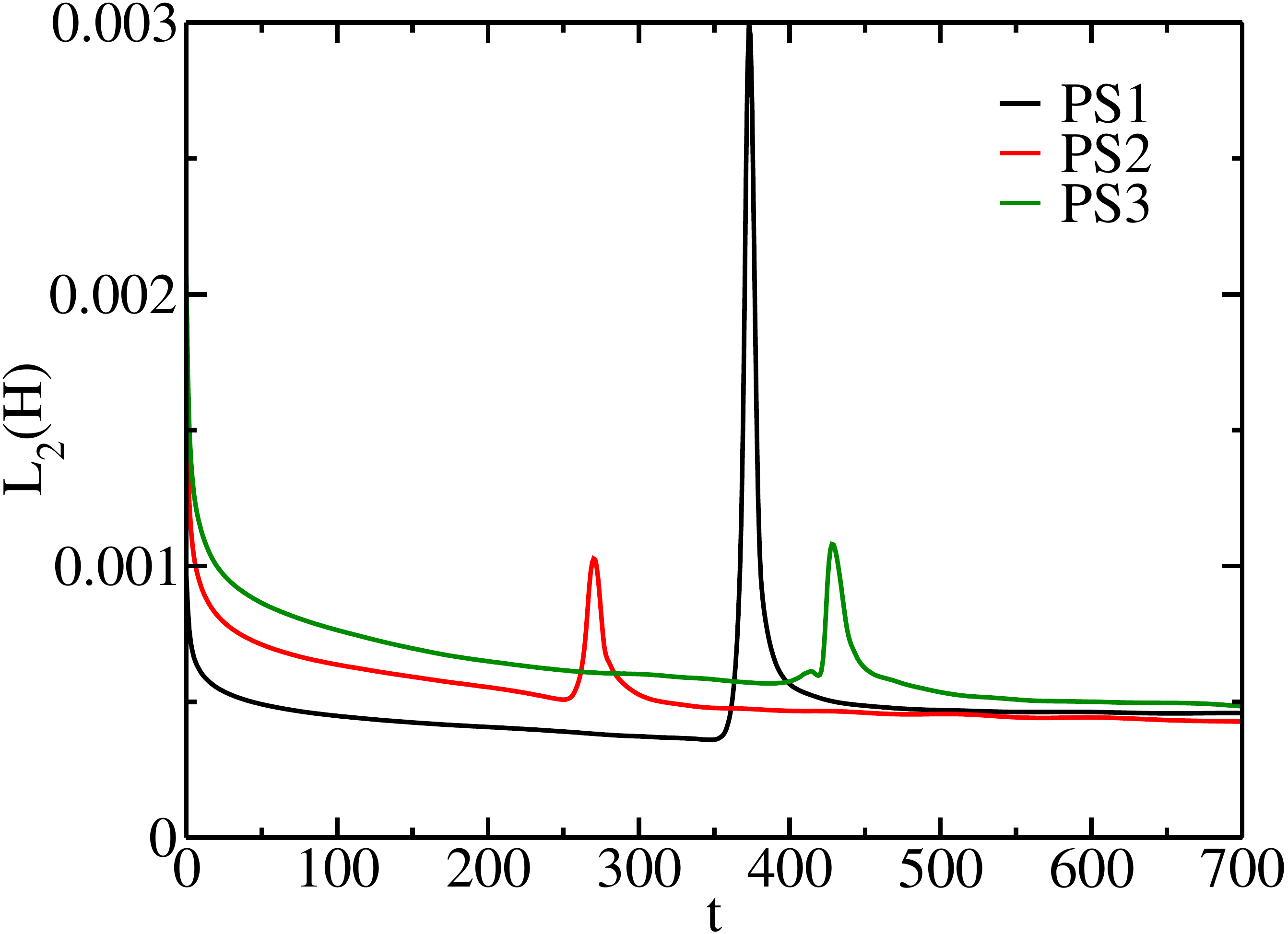}
\caption{$L_2$-norm of the Hamiltonian constraint for PS1-PS1, PS2-PS2 and PS3-PS3.
}
\label{fig10}
\end{figure}

We briefly comment here on the standard analyses we carried out to assess the quality of our simulations with the {\tt Einstein Toolkit}. In Fig.~\ref{fig10} we show the behaviour of the $L_{2}$-norm of the Hamiltonian constraint. The floor of the violation of this constraint is at $\sim 4\times 10^{-4}$ during most of the simulation. A significant peak in the violations of the constraint appears when the BH forms, but the errors quickly decrease to the pre-collision values and remain stable during the rest of the simulation.

\begin{figure}[t!]
\centering
 \includegraphics[height=2.5in]{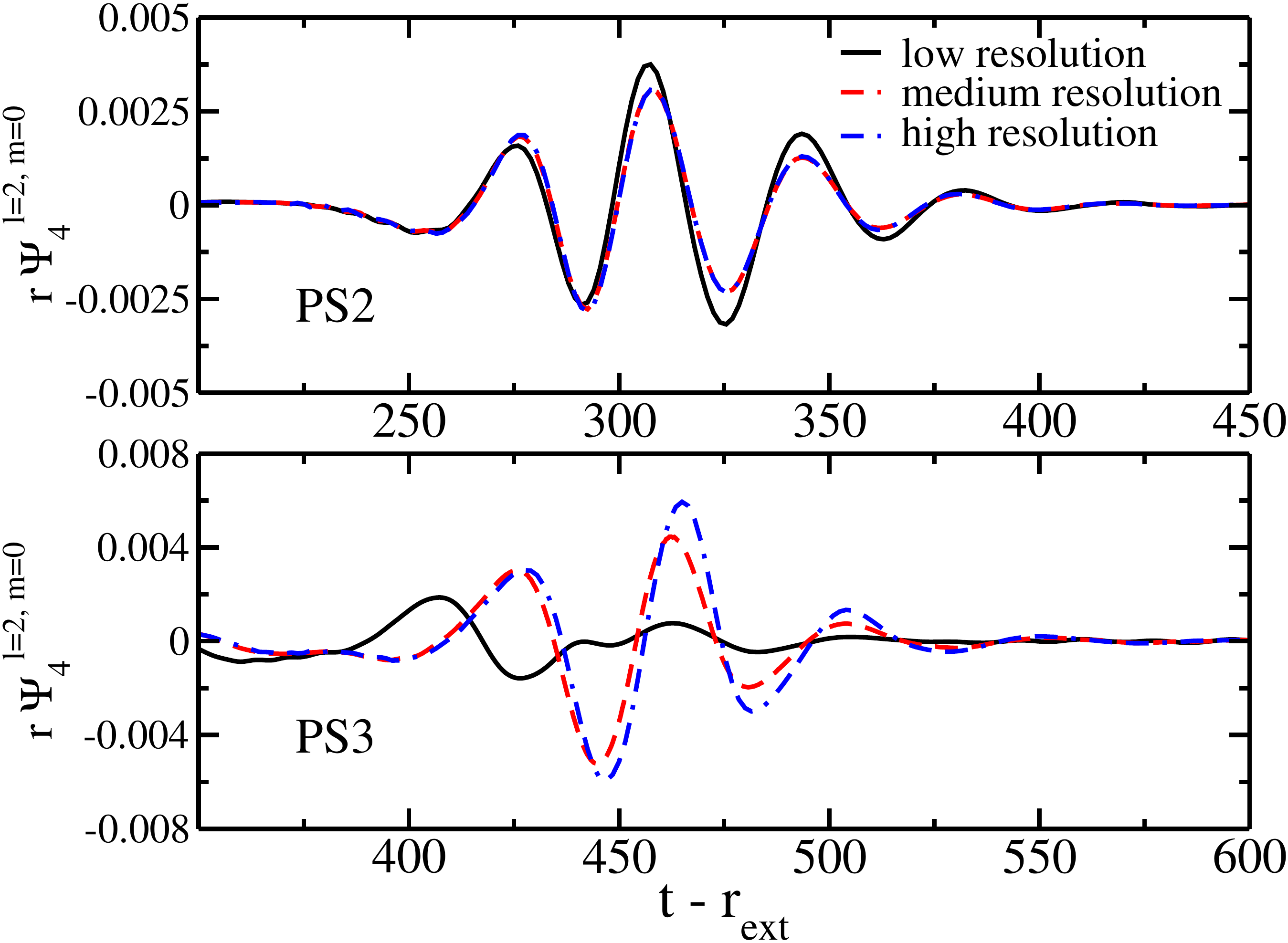}  
\caption{Resolution comparison for the gravitational wave extracted at $r_{\rm ext}=120$ for PS2-PS2 (top panel) and PS3-PS3 (bottom panel).
}
\label{fig11}
\end{figure}

In Fig.~\ref{fig11} we plot the gravitational wave of the PS2-PS2 and the PS3-PS3 cases for three different resolutions. The grid structure is $\lbrace(512, 64, 64, 32, 32, 8)$, $(8, 4, 2, 1, 0.5, 0.25)\rbrace$ and $\lbrace(512, 64, 64, 32, 32, 8)$, $(16, 8, 4, 2, 1, 0.5)\rbrace$ for medium and  low resolution, respectively.  The high resolution is the one used in the simulations shown in this paper. For PS2-PS2, the results for the three resolutions converge. For the PS3-PS3 model the low resolution is not good enough to allow us to extract the gravitational waveform. 
%The medium- and high resolution  provide adequate results but the convergence of the waveform is not as clear as in the PS2-PS2 case. \tf{Shall we add something to avoid possible doubts?} \nsg{I would change the last sentence to: " 
The medium and high resolutions provide adequate results showing that they are in the convergence regime. 

\begin{figure}
 \begin{center}
  \includegraphics[scale=0.3]{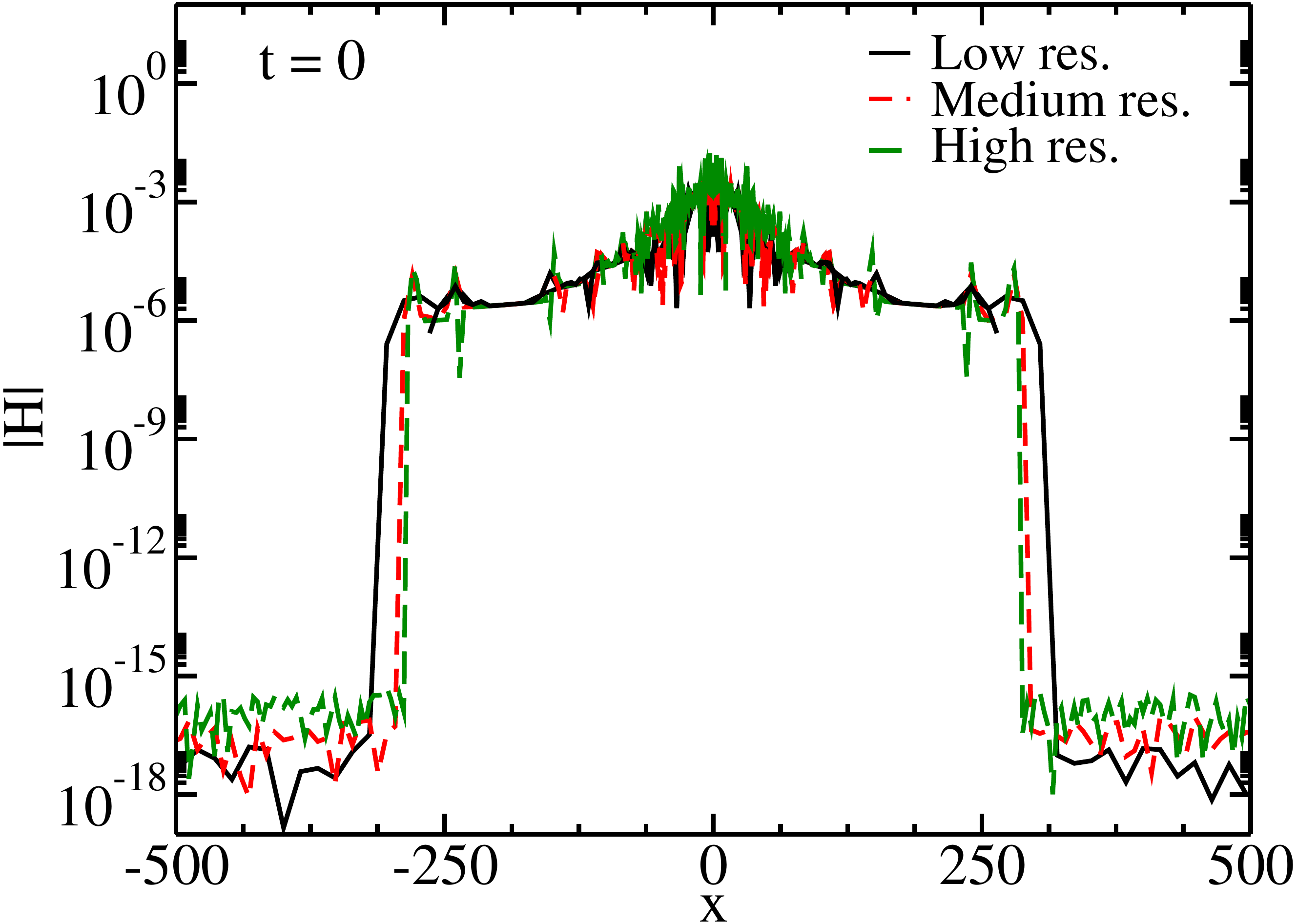}
 \includegraphics[scale=0.3]{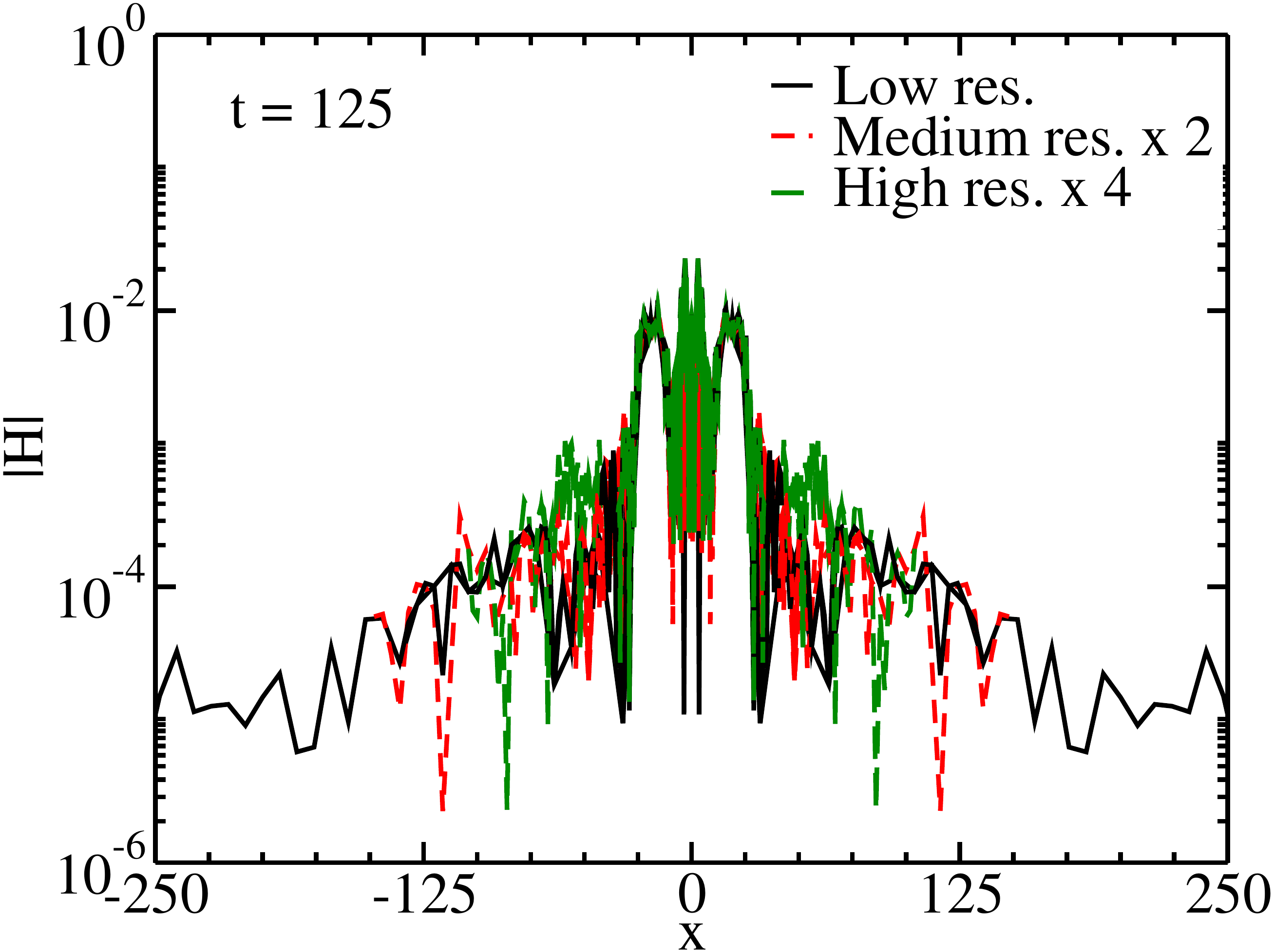}
  \includegraphics[scale=0.3]{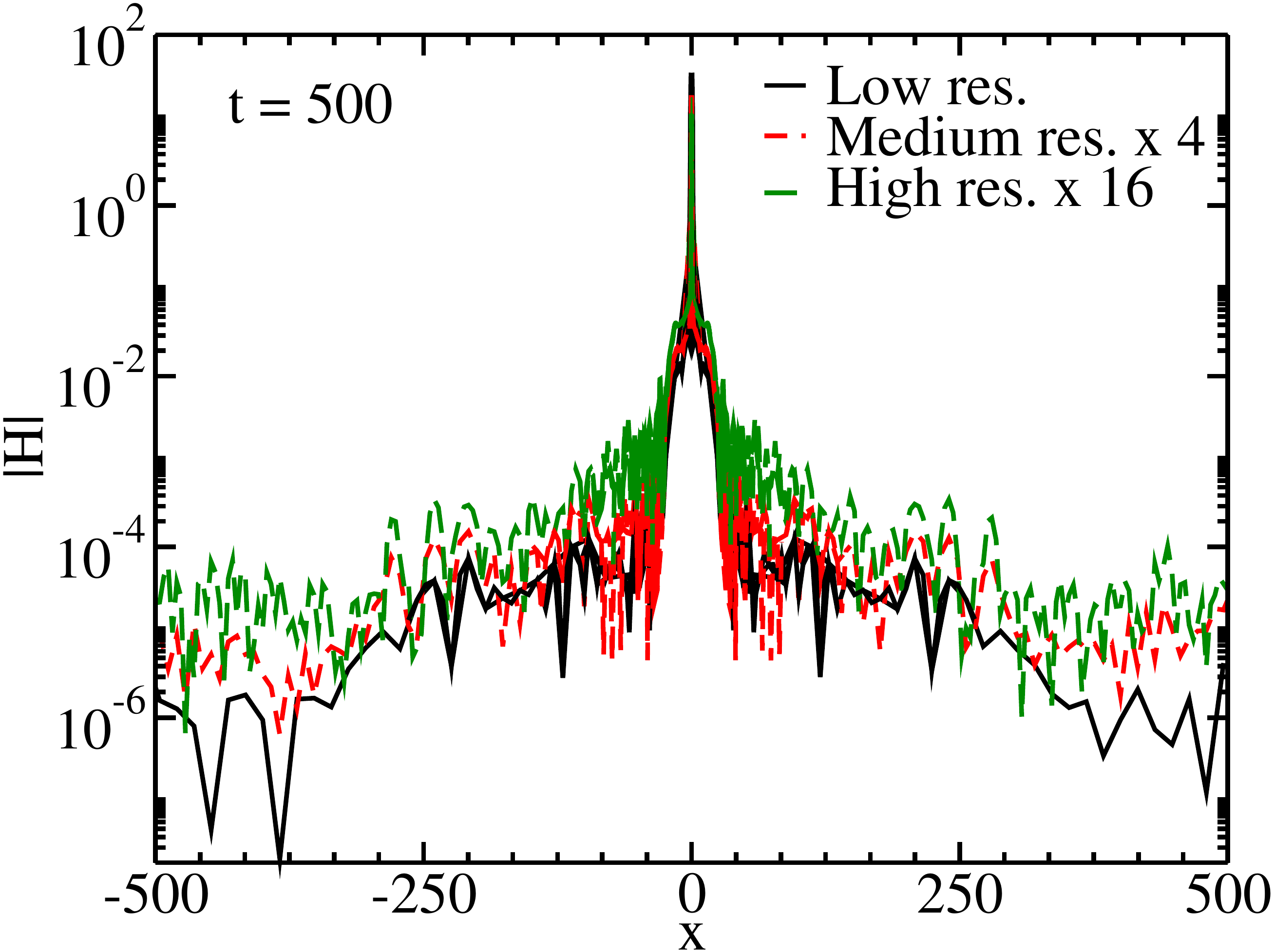}
 \caption{Hamiltonian constraint for PS2-PS2 at $t=0$ (top panel), $t=125$ (middle panel) and $t=500$ (bottom panel).}
 \label{fig:conv}
 \end{center}
 \end{figure}

Finally, in Fig~\ref{fig:conv} we plot the Hamiltonian constraint of the head-on collision of the PS2-PS2 model at $t=0$, $t=125$, and $t=500$. At $t=0$ the constraint violations do not converge with resolution (our initial data do not satisfy the constraints). At $t=125$ we obtain first-order convergence and, when the BH forms, a convergence of around second order is achieved. Therefore, we conclude that during the simulation we obtain between first- and second-order convergence. We note that the convergence order is greatly influenced by the linear second-order interpolation from the spherical grid of the Proca star solutions to the Cartesian grid we use to perform the evolutions with the \texttt{Einstein Toolkit}.

\end{appendix}

%\newpage

\bibliography{num-rel}

\begin{thebibliography}{54}
\expandafter\ifx\csname natexlab\endcsname\relax\def\natexlab#1{#1}\fi
\expandafter\ifx\csname bibnamefont\endcsname\relax
  \def\bibnamefont#1{#1}\fi
\expandafter\ifx\csname bibfnamefont\endcsname\relax
  \def\bibfnamefont#1{#1}\fi
\expandafter\ifx\csname citenamefont\endcsname\relax
  \def\citenamefont#1{#1}\fi
\expandafter\ifx\csname url\endcsname\relax
  \def\url#1{\texttt{#1}}\fi
\expandafter\ifx\csname urlprefix\endcsname\relax\def\urlprefix{URL }\fi
\providecommand{\bibinfo}[2]{#2}
\providecommand{\eprint}[2][]{\url{#2}}

\bibitem[{\citenamefont{{Abbott} et~al.}(2016)\citenamefont{{Abbott}, {Abbott},
  {Abbott}, {Abernathy}, {Acernese}, {Ackley}, {Adams}, {Adams}, {Addesso},
  {Adhikari} et~al.}}]{Abbott2016}
\bibinfo{author}{\bibfnamefont{B.~P.} \bibnamefont{{Abbott}}},
  \bibinfo{author}{\bibfnamefont{R.}~\bibnamefont{{Abbott}}},
  \bibinfo{author}{\bibfnamefont{T.~D.} \bibnamefont{{Abbott}}},
  \bibinfo{author}{\bibfnamefont{M.~R.} \bibnamefont{{Abernathy}}},
  \bibinfo{author}{\bibfnamefont{F.}~\bibnamefont{{Acernese}}},
  \bibinfo{author}{\bibfnamefont{K.}~\bibnamefont{{Ackley}}},
  \bibinfo{author}{\bibfnamefont{C.}~\bibnamefont{{Adams}}},
  \bibinfo{author}{\bibfnamefont{T.}~\bibnamefont{{Adams}}},
  \bibinfo{author}{\bibfnamefont{P.}~\bibnamefont{{Addesso}}},
  \bibinfo{author}{\bibfnamefont{R.~X.} \bibnamefont{{Adhikari}}},
  \bibnamefont{et~al.}, \bibinfo{journal}{Physical Review Letters}
  \textbf{\bibinfo{volume}{116}}, \bibinfo{eid}{061102} (\bibinfo{year}{2016}),
  \eprint{1602.03837}.

\bibitem[{\citenamefont{Abbott et~al.}(2016)}]{Abbott:2016nmj}
\bibinfo{author}{\bibfnamefont{B.~P.} \bibnamefont{Abbott}}
  \bibnamefont{et~al.} (\bibinfo{collaboration}{Virgo, LIGO Scientific}),
  \bibinfo{journal}{Phys. Rev. Lett.} \textbf{\bibinfo{volume}{116}},
  \bibinfo{pages}{241103} (\bibinfo{year}{2016}), \eprint{1606.04855}.

\bibitem[{\citenamefont{Abbott et~al.}(2017{\natexlab{a}})}]{Abbott:2017vtc}
\bibinfo{author}{\bibfnamefont{B.~P.} \bibnamefont{Abbott}}
  \bibnamefont{et~al.} (\bibinfo{collaboration}{VIRGO, LIGO Scientific}),
  \bibinfo{journal}{Phys. Rev. Lett.} \textbf{\bibinfo{volume}{118}},
  \bibinfo{pages}{221101} (\bibinfo{year}{2017}{\natexlab{a}}),
  \eprint{1706.01812}.

\bibitem[{\citenamefont{Abbott et~al.}(2017{\natexlab{b}})}]{Abbott:2017oio}
\bibinfo{author}{\bibfnamefont{B.~P.} \bibnamefont{Abbott}}
  \bibnamefont{et~al.} (\bibinfo{collaboration}{Virgo, LIGO Scientific}),
  \bibinfo{journal}{Phys. Rev. Lett.} \textbf{\bibinfo{volume}{119}},
  \bibinfo{pages}{141101} (\bibinfo{year}{2017}{\natexlab{b}}),
  \eprint{1709.09660}.

\bibitem[{\citenamefont{Abbott et~al.}(2017{\natexlab{c}})}]{Abbott:2017gyy}
\bibinfo{author}{\bibfnamefont{B.~P.} \bibnamefont{Abbott}}
  \bibnamefont{et~al.} (\bibinfo{collaboration}{Virgo, LIGO Scientific}),
  \bibinfo{journal}{Astrophys. J.} \textbf{\bibinfo{volume}{851}},
  \bibinfo{pages}{L35} (\bibinfo{year}{2017}{\natexlab{c}}),
  \eprint{1711.05578}.

\bibitem[{\citenamefont{Abbott
  et~al.}(2017{\natexlab{d}})}]{TheLIGOScientific:2017qsa}
\bibinfo{author}{\bibfnamefont{B.}~\bibnamefont{Abbott}} \bibnamefont{et~al.}
  (\bibinfo{collaboration}{Virgo, LIGO Scientific}), \bibinfo{journal}{Phys.
  Rev. Lett.} \textbf{\bibinfo{volume}{119}}, \bibinfo{pages}{161101}
  (\bibinfo{year}{2017}{\natexlab{d}}), \eprint{1710.05832}.

\bibitem[{\citenamefont{Barack et~al.}(2018)}]{Barack:2018yly}
\bibinfo{author}{\bibfnamefont{L.}~\bibnamefont{Barack}} \bibnamefont{et~al.}
  (\bibinfo{year}{2018}), \eprint{1806.05195}.

\bibitem[{\citenamefont{Johnson-Mcdaniel
  et~al.}(2018)\citenamefont{Johnson-Mcdaniel, Mukherjee, Kashyap, Ajith,
  Del~Pozzo, and Vitale}}]{Johnson-McDaniel:2018uvs}
\bibinfo{author}{\bibfnamefont{N.~K.} \bibnamefont{Johnson-Mcdaniel}},
  \bibinfo{author}{\bibfnamefont{A.}~\bibnamefont{Mukherjee}},
  \bibinfo{author}{\bibfnamefont{R.}~\bibnamefont{Kashyap}},
  \bibinfo{author}{\bibfnamefont{P.}~\bibnamefont{Ajith}},
  \bibinfo{author}{\bibfnamefont{W.}~\bibnamefont{Del~Pozzo}},
  \bibnamefont{and} \bibinfo{author}{\bibfnamefont{S.}~\bibnamefont{Vitale}}
  (\bibinfo{year}{2018}), \eprint{1804.08026}.

\bibitem[{\citenamefont{Schunck and Mielke}(2003)}]{Schunck:2003kk}
\bibinfo{author}{\bibfnamefont{F.~E.} \bibnamefont{Schunck}} \bibnamefont{and}
  \bibinfo{author}{\bibfnamefont{E.~W.} \bibnamefont{Mielke}},
  \bibinfo{journal}{Class. Quant. Grav.} \textbf{\bibinfo{volume}{20}},
  \bibinfo{pages}{R301} (\bibinfo{year}{2003}), \eprint{0801.0307}.

\bibitem[{\citenamefont{Seidel and Suen}(1991)}]{Seidel:1991zh}
\bibinfo{author}{\bibfnamefont{E.}~\bibnamefont{Seidel}} \bibnamefont{and}
  \bibinfo{author}{\bibfnamefont{W.~M.} \bibnamefont{Suen}},
  \bibinfo{journal}{Phys. Rev. Lett.} \textbf{\bibinfo{volume}{66}},
  \bibinfo{pages}{1659} (\bibinfo{year}{1991}).

\bibitem[{\citenamefont{Liebling and Palenzuela}(2012)}]{Liebling:2012fv}
\bibinfo{author}{\bibfnamefont{S.~L.} \bibnamefont{Liebling}} \bibnamefont{and}
  \bibinfo{author}{\bibfnamefont{C.}~\bibnamefont{Palenzuela}},
  \bibinfo{journal}{Living Rev. Rel.} \textbf{\bibinfo{volume}{15}},
  \bibinfo{pages}{6} (\bibinfo{year}{2012}), \eprint{1202.5809}.

\bibitem[{\citenamefont{Palenzuela et~al.}(2007)\citenamefont{Palenzuela,
  Olabarrieta, Lehner, and Liebling}}]{palenzuela2007head}
\bibinfo{author}{\bibfnamefont{C.}~\bibnamefont{Palenzuela}},
  \bibinfo{author}{\bibfnamefont{I.}~\bibnamefont{Olabarrieta}},
  \bibinfo{author}{\bibfnamefont{L.}~\bibnamefont{Lehner}}, \bibnamefont{and}
  \bibinfo{author}{\bibfnamefont{S.~L.} \bibnamefont{Liebling}},
  \bibinfo{journal}{Physical Review D} \textbf{\bibinfo{volume}{75}},
  \bibinfo{pages}{064005} (\bibinfo{year}{2007}).

\bibitem[{\citenamefont{Cardoso et~al.}(2016)\citenamefont{Cardoso, Hopper,
  Macedo, Palenzuela, and Pani}}]{cardoso2016gravitational}
\bibinfo{author}{\bibfnamefont{V.}~\bibnamefont{Cardoso}},
  \bibinfo{author}{\bibfnamefont{S.}~\bibnamefont{Hopper}},
  \bibinfo{author}{\bibfnamefont{C.~F.} \bibnamefont{Macedo}},
  \bibinfo{author}{\bibfnamefont{C.}~\bibnamefont{Palenzuela}},
  \bibnamefont{and} \bibinfo{author}{\bibfnamefont{P.}~\bibnamefont{Pani}},
  \bibinfo{journal}{Physical review D} \textbf{\bibinfo{volume}{94}},
  \bibinfo{pages}{084031} (\bibinfo{year}{2016}).

\bibitem[{\citenamefont{Brito et~al.}(2015)\citenamefont{Brito, Cardoso, and
  Okawa}}]{brito2015accretion}
\bibinfo{author}{\bibfnamefont{R.}~\bibnamefont{Brito}},
  \bibinfo{author}{\bibfnamefont{V.}~\bibnamefont{Cardoso}}, \bibnamefont{and}
  \bibinfo{author}{\bibfnamefont{H.}~\bibnamefont{Okawa}},
  \bibinfo{journal}{Physical review letters} \textbf{\bibinfo{volume}{115}},
  \bibinfo{pages}{111301} (\bibinfo{year}{2015}).

\bibitem[{\citenamefont{Brito et~al.}(2016{\natexlab{a}})\citenamefont{Brito,
  Cardoso, Macedo, Okawa, and Palenzuela}}]{brito2016interaction}
\bibinfo{author}{\bibfnamefont{R.}~\bibnamefont{Brito}},
  \bibinfo{author}{\bibfnamefont{V.}~\bibnamefont{Cardoso}},
  \bibinfo{author}{\bibfnamefont{C.~F.} \bibnamefont{Macedo}},
  \bibinfo{author}{\bibfnamefont{H.}~\bibnamefont{Okawa}}, \bibnamefont{and}
  \bibinfo{author}{\bibfnamefont{C.}~\bibnamefont{Palenzuela}},
  \bibinfo{journal}{Physical Review D} \textbf{\bibinfo{volume}{93}},
  \bibinfo{pages}{044045} (\bibinfo{year}{2016}{\natexlab{a}}).

\bibitem[{\citenamefont{Helfer et~al.}(2018)\citenamefont{Helfer, Lim, Garcia,
  and Amin}}]{helfer2018gravitational}
\bibinfo{author}{\bibfnamefont{T.}~\bibnamefont{Helfer}},
  \bibinfo{author}{\bibfnamefont{E.~A.} \bibnamefont{Lim}},
  \bibinfo{author}{\bibfnamefont{M.~A.} \bibnamefont{Garcia}},
  \bibnamefont{and} \bibinfo{author}{\bibfnamefont{M.~A.} \bibnamefont{Amin}},
  \bibinfo{journal}{arXiv preprint arXiv:1802.06733}  (\bibinfo{year}{2018}).

\bibitem[{\citenamefont{Bezares
  et~al.}(2017{\natexlab{a}})\citenamefont{Bezares, Palenzuela, and
  Bona}}]{Bezares:2017mzk}
\bibinfo{author}{\bibfnamefont{M.}~\bibnamefont{Bezares}},
  \bibinfo{author}{\bibfnamefont{C.}~\bibnamefont{Palenzuela}},
  \bibnamefont{and} \bibinfo{author}{\bibfnamefont{C.}~\bibnamefont{Bona}},
  \bibinfo{journal}{Phys. Rev.} \textbf{\bibinfo{volume}{D95}},
  \bibinfo{pages}{124005} (\bibinfo{year}{2017}{\natexlab{a}}),
  \eprint{1705.01071}.

\bibitem[{\citenamefont{Palenzuela et~al.}(2017)\citenamefont{Palenzuela, Pani,
  Bezares, Cardoso, Lehner, and Liebling}}]{Palenzuela:2017kcg}
\bibinfo{author}{\bibfnamefont{C.}~\bibnamefont{Palenzuela}},
  \bibinfo{author}{\bibfnamefont{P.}~\bibnamefont{Pani}},
  \bibinfo{author}{\bibfnamefont{M.}~\bibnamefont{Bezares}},
  \bibinfo{author}{\bibfnamefont{V.}~\bibnamefont{Cardoso}},
  \bibinfo{author}{\bibfnamefont{L.}~\bibnamefont{Lehner}}, \bibnamefont{and}
  \bibinfo{author}{\bibfnamefont{S.}~\bibnamefont{Liebling}},
  \bibinfo{journal}{Phys. Rev.} \textbf{\bibinfo{volume}{D96}},
  \bibinfo{pages}{104058} (\bibinfo{year}{2017}), \eprint{1710.09432}.

\bibitem[{\citenamefont{Dietrich et~al.}(2018)\citenamefont{Dietrich, Ossokine,
  and Clough}}]{dietrich2018full}
\bibinfo{author}{\bibfnamefont{T.}~\bibnamefont{Dietrich}},
  \bibinfo{author}{\bibfnamefont{S.}~\bibnamefont{Ossokine}}, \bibnamefont{and}
  \bibinfo{author}{\bibfnamefont{K.}~\bibnamefont{Clough}},
  \bibinfo{journal}{arXiv preprint arXiv:1807.06959}  (\bibinfo{year}{2018}).

\bibitem[{\citenamefont{Clough et~al.}(2018)\citenamefont{Clough, Dietrich, and
  Niemeyer}}]{clough2018axion}
\bibinfo{author}{\bibfnamefont{K.}~\bibnamefont{Clough}},
  \bibinfo{author}{\bibfnamefont{T.}~\bibnamefont{Dietrich}}, \bibnamefont{and}
  \bibinfo{author}{\bibfnamefont{J.~C.} \bibnamefont{Niemeyer}},
  \bibinfo{journal}{Physical Review D} \textbf{\bibinfo{volume}{98}},
  \bibinfo{pages}{083020} (\bibinfo{year}{2018}).

\bibitem[{\citenamefont{Bezares and Palenzuela}(2018)}]{Bezares:2018qwa}
\bibinfo{author}{\bibfnamefont{M.}~\bibnamefont{Bezares}} \bibnamefont{and}
  \bibinfo{author}{\bibfnamefont{C.}~\bibnamefont{Palenzuela}},
  \bibinfo{journal}{Classical and Quantum Gravity}  (\bibinfo{year}{2018}).

\bibitem[{\citenamefont{Palenzuela et~al.}(2008)\citenamefont{Palenzuela,
  Lehner, and Liebling}}]{palenzuela2008orbital}
\bibinfo{author}{\bibfnamefont{C.}~\bibnamefont{Palenzuela}},
  \bibinfo{author}{\bibfnamefont{L.}~\bibnamefont{Lehner}}, \bibnamefont{and}
  \bibinfo{author}{\bibfnamefont{S.~L.} \bibnamefont{Liebling}},
  \bibinfo{journal}{Physical Review D} \textbf{\bibinfo{volume}{77}},
  \bibinfo{pages}{044036} (\bibinfo{year}{2008}).

\bibitem[{\citenamefont{Bezares
  et~al.}(2017{\natexlab{b}})\citenamefont{Bezares, Palenzuela, and
  Bona}}]{bezares2017final}
\bibinfo{author}{\bibfnamefont{M.}~\bibnamefont{Bezares}},
  \bibinfo{author}{\bibfnamefont{C.}~\bibnamefont{Palenzuela}},
  \bibnamefont{and} \bibinfo{author}{\bibfnamefont{C.}~\bibnamefont{Bona}},
  \bibinfo{journal}{Physical Review D} \textbf{\bibinfo{volume}{95}},
  \bibinfo{pages}{124005} (\bibinfo{year}{2017}{\natexlab{b}}).

\bibitem[{\citenamefont{Herdeiro and Radu}(2014)}]{Herdeiro:2014goa}
\bibinfo{author}{\bibfnamefont{C.~A.~R.} \bibnamefont{Herdeiro}}
  \bibnamefont{and} \bibinfo{author}{\bibfnamefont{E.}~\bibnamefont{Radu}},
  \bibinfo{journal}{Phys.Rev.Lett.} \textbf{\bibinfo{volume}{112}},
  \bibinfo{pages}{221101} (\bibinfo{year}{2014}), \eprint{1403.2757}.

\bibitem[{\citenamefont{Herdeiro and Radu}(2015)}]{Herdeiro:2015gia}
\bibinfo{author}{\bibfnamefont{C.}~\bibnamefont{Herdeiro}} \bibnamefont{and}
  \bibinfo{author}{\bibfnamefont{E.}~\bibnamefont{Radu}},
  \bibinfo{journal}{Class. Quant. Grav.} \textbf{\bibinfo{volume}{32}},
  \bibinfo{pages}{144001} (\bibinfo{year}{2015}), \eprint{1501.04319}.

\bibitem[{\citenamefont{Brito et~al.}(2016{\natexlab{b}})\citenamefont{Brito,
  Cardoso, Herdeiro, and Radu}}]{Brito:2015pxa}
\bibinfo{author}{\bibfnamefont{R.}~\bibnamefont{Brito}},
  \bibinfo{author}{\bibfnamefont{V.}~\bibnamefont{Cardoso}},
  \bibinfo{author}{\bibfnamefont{C.~A.~R.} \bibnamefont{Herdeiro}},
  \bibnamefont{and} \bibinfo{author}{\bibfnamefont{E.}~\bibnamefont{Radu}},
  \bibinfo{journal}{Phys. Lett.} \textbf{\bibinfo{volume}{B752}},
  \bibinfo{pages}{291} (\bibinfo{year}{2016}{\natexlab{b}}),
  \eprint{1508.05395}.

\bibitem[{\citenamefont{Herdeiro et~al.}(2016)\citenamefont{Herdeiro, Radu, and
  Runarsson}}]{Herdeiro:2016tmi}
\bibinfo{author}{\bibfnamefont{C.}~\bibnamefont{Herdeiro}},
  \bibinfo{author}{\bibfnamefont{E.}~\bibnamefont{Radu}}, \bibnamefont{and}
  \bibinfo{author}{\bibfnamefont{H.}~\bibnamefont{Runarsson}},
  \bibinfo{journal}{Class. Quant. Grav.} \textbf{\bibinfo{volume}{33}},
  \bibinfo{pages}{154001} (\bibinfo{year}{2016}), \eprint{1603.02687}.

\bibitem[{\citenamefont{García and Landea}(2016)}]{Garcia:2016ldc}
\bibinfo{author}{\bibfnamefont{F.}~\bibnamefont{García}} \bibnamefont{and}
  \bibinfo{author}{\bibfnamefont{I.~S.} \bibnamefont{Landea}}
  (\bibinfo{year}{2016}), \eprint{1608.00011}.

\bibitem[{\citenamefont{Duarte and Brito}(2016)}]{Duarte:2016lig}
\bibinfo{author}{\bibfnamefont{M.}~\bibnamefont{Duarte}} \bibnamefont{and}
  \bibinfo{author}{\bibfnamefont{R.}~\bibnamefont{Brito}},
  \bibinfo{journal}{Phys. Rev.} \textbf{\bibinfo{volume}{D94}},
  \bibinfo{pages}{064055} (\bibinfo{year}{2016}), \eprint{1609.01735}.

\bibitem[{\citenamefont{Minamitsuji}(2018)}]{Minamitsuji:2018kof}
\bibinfo{author}{\bibfnamefont{M.}~\bibnamefont{Minamitsuji}},
  \bibinfo{journal}{Phys. Rev.} \textbf{\bibinfo{volume}{D97}},
  \bibinfo{pages}{104023} (\bibinfo{year}{2018}), \eprint{1805.09867}.

\bibitem[{\citenamefont{Cunha et~al.}(2017)\citenamefont{Cunha, Font, Herdeiro,
  Radu, Sanchis-Gual, and Zilhão}}]{Cunha:2017wao}
\bibinfo{author}{\bibfnamefont{P.~V.~P.} \bibnamefont{Cunha}},
  \bibinfo{author}{\bibfnamefont{J.~A.} \bibnamefont{Font}},
  \bibinfo{author}{\bibfnamefont{C.}~\bibnamefont{Herdeiro}},
  \bibinfo{author}{\bibfnamefont{E.}~\bibnamefont{Radu}},
  \bibinfo{author}{\bibfnamefont{N.}~\bibnamefont{Sanchis-Gual}},
  \bibnamefont{and} \bibinfo{author}{\bibfnamefont{M.}~\bibnamefont{Zilhão}},
  \bibinfo{journal}{Phys. Rev.} \textbf{\bibinfo{volume}{D96}},
  \bibinfo{pages}{104040} (\bibinfo{year}{2017}), \eprint{1709.06118}.

\bibitem[{\citenamefont{Witek et~al.}(2013)\citenamefont{Witek, Cardoso,
  Ishibashi, and Sperhake}}]{witek2013superradiant}
\bibinfo{author}{\bibfnamefont{H.}~\bibnamefont{Witek}},
  \bibinfo{author}{\bibfnamefont{V.}~\bibnamefont{Cardoso}},
  \bibinfo{author}{\bibfnamefont{A.}~\bibnamefont{Ishibashi}},
  \bibnamefont{and} \bibinfo{author}{\bibfnamefont{U.}~\bibnamefont{Sperhake}},
  \bibinfo{journal}{Physical Review D} \textbf{\bibinfo{volume}{87}},
  \bibinfo{pages}{043513} (\bibinfo{year}{2013}).

\bibitem[{\citenamefont{Zilh\~ao et~al.}(2015)\citenamefont{Zilh\~ao, Witek,
  and Cardoso}}]{Zilhao:2015tya}
\bibinfo{author}{\bibfnamefont{M.}~\bibnamefont{Zilh\~ao}},
  \bibinfo{author}{\bibfnamefont{H.}~\bibnamefont{Witek}}, \bibnamefont{and}
  \bibinfo{author}{\bibfnamefont{V.}~\bibnamefont{Cardoso}},
  \bibinfo{journal}{Class. Quant. Grav.} \textbf{\bibinfo{volume}{32}},
  \bibinfo{pages}{234003} (\bibinfo{year}{2015}), \eprint{1505.00797}.

\bibitem[{\citenamefont{Rosa and Dolan}(2012)}]{Rosa:2011my}
\bibinfo{author}{\bibfnamefont{J.~G.} \bibnamefont{Rosa}} \bibnamefont{and}
  \bibinfo{author}{\bibfnamefont{S.~R.} \bibnamefont{Dolan}},
  \bibinfo{journal}{Phys. Rev.} \textbf{\bibinfo{volume}{D85}},
  \bibinfo{pages}{044043} (\bibinfo{year}{2012}), \eprint{1110.4494}.

\bibitem[{\citenamefont{East and Pretorius}(2017)}]{east2017superradiant1}
\bibinfo{author}{\bibfnamefont{W.~E.} \bibnamefont{East}} \bibnamefont{and}
  \bibinfo{author}{\bibfnamefont{F.}~\bibnamefont{Pretorius}},
  \bibinfo{journal}{Physical review letters} \textbf{\bibinfo{volume}{119}},
  \bibinfo{pages}{041101} (\bibinfo{year}{2017}).

\bibitem[{\citenamefont{East}(2017)}]{east2017superradiant2}
\bibinfo{author}{\bibfnamefont{W.~E.} \bibnamefont{East}},
  \bibinfo{journal}{Physical Review D} \textbf{\bibinfo{volume}{96}},
  \bibinfo{pages}{024004} (\bibinfo{year}{2017}).

\bibitem[{\citenamefont{Herdeiro and Radu}(2017)}]{Herdeiro:2017phl}
\bibinfo{author}{\bibfnamefont{C.~A.~R.} \bibnamefont{Herdeiro}}
  \bibnamefont{and} \bibinfo{author}{\bibfnamefont{E.}~\bibnamefont{Radu}},
  \bibinfo{journal}{Phys. Rev. Lett.} \textbf{\bibinfo{volume}{119}},
  \bibinfo{pages}{261101} (\bibinfo{year}{2017}), \eprint{1706.06597}.

\bibitem[{\citenamefont{Sanchis-Gual et~al.}(2017)\citenamefont{Sanchis-Gual,
  Herdeiro, Radu, Degollado, and Font}}]{sanchis2017numerical}
\bibinfo{author}{\bibfnamefont{N.}~\bibnamefont{Sanchis-Gual}},
  \bibinfo{author}{\bibfnamefont{C.}~\bibnamefont{Herdeiro}},
  \bibinfo{author}{\bibfnamefont{E.}~\bibnamefont{Radu}},
  \bibinfo{author}{\bibfnamefont{J.~C.} \bibnamefont{Degollado}},
  \bibnamefont{and} \bibinfo{author}{\bibfnamefont{J.~A.} \bibnamefont{Font}},
  \bibinfo{journal}{Physical Review D} \textbf{\bibinfo{volume}{95}},
  \bibinfo{pages}{104028} (\bibinfo{year}{2017}).

\bibitem[{\citenamefont{Seidel and Suen}(1990)}]{Seidel:1990jh}
\bibinfo{author}{\bibfnamefont{E.}~\bibnamefont{Seidel}} \bibnamefont{and}
  \bibinfo{author}{\bibfnamefont{W.-M.} \bibnamefont{Suen}},
  \bibinfo{journal}{Phys. Rev.} \textbf{\bibinfo{volume}{D42}},
  \bibinfo{pages}{384} (\bibinfo{year}{1990}).

\bibitem[{\citenamefont{Balakrishna et~al.}(1998)\citenamefont{Balakrishna,
  Seidel, and Suen}}]{Balakrishna:1997ej}
\bibinfo{author}{\bibfnamefont{J.}~\bibnamefont{Balakrishna}},
  \bibinfo{author}{\bibfnamefont{E.}~\bibnamefont{Seidel}}, \bibnamefont{and}
  \bibinfo{author}{\bibfnamefont{W.-M.} \bibnamefont{Suen}},
  \bibinfo{journal}{Phys. Rev.} \textbf{\bibinfo{volume}{D58}},
  \bibinfo{pages}{104004} (\bibinfo{year}{1998}), \eprint{gr-qc/9712064}.

\bibitem[{\citenamefont{Guzman}(2004)}]{Guzman:2004jw}
\bibinfo{author}{\bibfnamefont{F.~S.} \bibnamefont{Guzman}},
  \bibinfo{journal}{Phys. Rev.} \textbf{\bibinfo{volume}{D70}},
  \bibinfo{pages}{044033} (\bibinfo{year}{2004}), \eprint{gr-qc/0407054}.

\bibitem[{\citenamefont{Escorihuela-Tom{\`a}s
  et~al.}(2017)\citenamefont{Escorihuela-Tom{\`a}s, Sanchis-Gual, Degollado,
  and Font}}]{escorihuela2017quasistationary}
\bibinfo{author}{\bibfnamefont{A.}~\bibnamefont{Escorihuela-Tom{\`a}s}},
  \bibinfo{author}{\bibfnamefont{N.}~\bibnamefont{Sanchis-Gual}},
  \bibinfo{author}{\bibfnamefont{J.~C.} \bibnamefont{Degollado}},
  \bibnamefont{and} \bibinfo{author}{\bibfnamefont{J.~A.} \bibnamefont{Font}},
  \bibinfo{journal}{Physical Review D} \textbf{\bibinfo{volume}{96}},
  \bibinfo{pages}{024015} (\bibinfo{year}{2017}).

\bibitem[{\citenamefont{Di~Giovanni et~al.}(2018)\citenamefont{Di~Giovanni,
  Sanchis-Gual, Herdeiro, and Font}}]{di2018dynamical}
\bibinfo{author}{\bibfnamefont{F.}~\bibnamefont{Di~Giovanni}},
  \bibinfo{author}{\bibfnamefont{N.}~\bibnamefont{Sanchis-Gual}},
  \bibinfo{author}{\bibfnamefont{C.~A.~R.} \bibnamefont{Herdeiro}},
  \bibnamefont{and} \bibinfo{author}{\bibfnamefont{J.~A.} \bibnamefont{Font}},
  \bibinfo{journal}{Phys. Rev.} \textbf{\bibinfo{volume}{D98}},
  \bibinfo{pages}{064044} (\bibinfo{year}{2018}), \eprint{1803.04802}.

\bibitem[{\citenamefont{Seidel and Suen}(1994)}]{seidel1994formation}
\bibinfo{author}{\bibfnamefont{E.}~\bibnamefont{Seidel}} \bibnamefont{and}
  \bibinfo{author}{\bibfnamefont{W.-M.} \bibnamefont{Suen}},
  \bibinfo{journal}{Physical review letters} \textbf{\bibinfo{volume}{72}},
  \bibinfo{pages}{2516} (\bibinfo{year}{1994}).

\bibitem[{\citenamefont{Shibata et~al.}(2008)\citenamefont{Shibata, Okawa, and
  Yamamoto}}]{shibata2008high}
\bibinfo{author}{\bibfnamefont{M.}~\bibnamefont{Shibata}},
  \bibinfo{author}{\bibfnamefont{H.}~\bibnamefont{Okawa}}, \bibnamefont{and}
  \bibinfo{author}{\bibfnamefont{T.}~\bibnamefont{Yamamoto}},
  \bibinfo{journal}{Physical Review D} \textbf{\bibinfo{volume}{78}},
  \bibinfo{pages}{101501} (\bibinfo{year}{2008}).

\bibitem[{\citenamefont{Toolkit}(2012)}]{toolkit2012open}
\bibinfo{author}{\bibfnamefont{E.}~\bibnamefont{Toolkit}},
  \bibinfo{journal}{URL http://einsteintoolkit. org}  (\bibinfo{year}{2012}).

\bibitem[{\citenamefont{L{\"o}ffler}(2012)}]{loffler2012f}
\bibinfo{author}{\bibfnamefont{F.}~\bibnamefont{L{\"o}ffler}},
  \bibinfo{journal}{Classical Quantum Gravity} \textbf{\bibinfo{volume}{29}},
  \bibinfo{pages}{115001} (\bibinfo{year}{2012}).

\bibitem[{\citenamefont{Brown et~al.}(2009)\citenamefont{Brown, Diener,
  Sarbach, Schnetter, and Tiglio}}]{brown2009turduckening}
\bibinfo{author}{\bibfnamefont{D.}~\bibnamefont{Brown}},
  \bibinfo{author}{\bibfnamefont{P.}~\bibnamefont{Diener}},
  \bibinfo{author}{\bibfnamefont{O.}~\bibnamefont{Sarbach}},
  \bibinfo{author}{\bibfnamefont{E.}~\bibnamefont{Schnetter}},
  \bibnamefont{and} \bibinfo{author}{\bibfnamefont{M.}~\bibnamefont{Tiglio}},
  \bibinfo{journal}{Physical Review D} \textbf{\bibinfo{volume}{79}},
  \bibinfo{pages}{044023} (\bibinfo{year}{2009}).

\bibitem[{\citenamefont{Reisswig et~al.}(2011)\citenamefont{Reisswig, Ott,
  Sperhake, and Schnetter}}]{reisswig2011gravitational}
\bibinfo{author}{\bibfnamefont{C.}~\bibnamefont{Reisswig}},
  \bibinfo{author}{\bibfnamefont{C.~D.} \bibnamefont{Ott}},
  \bibinfo{author}{\bibfnamefont{U.}~\bibnamefont{Sperhake}}, \bibnamefont{and}
  \bibinfo{author}{\bibfnamefont{E.}~\bibnamefont{Schnetter}},
  \bibinfo{journal}{Physical Review D} \textbf{\bibinfo{volume}{83}},
  \bibinfo{pages}{064008} (\bibinfo{year}{2011}).

\bibitem[{web(2018)}]{webpage}
\emph{\bibinfo{title}{Head on collisions videos}}, \bibinfo{howpublished}{\url{
  http://gravitation.web.ua.pt/node/1522}} (\bibinfo{year}{2018}).

\bibitem[{\citenamefont{Leaver}(1985)}]{Leaver:1985ax}
\bibinfo{author}{\bibfnamefont{E.}~\bibnamefont{Leaver}},
  \bibinfo{journal}{Proc.Roy.Soc.Lond.} \textbf{\bibinfo{volume}{A402}},
  \bibinfo{pages}{285} (\bibinfo{year}{1985}).

\bibitem[{\citenamefont{Papadopoulos and Font}(1999)}]{papadopoulos1999matter}
\bibinfo{author}{\bibfnamefont{P.}~\bibnamefont{Papadopoulos}}
  \bibnamefont{and} \bibinfo{author}{\bibfnamefont{J.~A.} \bibnamefont{Font}},
  \bibinfo{journal}{Physical Review D} \textbf{\bibinfo{volume}{59}},
  \bibinfo{pages}{044014} (\bibinfo{year}{1999}).

\bibitem[{\citenamefont{{Papadopoulos} and {Font}}(2001)}]{PF2}
\bibinfo{author}{\bibfnamefont{P.}~\bibnamefont{{Papadopoulos}}}
  \bibnamefont{and} \bibinfo{author}{\bibfnamefont{J.~A.}
  \bibnamefont{{Font}}}, \bibinfo{journal}{\prd} \textbf{\bibinfo{volume}{63}},
  \bibinfo{eid}{044016} (\bibinfo{year}{2001}), \eprint{gr-qc/0009024}.

\bibitem[{\citenamefont{{Sun} and {Price}}(1990)}]{sun-price-1990}
\bibinfo{author}{\bibfnamefont{Y.}~\bibnamefont{{Sun}}} \bibnamefont{and}
  \bibinfo{author}{\bibfnamefont{R.~H.} \bibnamefont{{Price}}},
  \bibinfo{journal}{\prd} \textbf{\bibinfo{volume}{41}}, \bibinfo{pages}{2492}
  (\bibinfo{year}{1990}).

\end{thebibliography}

\end{document}